\documentclass[12pt]{article}
\usepackage[slantedGreek]{mathptmx}
\usepackage{fullpage}
\usepackage{amssymb, amsmath,bm}
\usepackage{caption, rotating}
\usepackage[footnotesize]{subfigure}

\newcommand{\diff}{\text{d}}
\newcommand{\ua}{^\alpha {} }
\newcommand{\ub}{^\beta {} }

\numberwithin{equation}{section}
\usepackage[numbers,square,comma,sort&compress]{natbib}
\graphicspath{{./figures/}} 

\title{Nonlinear soft-tissue elasticity, remodeling, and degradation described by an extended Finsler geometry} 

\author{J.D. Clayton$^{1}$\\ \\
$^1$Terminal Effects Division, Army Research Directorate
\\
DEVCOM ARL, Aberdeen, MD 21005, USA 
\\
*Email: john.d.clayton1.civ@army.mil
}

\date{}

\begin{document}
\graphicspath{{figures/}}
\maketitle

\begin{abstract}
A continuum mechanical theory incorporating an extension of Finsler geometry is
formulated for fibrous soft solids.
Especially if of biologic origin, such solids are nonlinear elastic with evolving microstructures. For example, elongated cells or collagen fibers can stretch and rotate independently of motions of their embedding matrix.
Here, a director vector or internal state vector, not always of unit length,
in generalized Finsler space relates to a physical mechanism, with possible preferred direction and intensity, in the microstructure.
Classical Finsler geometry is extended to accommodate
multiple director vectors (i.e., multiple fibers in both a differential-geometric and physical sense) at each point on the base
manifold. A metric tensor can depend on the ensemble of director vector fields.
Residual or remnant strains from biologic growth, remodeling, and degradation manifest as non-affine fiber and matrix stretches.
These remnant stretch fields are quantified by internal state vectors and a corresponding, generally non-Euclidean, metric tensor.
Euler-Lagrange equations derived from a variational principle
yield equilibrium configurations satisfying balances of forces from 
 elastic energy, remodeling and cohesive energies, and external chemical-biological interactions. Given certain assumptions, the model can reduce to a representation in Riemannian geometry. Residual stresses that emerge from a non-Euclidean material metric in the Riemannian setting are implicitly included in the Finslerian setting.
The theory is used to study stress and damage in the ventricle (heart muscle) expanding or contracting under internal and external pressure. 
Remnant strains from remodeling can reduce stress concentrations and mitigate tissue damage under severe loading.

 \end{abstract}
\noindent \textbf{Key words}: soft condensed matter; nonlinear elasticity; differential geometry; Finsler geometry; phase field; biological tissue; ventricle; residual stress; remodeling; fracture\\ \\
\noindent \textbf{Mathematics Subject Classification (MSC) 2020}: 53Z05 (primary), 53B40, 74B20

\noindent

 \tableofcontents

\section{Introduction}
Biological tissues exhibit complex microstructures containing
a mixture of cells, collagen and elastin fibers, ground substance, blood vessels,  
and various liquids, for example blood and interstitial fluids. Examples include
skin, muscle, cartilage, tendons, and internal organs such as the heart, liver, kidneys, and lungs. 
Background from a mechanics perspective can be found in books on the subject \cite{fung1990,fung1993,cowin2007,epstein2012,goriely2017}.
In some cases, cells comprise ``fibers'' independent
from the collagen network, for example myocytes containing contractile elements of skeletal and cardiac muscle \cite{fung1993,cowin2007,holz2009}.

Mechanical behaviors can include nonlinear elasticity, viscoelasticity, and if loading is severe enough, ``damage'' in the form of degradation, possibly contributing to remnant strain.
Transient responses can 
be affected by relative flow of fluids with respect to motion of solid constituents, as in poromechanics,
though the latter can be difficult to discern from viscoelasticity. Mechanisms are generally anisotropic, but some tissues show nearly isotropic response (e.g., liver and lung parenchyma) depending on dimensions of specimens tested. Nonlinear elastic models are surveyed in a review article \cite{chagnon2015},
and representative viscoelastic or dissipative rate-dependent models for soft tissues are described elsewhere \cite{fung1993,rubin2002,freed2014,gultekin2016}.  Poromechanical or mixture-theoretical \cite{bowen1976} approaches have been devoted to cartilage, skin, liver, heart, and lung \cite{mow1980,oomens1987,yang1991,ricken2010,claytonIJES2022,irwin2024}.
Models for tissue damage, whether resolved at the homogenized continuum scale (mm) or the fiber scale ($\mu$m),
tend to be phenomenological \cite{rodriguez2006,alastrue2007,balzani2012,li2016,hamed2018,li2019,claytonBM2020}, with a basis in continuum damage mechanics \cite{simo1987}.
Gradient-regularized theories such as phase-field models
conceived originally for fracture in brittle elastic solids (e.g., \cite{bourdin2008,claytonIJF2014}) have been adapted to fracture, tearing, or rupture of soft tissues \cite{raina2015,gultekin2019}.

Other physical phenomena affecting mechanical response include
activation, growth, remodeling, and aging.
An example of activation is switching from passive to active (i.e., stimulated) states of muscle fibers under contraction. Modeling this phenomenon \cite{fung1993,paetsch2012,stalhand2016} is outside the present scope; only passive tissue states are considered here. 
Growth, remodeling, and aging can be classified as follows \cite{epstein2012}.
Growth involves an increase in local mass, or in mass density, depending on whether the material system is viewed as ``open'' or ``closed'' and what configuration is witnessed by the observer \cite{kuhl2003}. Diverse and sophisticated mathematical models exist \cite{rodriguez1994,lubarda2002,ambrosi2002,garikipati2004,volokh2006,epstein2007,yavari2010,goriely2017}.
The converse of growth is called resorption, and surface growth is called accretion; see for example Refs.~\cite{sozio2017,sozio2019}.
Remodeling involves local reorientation and/or distortion of material without change in mass or material properties \cite{epstein2012}.
Examples include alignment of bone structure to support compressive load \cite{epstein2012}
and alignment of collagen fibers in skin and arteries to support tensile load \cite{yang2015,hamed2019}.
Models with a continuum mechanical basis exist \cite{menzel2005,epstein2007,melnik2013,hamed2019,kumar2023}.
Aging, in contrast, is a transient change in local material properties \cite{epstein2012}, for example the familiar loss of elasticity
and embrittlement of living tissues with time. The aging process, for which descriptive theoretical notions have been forwarded \cite{epstein2009,epstein2015}, occurs at time scales much longer than considered in this work.

Growth and remodeling are frequent sources of residual stress, that is, stress remaining in the body when tractions are removed from its external boundaries.
Correspondingly, the material contains remnant, residual, or internal strains, terms used interchangeably herein to label anelastic deformation.
Different approaches have been applied to characterize residual stress or existence of stress-free configurations, as examined in detail elsewhere \cite{klarbring2007,yavari2010,lychev2021}.
In perhaps the simplest approach, which is not possible for all physical problems,
a residually stressed configuration is achieved by changing the topology of
a compatible stress-free configuration. This approach has been widely used
for arteries \cite{chuong1986,holz2000}: a residually stressed state for the artery is achieved by deforming and bonding a  stress-free radially cut-open and simply connected sector into a residually stressed and multiply-connected hollow cylinder. Circumferential cuts may be needed in addition to a radial cut to more closely realize a stress-free state \cite{omens2003}.

A second approach uses a multiplicative decomposition of the deformation gradient into
elastic and anelastic parts, where the latter may be due to growth, remodeling, or both \cite{rodriguez1994,lubarda2002,epstein2007,kumar2023}.
This implies a generally anholonomic intermediate configuration \cite{claytonDGKC2014}, whereupon the balance of linear momentum includes an extra term associated with non-integrability of elastic or anelastic deformation ``gradients'' \cite{noll1967,claytonMMS2012}.
Such a decomposition has origins in geometrically nonlinear elastoplasticity theory for crystalline
solids \cite{bilby1957,claytonNMC2011,steinmann2015}.
The torsion of the Weitzenbock connection from anholonomic deformation is physically associated with line defects such as dislocations in crystals. Generalizations of this connection \cite{minagawa1979,claytonPM2005,yavari2014,sanda2021} can describe other sources of incompatibility.

A third approach, not necessarily at odds with the second approach,
introduces a metric on the locally unstressed material manifold with non-vanishing curvature tensor: 
geometry of this space is Riemannian. This viewpoint, with origins in anelasticity \cite{eckart1948} predating the multiplicative decomposition, was used to quantify residual stress in the heart and arteries \cite{taka1990,taka2013} and more generally describe bulk growth in solids and shells \cite{yavari2010,sadik2016}.
As clarified elsewhere \cite{klarbring2007}, the Riemannian reference configuration is covered by continuous coordinate chart(s) (as opposed to anholonomic coordinates) that would exist even in the absence of curvature, as it is comprised of the same set of material points that make up the body in the current configuration embedded in Euclidean ambient space. A multiplicative decomposition can be introduced \cite{taka1990,kumar2023}, but it is not always necessary.

The present article develops and applies a generalized Finsler-geometric continuum theory to describe
nonlinear elasticity, remodeling, and degradation in fibrous soft tissues. 
Attention is restricted to quasi-static problems in the isothermal setting.
Viscoelasticity is not modeled explicitly: time scales for loading are considered either long or short enough that the
material behaves in either a glassy or configurationally relaxed manner \cite{simo1987,claytonBM2020}.
Remodeling is typically thought to be slow relative to characteristic times for acoustic waves, heat conduction, and even viscoelasticity \cite{menzel2005}, but experiments on skin at strain rates of $10^{-2}$/s suggest non-affine reorientation of collagen fibers.
Meant by ``non-affine'' is deformation of the microstructure in a local volume that
does not match the average (affine) deformation gradient of the continuum body over that local volume.

As discussed in a review article \cite{claytonMMS2022}, ideas from Finsler geometry seem to have first been applied toward continuum mechanics of solids in the context of ferromagnetic crystals \cite{amari1962}. In that
work, governing equations were derived using a variational principle, but no boundary value problems were solved.
Others \cite{ikeda1972,ikeda1973,bejancu1990} postulated geometry and kinematics for deformable continua but did not derive physical balance laws. Another series  \cite{saczuk1996,saczuk1997,stumpf2000,saczuk2001,stumpf2002,saczuk2003} developed a complete thermodynamic theory, but only two of these \cite{saczuk1996,stumpf2000} solved a single problem, and in that case kinematic fields seem to have been specified a priori, without regard to the balance equations for momentum and energy conservation. In contrast, a more recent theory of Finsler-geometric continuum mechanics \cite{claytonTR2016,claytonJGP2017} has been used to
solve, analytically or numerically, a number of boundary value problems for nonlinear elastic solids with auxiliary physical behaviors including fracture, cavitation, twinning, phase transitions, and ferromagnetism in crystals \cite{claytonZAMP2017,claytonIJF2017,claytonIJGMMP2018,claytonCMT2018,claytonMMS2022}.
A version of the Finsler-geometric theory for
mixtures of solid and fluid constituents was most recently applied \cite{claytonPRE2024,claytonPRE2024a} to study shock 
compression and tissue damage in liver, skeletal muscle, and lung, in the absence of remodeling or residual stress.
The framework has also been used to study anisotropic elasticity and tearing of skin \cite{claytonSYMM2023}.
Skin was modeled as an orthotropic 2-D sheet, and components of a single ``director'' vector of generalized Finsler geometry monitored degradation of the tissue at each point on the base manifold, in directions parallel and perpendicular to Langer's lines \cite{ni2012,jood2018}. 

Some similarities exist between this theory \cite{claytonJGP2017,claytonSYMM2023} and others that incorporate vector bundles or fiber bundles \cite{nguyen2022,sanda2024}.
Differing are theories invoking groupoid concepts \cite{epstein2007,leon2021,epstein2021} to study inhomogeneity, which in turn, can relate to residual stress \cite{noll1967}.
Another application of a Finsler metric invoked a discrete bond model, rather than continuum elasticity, to
describe soft tissues with oriented collagen fibers \cite{takano2017,mitsuhashi2018} and rubbery polymers with oriented molecular chains \cite{koibuchi2019}. Differential-geometric concepts (e.g., smooth manifolds and connections) were not used therein; governing equations were derived from sums entering a discrete Hamiltonian with contributions depending locally on a Finsler metric.

In classical Finsler geometry, the metric tensor is a positively homogeneous function of degree zero of an independent vector field, referred to here as a ``director vector'' or more broadly, an ``internal state vector''. The local fiber dimension matches the dimension of the base manifold; more general fiber-bundle treatments allow fiber and base dimensions to be different \cite{ikeda1972,bejancu1990}. Also relaxed in the present theory, and some others \cite{bejancu1990}, is the restriction of zero-order homogeneity of the metric, meaning both length and direction of the internal state vector are important.
Similarities and differences between this class of theory and micropolar or micromorphic models \cite{toupin1964,eremeyev2020} are discussed elsewhere \cite{claytonJGP2017,claytonMMS2022}.
Herein, like the original and most recent work \cite{claytonTR2016,claytonJGP2017,claytonSYMM2023}, no (e.g., multiplicative) decomposition of the deformation gradient is used.  The theory does however allow decompositions of deformation gradients when warranted by physics so described \cite{claytonZAMP2017,claytonMMS2022}.

Recent work \cite{claytonSYMM2023} did not address growth and remodeling, nor was the theory therein capable of tracking behaviors of multiple fiber families of arbitrary lengths and orientations.  These deficiencies are alleviated in the present work that extends the description to multiple fiber families at each point on the base manifold, following an early, purely kinematic treatment \cite{ikeda1973}.  
Energy density can depend on strain, internal state (e.g., remodeling energy \cite{kumar2023}
and cohesive energy for ruptuer), and its material gradient (e.g., energy of internal boundaries and fracture surfaces).
A variational principle yields Euler-Lagrange equations for linear momentum conservation and internal state equilibrium.
The latter include driving forces from strain energy and internal-state energy, often competing \cite{kumar2023},
as well as second-gradients of internal state and microscopic ``body forces'' due to chemical and biological interactions not captured by the material's energy density.  The internal state equilibrium equations are similar to quasi-static versions of Allen-Cahn or Ginzburg-Landau equations encountered in phase-field theory, and microscopic body forces have been introduced in a similar context \cite{fried1993,fried1994,gurtin1996}.
Fiber coordinates evolve independently of the overall continuum (e.g., non-affine microstructure changes), as dictated by solutions to the equilibrium equations at each load increment. Damage is captured by allowance of the energy density to reduce as fibers are overly stretched or sheared, with gradient contributions linked to fracture surface energy as in phase-field mechanics \cite{bourdin2008,claytonIJF2014,gultekin2019}. Degraded fiber length can contribute to remnant strain. 

An application demonstrates utility of the theory and provides physical insight into coupling among stress, remodeling, and degradation. The left ventricle is subjected to internal and possible external pressure. The true geometry and microstructure are complex, the former approximately a truncated ellipsoid of revolution \cite{guccione1995,holz2009}. The layered structure consists of endocardium, myocardium, and epicardium.
Muscle fibers (cells or myocytes) vary in orientation through the thickness, as does geometry of the collagen fiber network in the ground substance of the interstitial volume.  Properties are typically modeled as locally orthotropic \cite{holz2009,gultekin2016}, but directions and degree of orthotropy vary heterogeneously throughout the ventricle,
as does stiffness \cite{novak1994}. 
Numerical simulations \cite{guccione1995,walker2005,hopkins2020}, outside the present scope, are needed to model the geometry and microstructure, at the $\mu$m scale, with high fidelity.

Analytical or semi-analytical models approximated the ventricle or a section of it as a thick-walled cylinder \cite{guccione1991} or thick-walled sphere \cite{janz1976,taka1990}. 
The latter representation is used herein.  Residual
strains, in the form of a material metric with non-vanishing curvature in the Riemannian geometric viewpoint,  
thought to arise from growth or remodeling, are introduced a priori in some calculations \cite{taka1990}. 
Two families of fibers are defined in the Finsler-geometric sense.
One family measures residual strain contributions from remodeling, with components addressing
remnant stretching in locally orthogonal radial, polar, and azimuthal directions.  The radial direction would approximate fibers of the collagen network, whereas the angular directions (i.e., tangential or transverse directions) depict mean orientations of myocytes as an orthogonal net \cite{janz1976}.
As in a previous idealization \cite{taka1990}, elastic anisotropy, layered microstructures, and heterogeneous fiber orientations and material properties are not addressed out of simplicity.  Even so, the model is capable of matching experimental pressure-volume data \cite{spotnitz1966}.
Furthermore, the possibility of residual strain redistribution in the context of incremental remodeling during very slow loading is addressed newly modeled here, as is degradation from internal overpressure. A second geometric fiber family depicts the latter damage processes, also idealized as isotropic. In this idealization, only a single arbitrary component of the internal state vector need be used, its magnitude measuring the local intensity of damage \cite{claytonJGP2017,claytonMMS2022}.
Prior finite-element studies of ventricular trauma and tissue failure \cite{walker2005,forsell2011}
did not consider effects of residual strain on damage.
In contrast, finite element simulations of arterial expansion and dissection from overpressure
showed an increase in tissue damage with axial residual strain \cite{balzani2012} and a decrease with circumferential residual strain \cite{wang2017}. At the $\mu$m scale, mechanisms observed in failure of ventricular tissue
include splitting of ``matrix'' between muscle cells (primary), muscle fiber rupture, and collagen fiber pull-out \cite{gasser2009}. Remnant strains are also noted in ventricle and arterial walls at overpressure (e.g., hypertension) \cite{balzani2012}. 


This article is organized as follows.
The general continuum framework 
is developed in \S2, extending prior work \cite{claytonMMS2022,claytonSYMM2023} to accommodate multiple fiber families at each point on the base manifold.
A requisite extension of Rund's theorem \cite{rund1975}, first adapted to a generalized Finsler
geometry with a single fiber family \cite{claytonJGP2017,claytonMMS2022}, is further proven for multiple fiber families in Appendix A. Variational derivatives used in derivation of Euler-Lagrange equations are defined in Appendix B.
Energy potentials are constructed, and resultant governing equations derived, in \S3.
Given certain assumptions, the Finslerian representation can be recast to a representation in Riemannian geometry as shown in \S4. Such identification clarifies possibilities of residual stresses from remodeling, or a priori growth, that exist in both representations.
The model is used study the pressurized left ventricle, with remnant strain possibly evolving due to remodeling, and muscle damaged from extreme stretching or shearing, in \S5. 
Conclusions are in \S6.
Notation follows prior works \cite{claytonMMS2022,claytonSYMM2023}. Vectors and higher-order tensors are written in bold font, scalars in italics. When the index notation is used, subscripted and superscripted Roman letters (e.g., $a,b, \ldots,i,j,\ldots$ and  $A, B, \ldots, I,J \ldots$) follow the Einstein summation convention of tensor analysis. Greek subscripts and superscripts are not summed unless noted explicitly with the summation sign $\underset{\alpha}{\sum} (\cdot)_\alpha$.

\section{Nonlinear structured solids in generalized Finsler geometry}

\subsection{Reference configuration geometry}
A fiber bundle depiction of the reference configuration of a body with microstructure
extends prior treatments \cite{bejancu1990,claytonMMS2022,claytonSYMM2023} to allow multiple
fibers, in a geometric sense, at each material point \cite{ikeda1973}.
The base manifold of dimension $n$ corresponding to the body, \textit{excluding} its microstructure, is
denoted by $\mathcal{M}$, with points $X$ assigned coordinates via chart(s) $\{X^A \}$, $A = 1,\ldots,n$.
It is typically assumed that the set of points comprising $\mathcal{M}$ can be parameterized by Euclidean coordinates,
though a metric tensor for the body \textit{with} microstructure, and corresponding
residual strains and stresses, need not have vanishing curvature tensor \cite{taka1990,klarbring2007,yavari2010,claytonSYMM2023}. 
At each $X$, a set of $r$ director vector(s) ${\bm D}_\alpha$, also called internal state vector(s), is assigned,
where $\alpha = 1, \ldots, r$. Charts of secondary coordinates over $\mathcal{M}$ are the internal state components
$\{D^K_\alpha \}$, where $K = 1, \ldots,m$. 

Denote by $\mathsf{Z} = (\mathcal{Z},\Pi,\mathcal{M}, \{ \mathcal{U}_\alpha \} )$ a fiber bundle of total
space $\mathcal{Z}$, with $\dim \mathcal{Z} = n + mr$ and $\Pi: {\mathcal Z} \rightarrow \mathcal{M}$ a projection. 
A chart on $\mathcal{Z}$ is $\{X^A,D^K_\alpha\}$.
The set of all fibers at $X$ is $\{ {\mathcal U}_\alpha \}$; each fiber ${\mathcal U}_\alpha = \Pi^{-1}_\alpha (X) $ is parameterized by corresponding $\{ D^K_\alpha \}$ and is a vector space of dimension $m$. 
The $\{D^K_\alpha \}$ fields are continuously differentiable with respect to $X^A$ to any necessary order. Coordinate indices $A$ and $K$ are dropped from notation when there is no chance for confusion.
Fiber numbers $\alpha$ are not coordinates and are not subject to the Einstein summation rules.

\subsubsection{Coordinates and transformations}
Let $\{ X, D_\alpha \}$ and $\{ \tilde{X}, \tilde{D}_\alpha \}$ denote two coordinate charts on $\mathcal Z$.
Transformation rules are
\begin{equation}
\label{eq:trans1}
\tilde{X}^A = \tilde{X}^A(X), \qquad \tilde{D}^J_\alpha (X,D) =  Q \ua^J_K (X) D^K_\alpha,
 \qquad (\alpha = 1,\ldots,r); \qquad
Q \ua^I_K \tilde{Q} \ua^K_J = \delta^I_J,
\end{equation}
where $Q \ua^I_K$ and $\tilde{Q} \ua^K_J$ are differentiable and mutual inverses.
An independent transformation law in the second of \eqref{eq:trans1} holds for each $\alpha = 1, \ldots, r$.
Write the total tangent bundle as $T \mathcal{Z}$ with holonomic bases $\{ \frac{\partial}{\partial X^A}, \frac{\partial}{\partial D^K_\alpha} \}$. The cotangent bundle $T^* \mathcal{Z}$ has holonomic bases $\{ d X^A, d D^K_\alpha \}$.
According to \eqref{eq:trans1} with chain-rule differentiation, holonomic bases transform on total space $\mathcal{Z}$
under coordinate changes $\{X,D_\alpha\} \rightarrow \{\tilde{X},\tilde{D}_\alpha\}$ on $\mathcal{M}$ as
follows (e.g., \cite{bejancu1990,bao2000,minguzzi2014}):
\begin{equation}
\label{eq:trans2}
\frac{\partial}{\partial \tilde{X}^A}  = \frac{\partial X^B}{\partial \tilde{X}^A}\frac{\partial}{\partial X^B}
+ \sum_{\alpha = 1}^r \frac{\partial D^K_\alpha }{\partial \tilde{X}^A  }  \frac{\partial}{\partial D^K_\alpha}
= \frac{\partial X^B}{\partial \tilde{X}^A}\frac{\partial}{\partial X^B}
+ \sum_{\alpha = 1}^r \frac{\partial \tilde{Q}\ua^K_J}{\partial \tilde{X}^A } \tilde{D}_\alpha^J  \frac{\partial}{\partial D^K_\alpha}, 
\end{equation}
\begin{equation}
\label{eq:trans2b}
\frac{\partial}{\partial \tilde{D}^J_\alpha}  = \frac{\partial X^B}{\partial \tilde{D}^J_\alpha }\frac{\partial}{\partial X^B} +
\frac{\partial D^K_\alpha }{\partial \tilde{D}^J_\alpha }\frac{\partial}{\partial D^K_\alpha}
= \tilde{Q}\ua^K_J \frac{\partial}{\partial D^K_\alpha},
\end{equation}
\begin{equation}
\label{eq:trans3}
d \tilde{X}^A =  \frac{\partial \tilde{X}^A}{\partial {X}^B} d X^B + \sum_{\alpha = 1}^r \frac{\partial \tilde{X}^A}{\partial {D}^K_\alpha} d D^K_\alpha
=  \frac{\partial \tilde{X}^A}{\partial {X}^B} d X^B,
\end{equation}
\begin{equation}
\label{eq:trans3b}
d \tilde{D}^J_\alpha  = \frac{\partial \tilde{D}^J_\alpha}{\partial {X}^B} dX^B +
\frac{\partial \tilde{D}^J_\alpha}{\partial {D}^K_\alpha} dD^K_\alpha
=  \frac{\partial {Q}\ua^J_K}{\partial {X}^B } D^K_\alpha d X^B + {Q}\ua^J_K d D^K_\alpha.
\end{equation}
Extending usual definitions of Finsler geometry \cite{bejancu1990,bao2000} to multiple fiber families ($r \geq 1$), the non-holonomic bases invoked on $\mathcal{Z}$ in lieu of $\{ \frac{\partial}{\partial X^A}  \}$ and
$\{ d D^K_\alpha \}$  are defined, respectively, as
\begin{equation}
\label{eq:nonhol}
\frac{\delta}{\delta X^A} = \frac{\partial}{\partial X^A} - \sum_{\alpha = 1}^r N 
\ua^K_A \frac{\partial}{\partial D^K_\alpha},
\qquad \delta D^K_\alpha = dD^K_\alpha + N \ua^K_B dX^B.
\end{equation} 
The nonlinear connection coefficients $N\ua^K_A(X,D)$ transform for $\{X,D_\alpha\} \rightarrow \{\tilde{X},\tilde{D}_\alpha\}$ as follows:
\begin{equation}
\label{eq:Ntrans}
\tilde{N}\ua^J_A = \left(Q\ua^J_K  N\ua^K_B - \frac{\partial {Q}\ua^J_K}{\partial {X}^B } D^K_\alpha \right) \frac{\partial X^B} {\partial \tilde{X}^A}, \qquad (\alpha = 1, \ldots, r).
\end{equation}
From \eqref{eq:trans1}--\eqref{eq:Ntrans}, non-holonomic bases transform conventionally, with standard inner products:
\begin{equation}
\label{eq:dtrans}
\frac{\delta }{\delta \tilde{X}^A} = \frac{\partial X^B }{\partial \tilde{X}^A} \frac{\delta }{\delta {X}^B},
\quad
\delta \tilde{D}^J_\alpha = Q\ua^J_K \delta D^K_\alpha; 
\qquad
\bigr{\langle} \frac{\delta}{\delta X^B}, dX^A \bigr{\rangle} = \delta^A_B, \quad
\bigr{\langle} \frac{\partial}{\partial D^K_\alpha}, \delta D^J_\alpha \bigr{\rangle}= \delta^J_K.
\end{equation}

Also defined elsewhere \cite{bejancu1990}, 
the tangent bundle $T \mathcal{Z}$ equipped with nonlinear connection suggests the
 orthogonal decomposition  $T \mathcal{Z} = V \mathcal{Z} \oplus H \mathcal{Z}$ into a vertical vector bundle $V \mathcal{Z}$ with local fields of frames $ \{ \frac{\partial}{\partial D^A_\alpha} \}$ ($\alpha = 1,\ldots,r$) and a horizontal distribution $H \mathcal{Z}$ with local field of frames $ \{ \frac{\delta}{\delta X^A} \}$.
Fiber dimensions of $\mathcal{H Z}$ and $\mathcal{VZ}$ are $n$ and $mr$. Henceforth,
as is sufficient for applications to materials physics that follow, take $m = n$, so indices
$A,B, \ldots$ and $I,J,\ldots$ can be used interchangeably in the Einstein summation convention.
Also, henceforth  prescribe in \eqref{eq:trans1} \cite{bao2000,minguzzi2014}
\begin{equation}
\label{eq:subseq}
Q\ua^A_B =  \frac{\partial \tilde{D}^A_\alpha}{\partial D^B_\alpha} =  \frac{\partial \tilde{X}^A}{\partial X^B}
= Q^A_B, \qquad (\forall \, \alpha = 1, \ldots , r).
\end{equation}

Let $f(X,D_\alpha)$ be a generic differentiable function of its arguments. The following condensed notation is introduced, along with special cases $f \rightarrow X$ and $f \rightarrow D_\alpha$ \cite{claytonJGP2017,claytonMMS2022}:
\begin{equation}
\label{eq:diffnot}
\partial_A f(X,D_\alpha) = \frac{\partial f (X,D_\alpha)}{\partial X^A},  \quad
\bar{\partial}_A^\alpha f(X,D_\alpha) = \frac{\partial f(X,D_\alpha)}{\partial D^A_\alpha},
\quad \delta_A (\cdot) = \frac{\delta (\cdot)}{\delta X^A}
= \partial_A (\cdot) - \sum_{\alpha = 1}^r N\ua_A^B \bar{\partial}_B^\alpha (\cdot);
\end{equation}
\begin{equation}
\label{eq:partialD}
\partial_B X^A  = \frac{\partial X^A }{ \partial X^B} = \delta^A_B, \quad \bar{\partial}_B^\alpha X^A  = 0; 
\qquad \partial_B D^A_\alpha  = \frac{\partial D^A_\alpha}{ \partial X^B}, \quad \bar{\partial}_B^\alpha D^A_\beta  = \delta^\alpha_\beta \delta^A_B .
\end{equation}

Nonlinear connection coefficients $N\ua^A_B(X,D_\alpha)$ should ideally obey \eqref{eq:trans1} and \eqref{eq:Ntrans}.  If $T \mathcal{Z}$ is restricted to locally flat sections \cite{bao2000,minguzzi2014}, then $N\ua^A_B = 0$ in a preferred coordinate chart $\{ X,D_\alpha \}$, but $\tilde{N}\ua^A_B$ in \eqref{eq:Ntrans} do not necessarily vanish in other arbitrary charts from heterogeneous transformations under which $\partial_B Q^J_K$ is nonzero.
Methods for construction of non-trivial nonlinear connections are discussed in Refs.~\cite{bejancu1990,claytonMMS2022,claytonSYMM2023}.

\subsubsection{Metric tensor}
The Sasaki metric tensor $\pmb{\mathcal G}$
\cite{yano1973,bao2000} enables a natural inner product of vectors over $\mathcal{Z}$, here extended to multiple fiber families $\alpha \geq 1$:
\begin{equation}
\label{eq:Sasaki}
\pmb{\mathcal{G}}(X,D_\alpha) = {\bm{G}}(X,D_\alpha) + \sum_{\alpha = 1}^r \check{\bm{G}}^\alpha (X,D_\alpha) = G_{AB}(X,D_\alpha) dX^A \otimes dX^B + \sum_{\alpha=1}^r\check{G}^\alpha_{AB}(X,D_\alpha) \delta D^A_\alpha \otimes \delta D^B_\alpha;
\end{equation}
\begin{equation}
\label{eq:Sasaki2}
\mathcal{G}_{AB} = G_{AB}  =
\bm{G} \left( \frac{\delta}{\delta X^A} , \frac{\delta}{\delta X^B} \right) = \check{G}^\alpha_{AB} = 
\check{\bm{G}}^\alpha \left( \frac{\partial}{\partial D^A_\alpha} , \frac{\partial}{\partial D^B_\alpha} \right) = \check{G}_{BA}^\alpha = G_{BA} = \mathcal{G}_{BA}, \, \, (\forall \alpha = 1,\ldots,r).
\end{equation}
Components of $\bm{G}$ and $\check{\bm{G}}^\alpha$ are equal for all $\alpha$ and simply referred to as $G_{AB}$ per convention \cite{bao2000}, but their bases span orthogonal subspaces.
Components $G_{AB}$ lower indices, and components $G^{AB}$ of $\bm{G}^{-1}$ raise indices, all
per usual rules of tensor analysis; determinant $G$ of $n \times n$ non-singular matrices of components of
$\bm{G}$ or $\check{\bm{G}}^\alpha$ is
\begin{equation}
\label{eq:detG}
G^{AB} G_{BC} = \delta^A_C; \qquad G(X,D_\alpha) = \det [G_{AB}(X,D_\alpha)] = \det [\check{G}^\alpha_{AB}(X,D_\alpha)], \quad (\forall \alpha = 1,\ldots,r).
\end{equation}

Denote a generic vector field over $\mathcal{Z}$ as $\bm{V} = V^A \frac{\delta}{\delta X^A} \in H \mathcal{Z}$.
The magnitude of $\bm{V}$ at $(X,D_\alpha)$ is $| \bm{V} | = \langle \bm{V},\pmb{\mathcal{G}} \bm{V} \rangle^{1/2} =
\langle \bm{V},\bm{G} \bm{V} \rangle^{1/2}
= |\bm{V} \cdot \bm{V}|^{1/2} = |V^A G_{AB} V^B |^{1/2} = |V^A V_A|^{1/2} \geq 0$, where $V^A$ and $G_{AB}$ are evaluated at $(X,D_\alpha)$.
When the Sasaki metric tensor is interpreted as a block diagonal $ (1+ r )n \times (1+r ) n$ matrix, the determinant of $\pmb{\mathcal{G}}$ is, by extension of the $r = 1$ case \cite{saczuk1996,saczuk1997,stumpf2000,claytonTR2016,claytonSYMM2023},
\begin{equation}
\label{eq:detGS}
\mathcal{G}(X,D_\alpha) = \det [G_{AB}(X,D_\alpha)] \prod_{\alpha = 1}^r \det [\check{G}^\alpha_{AB}(X,D_\alpha)] = | \det [G_{AB}(X,D_\alpha)] |^{1+r} = | G(X,D_\alpha) |^{1+r}.
\end{equation}

Write $\diff \bm{X}$ for a differential line element of $\mathcal{M}$ referred to horizontal basis $ \{ \frac{\delta}{\delta X^A} \}$. Write $\diff \bm{D}_\alpha$ for a line element of a single local fiber $\mathcal{U}_\alpha$ referred to a vertical basis $ \{ \frac{\partial}{\partial D^A_\alpha} \}$ (i.e., one value of $\alpha \in [1,r]$).  Squared lengths are, from \eqref{eq:Sasaki},
\begin{equation}
\label{eq:lengths}
|\diff \bm{X}|^2 = \langle \diff \bm{X}, \pmb{\mathcal{G}} \diff \bm{X} \rangle = G_{AB} \, \diff X^A \, \diff X^B,
\qquad
|\diff \bm{D}_\alpha|^2 = \langle \diff \bm{D}_\alpha, \pmb{\mathcal{G}} \diff \bm{D}_\alpha \rangle = G_{AB} \, \diff D^A_\alpha \, \diff D^B_\alpha.
\end{equation}
The volume element $\diff V$ and volume form $d \Omega$ of the $n$-dimensional base manifold $\mathcal{M}$, and
the area form $\Omega$ for its ($n-1$)-dimensional boundary $\partial \! \mathcal{M}$, are defined as \cite{rund1975}
\begin{equation}
\label{eq:volforms}
\diff V = \sqrt{G} \, \diff X^1 \diff X^2 \ldots \diff X^n, \qquad d \Omega = \sqrt{G} \, d X^1 \wedge d X^2 \wedge \ldots \wedge d X^n,
\end{equation}
\begin{equation}
\label{eq:areaform}
\Omega = \sqrt{B} \, d Y^1 \wedge \ldots \wedge d Y^{n-1}.
\end{equation}
Local coordinates $\{ X^A,Y^\mathfrak{a} \}$ on oriented hypersurface $\partial \! \mathcal{M}$ obey parametric equations $X^A = X^A(Y^\mathfrak{a})$, $(\mathfrak{a} = 1,\ldots,n-1)$, $B^A_\mathfrak{a} =
\frac{\partial X^A}{\partial Y^\mathfrak{a}}$, and $B = \det (B^A_\mathfrak{a} G_{AB} B^B_\mathfrak{b})$. 

\subsubsection{Linear connections}
Denote by $\nabla(\cdot)$ the covariant derivative. Generic coefficients of linear connections
$H^A_{BC}$ and $K^{\alpha A}_{BC}$ define horizontal gradients of basis vectors on $T \mathcal{Z}$.
Generic coefficients of linear connections $V^{A}_{BC}$ and $Y^{\alpha A}_{BC}$ define
vertical gradients. For
for vector bases $\{\frac{\delta}{\delta X^A}, \frac{\partial}{\partial D^A_\alpha} \}$ and their dual counterparts $\{d X^A, \delta D^A \}$ on $T^*\mathcal{Z}$, where $A,B = 1,\ldots,n$ and $\alpha,\beta = 1,\ldots,r$, gradients of these bases are
\begin{equation}
\label{eq:horizgrad}
\nabla_{\delta / \delta X^B} \frac{\delta}{\delta X^C} = H^A_{BC} \frac{\delta}{\delta X^A}, \qquad
\nabla_{\delta / \delta X^B} \frac{\partial}{\partial D^C_\alpha} = K^{\alpha A}_{BC} \frac{\partial}{\partial D^A_\alpha},
\end{equation}
\begin{equation}
\label{eq:vertgrad}
\nabla_{\partial / \partial D^B_\alpha} \frac{\partial}{\partial D^C_\beta} = \delta^\alpha_\beta V^{A}_{BC} \frac{\partial}{\partial D^A_\beta}, \qquad
\nabla_{\partial / \partial D^B_\alpha} \frac{\delta}{\delta X^C} = Y^{\alpha A}_{BC} \frac{\delta}{\delta X^A} ;
\end{equation}
and
\begin{equation}
\label{eq:horizgradd}
\nabla_{\delta / \delta X^B} \, d X^C = - H^C_{BA}  d X^A, \qquad
\nabla_{\delta / \delta X^B} \, \delta D^C_\alpha = - K^{\alpha C}_{BA}  \delta D^A_\alpha ,
\end{equation}
\begin{equation}
\label{eq:vertgradd}
\nabla_{\partial / \partial D^B_\alpha} \, \delta D^C_\beta = - \delta^\alpha_\beta V^{C}_{BA}  \delta D^A_\beta
, \qquad
\nabla_{\partial / \partial D^B_\alpha} \, d X^C = - Y^{\alpha C}_{BA} d X^A.
\end{equation}
An assumption inherent in \eqref{eq:vertgrad} and \eqref{eq:vertgradd}, which could be relaxed at the expense
of more coefficients, is all fiber families share the same vertical coefficients $\delta^\alpha_\beta V^A_{BC}$
with cross-gradients vanishing, meaning when $\alpha \neq \beta$, values of $V^A_{BC}$ are inconsequential.
This assumption is consistent with the independent transformation law for fiber families having $\alpha \neq \beta$ in \eqref{eq:trans1} and \eqref{eq:subseq}. 

Denote a generic, differentiable horizontal vector field by ${\bm{V}} = V^A \frac{\delta}{\delta X^A} \in H \mathcal{Z}$, and
denote a generic, differentiable vertical field on a basis spanning multiple fibers of the bundle (i.e., a linear combination of vectors over fibers of the vertical vector bundle) by ${\bm{W}} = \sum_{\alpha = 1}^r W^A_\alpha \frac{\partial}{\partial D^A_\alpha } \in V \mathcal{Z}$. Total covariant derivatives of fields ${\bm V}$ and ${\bm W}$ 
are defined as\footnote{The present \eqref{eq:excov} for $\nabla \bm{V}$ corrects misprinted basis vectors in eqs.~(21)$_2$ and (21)$_3$ of Ref.~\cite{claytonSYMM2023} for $r = 1$.}:
\begin{equation}
\label{eq:excov}
\begin{split}
\nabla \bm{V} & = \nabla_{\delta / \delta X^B} \bm{V} \otimes d X^B + \sum_{\alpha = 1}^r \nabla_{\partial / \partial D^B_\alpha} \bm{V} \otimes \delta D^B_\alpha \\ & = 
(\delta_B V^A + H^A_{BC} V^C) \frac{\delta}{\delta X^A} \otimes d X^B
+  \sum_{\alpha = 1}^r (\bar{\partial}_B^\alpha V^A + Y^{\alpha A}_{BC} V^C) \frac{\delta}{\delta X^A} \otimes \delta D^B_\alpha
\\ & = V^A_{\, |B} \frac{\delta}{\delta X^A} \otimes d X^B + \sum_{\alpha = 1}^r V^A|^\alpha_B \frac{\delta}{\delta X^A} \otimes \delta D^B_\alpha,
\end{split}
\end{equation}
\begin{equation}
\label{eq:excow}
\begin{split}
\nabla \bm{W} & = \nabla_{\delta / \delta X^B} \bm{W} \otimes d X^B + \sum_{\alpha = 1}^r \nabla_{\partial / \partial D^B_\alpha} \bm{W} \otimes \delta D^B_\alpha \\ & = 
\sum_{\alpha=1}^r (\delta_B W^A_\alpha + K^{\alpha A}_{BC} W_\alpha^C) \frac{\partial}{\partial D^A_\alpha} \otimes d X^B
+  \sum_{\alpha = 1}^r \sum_{\beta = 1}^r (\bar{\partial}_B^\alpha W^A_\beta + \delta^\alpha_\beta V^{A}_{BC} W^C_\beta) \frac{\partial}{\partial  D^A_\beta} \otimes \delta D^B_\alpha
\\ & = \sum_{\alpha = 1}^r W^A_{\alpha |B}  \frac{\partial}{\partial D^A_\alpha} \otimes d X^B + \sum_{\alpha = 1}^r \sum_{\beta = 1}^r  W^A_{\beta} |^\alpha_B  \frac{\partial}{\partial D^A_\beta} \otimes \delta D^B_\alpha.
\end{split}
\end{equation}
Horizontal covariant differentiation with respect to $\{X^B\}$ is
$(\cdot)_{|B}$.
Vertical covariant differentiation with respect to a single fiber ($\alpha$) coordinate chart
$\{D^B_\alpha\}$ is $(\cdot)|^\alpha_B$.

The horizontal covariant derivative of the horizontal part of the Sasaki metric, 
$\bm{G} = G_{AB} \, dX^A \otimes dX^B$
in \eqref{eq:Sasaki}, and determinant $G = \det (G_{AB})$ in \eqref{eq:detG}, a scalar density \cite{rund1975},
are, in components,
\begin{equation}
\label{eq:horizG1}
G_{AB|C} = \delta_C G_{AB} - H^D_{CA} G_{DB} -H^D_{CB} G_{AD} 
= \partial_C G_{AB} - \sum_{\alpha = 1}^r N\ua^D_C \bar{\partial}^\alpha_D G_{AB} - H^D_{CA} G_{DB} - H^D_{CB} G_{DA},
\end{equation}
\begin{equation}
\label{eq:Gids}
(\sqrt{G})_{|A} = 
\delta_A(\sqrt{G})  - \sqrt{G} H^B_{AB} =
\partial_A(\sqrt{G}) -  \sum_{\alpha = 1}^r N\ua^B_A \bar{\partial}^\alpha_B (\sqrt{G}) - \sqrt{G} H^B_{AB}.
\end{equation}

The following special linear connections are defined, all torsion-free.
Denoted by $\gamma^A_{BC}$ are Christoffel symbols of the second kind for the Levi-Civita connection from $G_{AB}$. Denoted by $C^{\alpha A}_{BC}$
are a generalization of coefficients of Cartan's tensor
to each fiber family $\alpha = 1,\ldots,r$.
Denoted by $\Gamma^A_{BC}$ are horizontal connection coefficients of Chern-Rund and Cartan. In coordinates $\{X^A,D^A_\alpha \}$,
\begin{equation}
\label{eq:LC1}
\gamma^A_{BC}={\textstyle{\frac{1}{2}}} G^{AD} (\partial_C G_{BD} + \partial_B G_{CD} - \partial_D G_{BC})
=G^{AD} \gamma_{BCD},
\end{equation}
\begin{equation}
\label{eq:Cartan1}
C^{\alpha A}_{BC}={\textstyle{\frac{1}{2}}} G^{AD} (\bar{\partial}^\alpha_C G_{BD} + \bar{\partial}^\alpha_B G_{CD} - \bar{\partial}^\alpha_D G_{BC})
=G^{AD} C^\alpha_{BCD},
\end{equation}
\begin{equation}
\label{eq:CR1}
\Gamma^A_{BC}={\textstyle{\frac{1}{2}}} G^{AD} (\delta_C G_{BD} + \delta_B G_{CD} - \delta_D G_{BC})
=G^{AD} \Gamma_{BCD}.
\end{equation}
The Chern-Rund-Cartan coefficients $\Gamma^A_{BC}$ are
metric-compatible for horizontal differentiation of $\bm{G} = G_{AB} \, d X^A \otimes d X^B $: $H^A_{BC} = \Gamma^A_{BC} \Rightarrow
G_{AB|C} = 0$ in \eqref{eq:horizG1}.
The coefficients of Cartan's tensor $C^{\alpha A}_{BC}$ are metric-compatible for vertical covariant differentiation
of $\bm{G}$: $Y^{\alpha A}_{BC} = C^{\alpha A}_{BC} \Rightarrow {G}_{AB}|^\alpha_C = 0$.

Applying \eqref{eq:LC1}, \eqref{eq:Cartan1}, and \eqref{eq:CR1}, traces of linear connections are related to gradients of $G$:
\begin{equation}
\label{eq:Gids2}
\partial_A ( \ln \sqrt{G}) =   \gamma^B_{AB}, \qquad
\bar{\partial}^\alpha_A ( \ln \sqrt{G}) 
=  C^{\alpha B}_{AB}, 
\qquad \delta_A (\ln \sqrt{G}) = 
 \textstyle{\frac{1}{2}} G^{BC}  {\delta}_A G_{CB} = \Gamma^B_{AB}.
\end{equation}
It follows that $H^A_{BC} = \Gamma^A_{BC} \Rightarrow
G_{|A} = 2 G (\ln \sqrt{G})_{|A} = 0$ and $Y^{\alpha A}_{BC} = C^{\alpha A}_{BC} 
\Rightarrow G|^\alpha_A = 2 G (\ln \sqrt{G})|^\alpha_A = 0$.

From \eqref{eq:Sasaki2}, Cartan's tensor, identical for $\alpha = 1, \ldots, r$, can equivalently be written
\begin{equation}
\label{eq:Cartan1check}
C^{\alpha A}_{BC}={\textstyle{\frac{1}{2}}} \check{G}^{AD}_\alpha (\bar{\partial}^\alpha_C \check{G}^\alpha_{BD} + \bar{\partial}^\alpha_B \check{G}^\alpha_{CD} - \bar{\partial}^\alpha_D \check{G}^\alpha_{BC})
=\check{G}^{AD}_\alpha C^\alpha_{BCD}.
\end{equation}
The vertical covariant derivative of a vertical contribution 
$\check{\bm G}^\alpha = \check{G}^\alpha_{AB} \, \delta D^A_\alpha \otimes \delta D^B_\alpha$ 
to $\pmb{\mathcal{G}}$ of \eqref{eq:Sasaki} 
and its determinant $\check{G}^\alpha = \det (\check{G}^\alpha_{AB})$ are, in components,
\begin{equation}
\label{eq:vertG1}
\check{G}^\alpha_{AB}|^\beta_C  
= \bar{\partial}^\beta_C \check{G}^\alpha_{AB} - \delta^\beta_\alpha V^D_{CA} \check{G}^\alpha_{DB} - \delta^\beta_\alpha V^D_{CB} \check{G}^\alpha_{DA},
\quad
(\sqrt{\check{G}^\alpha})|^\beta_{A} = \bar{\partial}^\beta_A (\sqrt{\check{G}^\alpha}) - \sqrt{\check{G}^\alpha} \delta^\beta_\alpha V^B_{AB}.
\end{equation}
Therefore, $\{ V^A_{BC} = C^{\beta A}_{BC} \, \forall \, \beta = 1, \ldots, r \} \Rightarrow
\check{G}^\alpha_{AB}|^\beta_C = 0$.
From this result, or since $\bar{\partial}^\beta_A \ln (\sqrt{\check{G}^\alpha}) = C^{\beta B}_{AB}$ in accordance
with the second of \eqref{eq:Gids2},
 $\check{G}^\alpha |^\beta_A = 0$ when $V^A_{BC} = C^{\beta A}_{BC} \, \forall \, \beta = 1, \ldots, r$.

\subsubsection{Divergence theorem}
Let $\mathcal{M}$, $\dim \mathcal{M} = n$, be the base space of generalized Finsler
bundle of total space $\mathcal{Z}$ with positively oriented boundary $\partial \! \mathcal{M}$.
Horizontal components of the metric, ${G}_{AB}$, and symmetric linear connection coefficients
$H^A_{BC} = H^A_{CB}$ are assigned such that $(\sqrt G)_{|A} = 0$, for example,
$H^A_{BC} = \Gamma^A_{BC}$. Assume $C^1$ functions $D_\alpha = D_\alpha (X)$
exist for all $X \in \mathcal{M}$ and all $\alpha = 1, \ldots, r$. Denote a differentiable vector field by ${\bm{V}}(X,D_\alpha)  = V^A (X,D_\alpha) \frac{\delta}{\delta X^A} \in H \mathcal{Z}$.
In a coordinate chart $\{X^A,D^A_\alpha \}$, Stokes' theorem is
\begin{equation}
\label{eq:stokesRef}
\int_{\mathcal{M}} \bigr{[} V^A_{|A} + \sum_{\alpha  = 1}^r (V^A C^{\alpha C}_{BC} + \bar{\partial}^\alpha_B V^A) D^B_{\alpha; \, A} \bigr{]}  \, d\Omega = \int_{\partial \! \mathcal{M}} V^A N_A \,\Omega ,
\end{equation}
where the horizontal covariant derivative is $V^A_{|A} = \delta_A V^A + H^B_{BA} V^A $, the definition $D^B_{\alpha; \, A} = \partial_A D^B_\alpha + N\ua^B_A$ with $\partial_A D^B_\alpha   = \partial D^B_\alpha / \partial X^A$, and $N_A$ is the unit outward normal component
of $\bm{N} = N_A \, dX^A$ to $\partial \! \mathcal{M}$.
The proof, newly derived in Appendix A, enriches that of Refs.~\cite{rund1975,claytonMMS2022} toward extended Finsler geometry with multiple fiber families (i.e., $r > 1$).

\subsection{Spatial configuration geometry}
The current or spatial configuration is represented identically to the reference configuration
with a change of variables and commensurate notation.
Lower-case symbols are used in spatial representations versus upper-case in reference representations, an exception being connections.
The base differentiable manifold is $\mathfrak{m}$ with $\dim \mathfrak{m} = n$.
The spatial image of a particle is $x \in \mathfrak{m}$, and a
coordinate chart on $\mathfrak{m}$ is $\{x^a\}$, $a = 1, \ldots, n$.
Each point is associated with a set of $r \geq 1$ director vectors ${\bm d}_\alpha$, where
$\alpha = 1, \ldots, r$. Secondary coordinates are chart(s) of components 
$\{ d^k_\alpha \}$ with $k = 1, \ldots, m$.

Denoted by $\mathsf{z} = (\mathfrak{z},\pi,\mathfrak{m}, \{ \mathfrak{u}_\alpha \} )$ is the fiber bundle for total
space $\mathfrak{z}$, with $\dim \mathfrak{z} = n + mr$ and $\pi: {\mathfrak{Z}} \rightarrow \mathfrak{m}$ the projection. 
A chart on $\mathfrak{z}$ is $\{x^a,d^k_\alpha\}$.
The set of all fibers at $x$ is $\{ {\mathfrak u}_\alpha \}$. Every fiber ${\mathfrak u}_\alpha = \pi^{-1}_\alpha (x) $ is covered by its $\{ d^k_\alpha \}$ and is a vector space of dimension $m$. 
The $\{d^k_\alpha\}$ fields are differentiable with respect to $x^a$ to any order needed. Coordinate indices $a$ and $k$ are dropped when there is no chance for confusion. Repeated fiber numbers $\alpha$ are not subject to Einstein summation.

\subsubsection{Coordinates and transformations}
Denote by $\{ x, d_\alpha \}$ and $\{ \tilde{x}, \tilde{d}_\alpha \}$ two coordinate charts on $\mathfrak z$.
Transformation rules are
\begin{equation}
\label{eq:trans1c}
\tilde{x}^a = \tilde{x}^a(x), \qquad \tilde{d}^j_\alpha (x,d) =  q\ua^j_k (x) d^k_\alpha,
 \qquad (\alpha = 1,\ldots,r); \qquad
q \ua^i_k \tilde{q} \ua^k_j = \delta^i_j,
\end{equation}
where $q \ua^i_k$ and $\tilde{q} \ua^k_j$ are differentiable inverses and an independent
transformation law applies for each fiber family $\alpha = 1, \ldots, r$.
The total tangent bundle $T \mathfrak{z}$ has holonomic bases 
$\{ \frac{\partial}{\partial x^a}, \frac{\partial}{\partial d^k_\alpha} \}$,
and $T^*\mathfrak{z}$ has $\{ d x^a, d d^k_\alpha \}$.
Under changes of coordinates $\{ x, d_\alpha \} \rightarrow \{ \tilde{x}, \tilde{d}_\alpha \}$,
\begin{equation}
\label{eq:trans2c}
\frac{\partial}{\partial \tilde{x}^a}  = \frac{\partial x^b}{\partial \tilde{x}^a}\frac{\partial}{\partial x^b}
+ \sum_{\alpha = 1}^r \frac{\partial d^k_\alpha }{\partial \tilde{x}^a  }  \frac{\partial}{\partial d^k_\alpha}
= \frac{\partial x^b}{\partial \tilde{x}^a}\frac{\partial}{\partial x^b}
+ \sum_{\alpha = 1}^r \frac{\partial \tilde{q}\ua^k_j}{\partial \tilde{x}^a } \tilde{d}_\alpha^j  \frac{\partial}{\partial d^k_\alpha}, 
\end{equation}
\begin{equation}
\label{eq:trans2bc}
\frac{\partial}{\partial \tilde{d}^j_\alpha}  = \frac{\partial x^b}{\partial \tilde{d}^j_\alpha }\frac{\partial}{\partial x^b} +
\frac{\partial d^k_\alpha }{\partial \tilde{d}^j_\alpha }\frac{\partial}{\partial d^k_\alpha}
= \tilde{q}\ua^k_j \frac{\partial}{\partial d^k_\alpha},
\end{equation}
\begin{equation}
\label{eq:trans3c}
d \tilde{x}^a =  \frac{\partial \tilde{x}^a}{\partial {x}^b} d x^b + \sum_{\alpha = 1}^r \frac{\partial \tilde{x}^a}{\partial {d}^k_\alpha} d d^k_\alpha
=  \frac{\partial \tilde{x}^a}{\partial {x}^b} d x^b,
\end{equation}
\begin{equation}
\label{eq:trans3bc}
d \tilde{d}^j_\alpha  = \frac{\partial \tilde{d}^j_\alpha}{\partial {x}^b} dx^b +
\frac{\partial \tilde{d}^j_\alpha}{\partial {d}^k_\alpha} dd^k_\alpha
=  \frac{\partial {q}\ua^j_k}{\partial {x}^b } d^k_\alpha d x^b + {q}\ua^j_k d d^k_\alpha.
\end{equation}
Spatial non-holonomic bases and nonlinear connection coefficients are defined as follows, obeying the subsequent transformation rules  for $\{x,d_\alpha\} \rightarrow \{\tilde{x},\tilde{d}_\alpha\}$:
\begin{equation}
\label{eq:nonholc}
\frac{\delta}{\delta x^a} = \frac{\partial}{\partial x^a} - \sum_{\alpha = 1}^r N 
\ua^k_a \frac{\partial}{\partial d^k_\alpha},
\qquad \delta d^k_\alpha = d d^k_\alpha + N \ua^k_b dx^b;
\end{equation} 
\begin{equation}
\label{eq:Ntransc}
\tilde{N}\ua^j_a = \left(q\ua^j_k  N\ua^k_b - \frac{\partial {q}\ua^j_k}{\partial {x}^b } d^k_\alpha \right) \frac{\partial x^b} {\partial \tilde{x}^a}, \qquad (\alpha = 1, \ldots, r);
\end{equation}
\begin{equation}
\label{eq:dtransc}
\frac{\delta }{\delta \tilde{x}^a} = \frac{\partial x^b }{\partial \tilde{x}^a} \frac{\delta }{\delta {x}^b},
\quad
\delta \tilde{d}^j_\alpha = q\ua^j_k \delta d^k_\alpha; 
\qquad
\bigr{\langle} \frac{\delta}{\delta x^b}, dx^a \bigr{\rangle} = \delta^a_b, \quad
\bigr{\langle} \frac{\partial}{\partial d^k_\alpha}, \delta d^j_\alpha \bigr{\rangle}= \delta^j_k.
\end{equation}
The orthogonal decomposition of total tangent bundle is $T \mathfrak{z} = V \mathfrak{z} \oplus H \mathfrak{z}$.
Henceforth, the horizontal distribution and vertical vector bundle have the same dimension $n = m$, and
for $f = f(x,d_\alpha)$,
\begin{equation}
\label{eq:subseqc}
q\ua^a_b =  \frac{\partial \tilde{d}^a_\alpha}{\partial d^b_\alpha} =  \frac{\partial \tilde{x}^a}{\partial x^b} = q^a_b, \qquad (\forall \, \alpha = 1, \ldots , r);
\end{equation}
\begin{equation}
\label{eq:diffnotc}
\partial_a f(x,d_\alpha) = \frac{\partial f (x,d_\alpha)}{\partial x^a},  \quad
\bar{\partial}_a^\alpha f(x,d_\alpha) = \frac{\partial f(x,d_\alpha)}{\partial d^a_\alpha};
\quad \delta_a (\cdot) = \frac{\delta (\cdot)}{\delta x^a}
= \partial_a (\cdot) - \sum_{\alpha = 1}^r N\ua_a^b \bar{\partial}_b^\alpha (\cdot);
\end{equation}
\begin{equation}
\label{eq:partialDc}
\partial_b x^a  = \frac{\partial x^a }{ \partial x^b} = \delta^a_b, \quad \bar{\partial}_b^\alpha x^a  = 0; 
\qquad \partial_b d^a_\alpha  = \frac{\partial d^a_\alpha}{ \partial x^b}, \quad \bar{\partial}_b^\alpha d^a_\beta  = \delta^\alpha_\beta \delta^a_b .
\end{equation}

\subsubsection{Metric tensor}
The spatial Sasaki metric tensor is $\pmb{\mathfrak g}$. Covariant components are $g_{ab}$ with
inverse components $g^{ab}$. The determinant of the horizontal part is $g = \det (g_{ab})$.
Explicitly,
\begin{equation}
\label{eq:Sasakic}
\pmb{\mathfrak{g}}(x,d_\alpha) = {\bm{g}}(x,d_\alpha) + \sum_{\alpha = 1}^r \check{\bm{g}}^\alpha (x,d_\alpha) = g_{ab}(x,d_\alpha) dx^a \otimes dx^b + \sum_{\alpha=1}^r\check{g}^\alpha_{ab}(x,d_\alpha) \delta d^a_\alpha \otimes \delta d^b_\alpha;
\end{equation}
\begin{equation}
\label{eq:Sasaki2c}
\mathfrak{g}_{ab} = g_{ab}  =
\bm{g} \left( \frac{\delta}{\delta x^A} , \frac{\delta}{\delta x^B} \right) = \check{g}^\alpha_{ab} = 
\check{\bm{g}}^\alpha \left( \frac{\partial}{\partial d^a_\alpha} , \frac{\partial}{\partial d^b_\alpha} \right) = \check{g}_{ba}^\alpha = g_{ba} = \mathfrak{g}_{ba}, \, \, \, (\forall \alpha = 1,\ldots,r);
\end{equation}
\begin{equation}
\label{eq:detGc}
g^{ab} g_{bc} = \delta^a_c; \qquad g(x,d_\alpha) = \det [g_{ab}(x,d_\alpha)] = \det [\check{g}^\alpha_{ab}(x,d_\alpha)], \quad (\forall \alpha = 1,\ldots,r).
\end{equation}
An equation analogous to \eqref{eq:detGS} can be introduced.
Squared line lengths of the base continuum and fibers, volume element, volume form, and area form are, respectively,
\begin{equation}
\label{eq:lengthsc}
|\diff \bm{x}|^2 = \langle \diff \bm{x}, \pmb{\mathfrak{g}} \, \diff \bm{x} \rangle = g_{ab} \, \diff x^a \, \diff x^b,
\qquad
|\diff \bm{d}_\alpha|^2 = \langle \diff \bm{d}_\alpha, \pmb{\mathfrak{g}} \, \diff \bm{d}_\alpha \rangle = g_{ab} \, \diff d^a_\alpha \, \diff d^b_\alpha;
\end{equation}
\begin{equation}
\label{eq:volformsc}
\diff v = \sqrt{g} \, \diff x^1 \diff x^2 \ldots \diff x^n, \qquad d \omega = \sqrt{g} \, d x^1 \wedge d x^2 \wedge \ldots \wedge d x^n, \qquad
\omega = \sqrt{b} \, d y^1 \wedge \ldots \wedge d y^{n-1}.
\end{equation}
Local coordinates $\{ x^a,y^\mathfrak{a} \}$ on $\partial \! \mathfrak{m}$ obey $x^a = x^a(y^\mathfrak{a})$, $(\mathfrak{a} = 1,\ldots,n-1)$, $b^a_\mathfrak{a} =
\frac{\partial x^a}{\partial y^\mathfrak{a}}$, and $b = \det (b^a_\mathfrak{a} g_{ab} b^b_\mathfrak{b})$.  

\subsubsection{Linear connections}
Let $\nabla(\cdot)$ denote covariant differentiation.
Gradients of basis vectors on $T \mathfrak{z}$ and $T^* \mathfrak{z}$ are furnished by 
linear connection coefficients 
$H^a_{bc}$, $K^{\alpha a}_{bc}$, $V^{a}_{bc}$ and $Y^{\alpha a}_{bc}$:
\begin{equation}
\label{eq:horizgradc}
\nabla_{\delta / \delta x^b} \frac{\delta}{\delta x^c} = H^a_{bc} \frac{\delta}{\delta x^a}, \qquad
\nabla_{\delta / \delta x^b} \frac{\partial}{\partial d^c_\alpha} = K^{\alpha a}_{bc} \frac{\partial}{\partial d^a_\alpha},
\end{equation}
\begin{equation}
\label{eq:vertgradc}
\nabla_{\partial / \partial d^b_\alpha} \frac{\partial}{\partial d^c_\beta} = \delta^\alpha_\beta V^{a}_{bc} \frac{\partial}{\partial d^a_\beta}, \qquad
\nabla_{\partial / \partial d^b_\alpha} \frac{\delta}{\delta x^c} = Y^{\alpha a}_{bc} \frac{\delta}{\delta x^a};
\end{equation}
\begin{equation}
\label{eq:horizgraddc}
\nabla_{\delta / \delta x^b} \, d x^c = - H^c_{ba}  d x^a, \qquad
\nabla_{\delta / \delta x^b} \, \delta d^c_\alpha = - K^{\alpha c}_{ba}  \delta d^a_\alpha ,
\end{equation}
\begin{equation}
\label{eq:vertgraddc}
\nabla_{\partial / \partial d^b_\alpha} \, \delta d^c_\beta = - \delta^\alpha_\beta V^{c}_{ba}  \delta d^a_\beta
, \qquad
\nabla_{\partial / \partial d^b_\alpha} \, d x^c = - Y^{\alpha c}_{ba} d x^a.
\end{equation}
Denote a horizontal vector field by ${\bm{V}} = V^a \frac{\delta}{\delta x^a} \in H \mathfrak{z}$, and denote
a vertical field on a basis spanning multiple fibers by ${\bm{W}} = \sum_{\alpha = 1}^r W^a_\alpha \frac{\partial}{\partial d^a_\alpha } \in V \mathfrak{z}$. Their (total) covariant derivatives are
\begin{equation}
\label{eq:excovc}
\begin{split}
\nabla \bm{V} & = \nabla_{\delta / \delta x^b} \bm{V} \otimes d x^b + \sum_{\alpha = 1}^r \nabla_{\partial / \partial d^b_\alpha} \bm{V} \otimes \delta d^b_\alpha 
= V^a_{\, |b} \frac{\delta}{\delta x^a} \otimes d x^b + \sum_{\alpha = 1}^r V^a|^\alpha_b \frac{\delta}{\delta x^a} \otimes \delta d^b_\alpha,
\end{split}
\end{equation}
\begin{equation}
\label{eq:excowc}
\begin{split}
\nabla \bm{W} & = \nabla_{\delta / \delta x^b} \bm{W} \otimes d x^b + \sum_{\alpha = 1}^r \nabla_{\partial / \partial d^b_\alpha} \bm{W} \otimes \delta d^b_\alpha 
= \sum_{\alpha = 1}^r W^a_{\alpha |b}  \frac{\partial}{\partial d^a_\alpha} \otimes d x^b + \sum_{\alpha = 1}^r \sum_{\beta = 1}^r  W^a_{\beta} |^\alpha_b  \frac{\partial}{\partial d^a_\beta} \otimes \delta d^b_\alpha.
\end{split}
\end{equation}
The horizontal covariant derivative with respect to $\{x^b\}$ is
$(\cdot)_{|b}$.
The vertical covariant derivative with respect to fiber ($\alpha$) chart
$\{d^b_\alpha\}$ is $(\cdot)|^\alpha_b$. 
Horizontal covariant derivatives of ${\bm g}$ and $g = \det (g_{ab})$ are
\begin{equation}
\label{eq:horizG1c}
g_{ab|c} = \delta_c g_{ab} - H^d_{ca} g_{db} -H^d_{cb} g_{ad}, \qquad
(\sqrt{g})_{|a} = 
\delta_a(\sqrt{g})  - \sqrt{g} H^b_{ab}.
\end{equation}

The Levi-Civita connection $\gamma^a_{bc}$, Cartan's tensors
$C^{\alpha a}_{bc}$, and horizontal Chern-Rund-Cartan connection $\Gamma^a_{bc}$ are defined as
\begin{equation}
\label{eq:LC1c}
\gamma^a_{bc}={\textstyle{\frac{1}{2}}} g^{ad} (\partial_c g_{bd} + \partial_b g_{cd} - \partial_d g_{bc})
=g^{ad} \gamma_{bcd},
\end{equation}
\begin{equation}
\label{eq:Cartan1c}
C^{\alpha a}_{bc}={\textstyle{\frac{1}{2}}} g^{ad} (\bar{\partial}^\alpha_c g_{bd} + \bar{\partial}^\alpha_b g_{cd} - \bar{\partial}^\alpha_d g_{bc})
=g^{ad} C^\alpha_{bcd},
\end{equation}
\begin{equation}
\label{eq:CR1c}
\Gamma^a_{bc}={\textstyle{\frac{1}{2}}} g^{ad} (\delta_c g_{bd} + \delta_b g_{cd} - \delta_d g_{bc})
=g^{ad} \Gamma_{bcd}.
\end{equation}
It follows that $H^a_{bc} = \Gamma^a_{bc} \Rightarrow
g_{ab|c} = 0 \Rightarrow g_{|c} = 0$ and
$Y^{\alpha a}_{bc} = C^{\alpha a}_{bc} \Rightarrow {g}_{ab}|^\alpha_c = 0 \Rightarrow g|^\alpha_c = 0$. Also,
\begin{equation}
\label{eq:Gids2c}
\partial_a ( \ln \sqrt{g}) =   \gamma^b_{ab}, \qquad
\bar{\partial}^\alpha_a ( \ln \sqrt{g}) 
=  C^{\alpha b}_{ab}, 
\qquad \delta_a (\ln \sqrt{g}) = 
 \textstyle{\frac{1}{2}} g^{bc}  {\delta}_a g_{cb} = \Gamma^b_{ab}.
\end{equation}

\subsubsection{Divergence theorem}
The spatial analog of \eqref{eq:stokesRef} and \eqref{eq:stokes} on manifold $\mathfrak{m}$ with
boundary \! $\partial \mathfrak{m}$ is as follows, where now
${\bm{V}}(x,d_\alpha)  = V^a (x,d_\alpha) \frac{\delta}{\delta x^a} \in H \mathfrak{z}$,
$d^b_{\alpha; \, a} = \partial_a d^b_\alpha + N\ua^b_a$, and $\bm{n} = n_a \, dx^a$
is the unit outward normal to $\partial \! \mathfrak{m}$:
\begin{equation}
\label{eq:stokesc}
\int_{\mathfrak{m}} \bigr{[} V^a_{|a} + \sum_{\alpha  = 1}^r (V^a C^{\alpha c}_{bc} + \bar{\partial}^\alpha_b V^a) d^b_{\alpha; \, a} \bigr{]}  \, d\omega = \int_{\partial \! \mathfrak{m}} V^a n_a \,\omega .
\end{equation}
The proof and its corollary in Appendix A apply with the change of reference to spatial coordinates.

\subsection{Continuum kinematics}
The theory of Refs.~\cite{claytonJGP2017,claytonMMS2022,claytonSYMM2023} is extended
to account for multiple fiber families $\alpha = 1, \ldots, r$.

\subsubsection{Deformation}
Denote the motion of a material particle by $\varphi: \mathcal{M} \rightarrow \mathfrak{m}$ with
inverse motion $\Phi: \mathfrak{m} \rightarrow \mathcal{M}$ such that $(\Phi \circ \varphi)(X) = X$. Motions are at least $C^3$ and one-to-one, and in coordinates \cite{bejancu1990,claytonZAMP2017},
\begin{equation}
\label{eq:varphiC}
x^a = \varphi^a (X), \qquad X^A = \Phi^A (x),  \qquad (a,A = 1, \ldots, n).
\end{equation}
The total motion of particles and the collective fiber bundle is $\Xi = (\varphi, \{ \theta_\alpha \}): \mathcal{Z} \rightarrow \mathfrak{z}$, where 
$\theta_\alpha$ is the motion of a single director $D_\alpha$ and $\alpha = 1, \ldots, r$. 
The fields $\bm{D}_\alpha$ and $\bm{d}_\alpha$ are referred to as referential and spatial director vector fields,
though none of these vectors need be of unit length.  They are also called internal state vectors.
Motions, of at least class $C^3$, for internal state vectors are
\begin{equation}
\label{eq:thetaC}
d^a_\alpha = \theta^a_\alpha (X,D_\alpha), \qquad D^A_\alpha = \Theta^A_\alpha (x,d_\alpha),  \qquad (a,A = 1, \ldots, n; \, \alpha = 1, \ldots, r).
\end{equation}
Although more general forms are conceivable, for brevity and sufficiency, $\theta_\alpha$ does not depend on
$D_\beta$ for $\beta \neq \alpha$, and similarly $\Theta_\alpha$ does not depend on $d_\beta$ for $\beta \neq \alpha$.

From \eqref{eq:varphiC} and \eqref{eq:thetaC}, transformation formulae for partial differentiation across configurations of a differentiable function $ h(x,d_\alpha): \mathfrak{z} \rightarrow \mathbb{R}$ are
\begin{equation}
\label{eq:diffC}
\frac{\partial ( h \circ \Xi )}{\partial X^A} = \frac{\partial  h}{\partial x^a} \frac{\partial \varphi^a}{\partial X^A}
+ \sum_{\alpha = 1}^r \frac{\partial h}{\partial d^a_\alpha} \frac{\partial \theta^a_\alpha }{\partial X^A},
\qquad
\frac{\partial ( h \circ \Xi )}{\partial D^A_\alpha} =   \frac{\partial h}{\partial d^a_\alpha } \frac{\partial \theta^a_\alpha }{\partial D^A_\alpha}.
\end{equation}
Invoke a ``push-forward'' operation for the nonlinear connection coefficients, extending eq. (4.23) of Ref.~\cite{claytonMMS2022}
and eq.~(4.1.7) of Ref.~\cite{bejancu1990}:
\begin{equation}
\label{eq:NrelateC}
N\ua^b_a \frac{\partial \varphi^a}{\partial X^A} = N\ua^B_A \frac{\partial \theta^b_\alpha}{\partial D^B_\alpha} - \frac{\partial \theta^b_\alpha}{ \partial X^A }, \qquad (\forall \, \alpha = 1, \ldots, r).
\end{equation}
It follows from \eqref{eq:diffnot} and \eqref{eq:varphiC}--\eqref{eq:NrelateC} that
\begin{equation}
\label{eq:NrelateC2}
\frac{\delta (h \circ \Xi) }{\delta X^A} =
\frac{\partial (h \circ \Xi) }{\partial X^A} - \sum_{\alpha = 1}^r N\ua^B_A \frac{\partial (h \circ \Xi) }{\partial D^B_\alpha}
=
  \frac{\delta h  }{\delta x^a}  \frac{\partial \varphi^a  }{ \partial X^A} =  \frac{\delta h  }{\delta x^a}  
\frac{\delta \varphi^a  }{ \delta X^A} =  \frac{\delta h  }{\delta x^a}  F^a_A.
\end{equation}

The deformation gradient $\bm{F}: H \mathcal{Z} \rightarrow H \mathfrak{z}$ is the two-point tensor field that follows, with inverse  $\bm{f}: H \mathfrak{z} \rightarrow H \mathcal{Z}$ such that 
$F^a_A(X) f^A_b(x(X)) = \delta^a_b$ and $F^a_A(X) f^B_a(x(X)) = \delta^B_A$:
\begin{equation}
\label{eq:defgradC}
\bm{F} = \frac{ \delta \pmb{\varphi}}{\delta \bm{X}} = \frac{\delta \varphi^a}{\delta X^A} \frac{\delta}{\delta x^a} \otimes
d X^A = \frac{\partial \varphi^a}{\partial X^A} \frac{\delta}{\delta x^a} \otimes d X^A,
\end{equation}
\begin{equation}
\label{eq:defgradiC}
\bm{f} = \frac{ \delta \pmb{\Phi}}{\delta \bm{x}} = \frac{\delta \Phi^A}{\delta x^a} \frac{\delta}{\delta X^A} \otimes
d x^a = \frac{\partial \Phi^A}{\partial x^a} \frac{\delta}{\delta X^A} \otimes d x^a.
\end{equation}
Regularity of \eqref{eq:varphiC} ensures $\det (F^a_A) > 0 $ and $\det (f^A_a) >0$, implied by invertibility of
\eqref{eq:defgradC} and \eqref{eq:defgradiC}.
Differential line elements of \eqref{eq:lengths} and \eqref{eq:lengthsc} are related as follows:
\begin{equation}
\label{eq:linetrans}
\diff \bm{x} = \diff x^a \frac{\delta}{\delta x^a} =  F^a_A \diff X^A \frac{\delta}{\delta x^a} = \bm{F} \diff \bm{X},
\qquad
\diff \bm{X} = \diff X^A \frac{\delta }{\delta X^A }  = f^A_a \diff x^a  \frac{\delta}{\delta X^A} = \bm{f} \diff \bm{x}.
\end{equation}
Squared line lengths are related by symmetric Lagrangian deformation tensor
$\bm{C} = C_{AB} dX^A \otimes dX^B$:
\begin{equation}
\label{eq:lengthC}
|\diff \bm{x}|^2 =  F^a_A g_{ab} F ^b_B  \diff X^A \diff X^B = C_{AB} \diff X^A \diff X^B
= \langle \diff \bm{X}, \bm{C} \diff \bm{X} \rangle,
\quad
C_{AB} = F^a_A g_{ab} F^b_B = G_{AC} C^C_B = C_{BA}.
\end{equation}
Volume elements and volume forms of \eqref{eq:volforms} and \eqref{eq:volformsc} are related by
\begin{equation}
\label{eq:voltrans}
\diff v = J \diff V = [ \det(F^a_A) \sqrt{g/G} ] \diff V,  \qquad \diff V = j \diff v = [\det(f^A_a) \sqrt{G/g}] \diff v,
\end{equation}
\begin{equation}
\label{eq:voltrans2}
\varphi^* d \omega = J d \Omega, \qquad  \Phi^* d \Omega = j d \omega;
\end{equation}
\begin{equation}
\label{eq:Jdet}
J = \det(F^a_A) \sqrt{g/G} = \sqrt{\det(C^A_B)} > 0, \qquad j = \det(f^A_a) \sqrt{G/g} = 1/J = J^{-1} > 0.
\end{equation}
 From \eqref{eq:horizgradc}, \eqref{eq:NrelateC}, and \eqref{eq:NrelateC2},
\begin{equation}
\label{eq:horiztranspf}
\nabla_{\delta / \delta X^A} \frac{\delta}{\delta x^c}  
=F^a_A H^b_{ac} \frac{\delta}{\delta x^b},
\qquad
\nabla_{\delta / \delta X^A} \frac{\partial}{\partial d^c_\alpha} = F^a_A K^{\alpha b}_{ac}  \frac{\partial}{\partial d^b_\alpha}.
\end{equation}

The divergence theorem of extended Finsler geometry, \eqref{eq:stokesRef} or \eqref{eq:stokes}, is used to derive Euler-Lagrange equations  in \S3.  
As shown in Appendix A (i.e., Theorem A.1) and Refs.~\cite{claytonMMS2022,rund1975}, proof of this theorem requires existence of $C^1$ functional relations, hereafter\footnote{
Class $C^1$ continuity is sufficient for derivation of \eqref{eq:stokes} and local balance laws of \S3.3.
Class $C^2$ continuity may be required for unique calculation of curvature (e.g., entering the second of \eqref{eq:ghatderiv}) and solution of \eqref{eq:micromom}, depending on respective forms of $G_{AB}$ and $\psi$.
Sufficient smoothness of $D_\alpha(X)$ and $d_\alpha(x)$ is stated at the outset of \S2.1 and \S2.2.} assumed at least $C^2$:
\begin{equation}
\label{eq:funcC}
D^A_\alpha = D^A_\alpha (X), \qquad d^a_\alpha = d^a_\alpha(x), \qquad (\alpha = 1, \ldots, r),
\end{equation}
where the second of \eqref{eq:funcC} is implied by the first under a consistent change of variables via  \eqref{eq:varphiC}. In precise notation, the 
following functional forms must exist from \eqref{eq:varphiC}, \eqref{eq:thetaC}, and \eqref{eq:funcC}:
\begin{equation}
\label{eq:thetaC2}
d^a_\alpha = \theta^a_\alpha (X,D_\alpha) = \hat{\theta}^a_\alpha(X,D_\alpha(X)) = \bar{\theta}^a_\alpha(X), 
\quad D^A_\alpha = \Theta^A_\alpha (x,d_\alpha) = \hat{\Theta}_\alpha^A (x,d_\alpha(x)) = \bar{\Theta}^A_\alpha (x).
\end{equation}
Implemented here, a canonical choice for ${\theta}^a_\alpha(X,D_\alpha)$, given coordinate fields $D^A_\alpha(X)$, is \cite{claytonZAMP2017,claytonMMS2022}
\begin{equation}
\label{eq:thetasimp}
d_\alpha = D_\alpha \circ \Phi \, \Leftrightarrow \, d_\alpha(x) = D_\alpha (\Phi (x)) \, \Rightarrow \, {\theta}^a_\alpha(D_\alpha(X)) = D^A_\alpha (X) \langle \delta d^a_\alpha, \frac{\partial}{\partial D^A_\alpha} \rangle_X =   D^A_\alpha (X)  \delta^a_A,
\end{equation}
where $\delta^a_A$ is interpreted as a shifter between $V \mathfrak{z}$ and $ V \mathcal{Z}$:  
$\delta^a_A = 1 \, \forall \, a=A, \delta^a_A = 0 \, \forall \, a \neq A$.
Invoking \eqref{eq:thetasimp}, $\partial_A \theta^a_\alpha (D_\alpha(X)) = 0$
by definitions of $\theta^a_\alpha = \theta^a_\alpha (D_\alpha (X))$ and $\partial_A(\cdot) = (\partial(\cdot)/\partial X^A)|_{D_\alpha =\text{const}}$ in \eqref{eq:diffnot}. Furthermore,
$\bar{\partial}^\alpha_A \theta^a_\alpha (D_\alpha(X)) = \delta^a_A$ by \eqref{eq:partialD} and \eqref{eq:thetasimp}.
It follows from \eqref{eq:defgradC} and \eqref{eq:thetasimp} that \eqref{eq:NrelateC} simplifies to $N\ua^a_b = N\ua^A_B f^B_b \delta^a_A$, and notably for the trivial case: $N\ua^A_B = 0 \Leftrightarrow N\ua^a_b = 0$.

Relations \eqref{eq:thetasimp} do not imply that director coordinates $d^a_\alpha(x)$ or
$D^A_\alpha(X)$ are constants at any $x$ or $X$ over a deformation history.
Rather, solutions to equilibrium equations derived in \S3 produce director (i.e., internal state) fields that evolve with boundary and external forcing conditions.
Director vectors do not generally obey a ``push-forward'' operation by
the motion $\varphi^a (X)$ with deformation gradient $F^a_A(X)$; their evolution is non-affine.

\subsubsection{Preferred connections and metric decompositions}

Proof of \eqref{eq:stokes} for any admissible $G_{AB}(X,D_\alpha)$
requires a symmetric linear connection horizontally compatible with $G_{AB}$: $H^A_{BC}= \Gamma^A_{BC}$
 of \eqref{eq:CR1}.  The simplest admissible choice of vertical coefficients is $V^A_{BC}=0$, an
 extended analog of the Chern-Rund connection \cite{bao2000}.  Setting $K^{\alpha A}_{BC} = H^A_{BC}$  and $Y^{\alpha A}_{BC} = V^A_{BC}$ for all $\alpha = 1, \ldots, r$ is
 reasonable considering \eqref{eq:subseq}, but not mandatory.
 Another pragmatic, and simple, choice is $K^{\alpha A}_{BC} = 0$, used
 later in \S5 and elsewhere for modeling ferromagnetism \cite{claytonMMS2022}.
 Let Sasaki metric $\pmb{\mathcal{G}}$ of \eqref{eq:Sasaki} be given, namely the field
of covariant components $G_{AB}(X,D_\alpha)$ of \eqref{eq:Sasaki2}.
Also assume Sasaki metric $\pmb{\mathfrak{g}}$ with  $g_{ab}(x,d_\alpha)$ of \eqref{eq:Sasaki2c} is known.
Then connection coefficients over $\mathcal{Z}$ and analogous connections over $\mathfrak{z}$ 
are prescribed as
\begin{equation}
\label{eq:connrec}
H^A_{BC} =  \Gamma^A_{BC}, \quad V^A_{BC} = Y^{\alpha A}_{BC} = 0;
 \quad H^a_{bc} = \Gamma^a_{bc}, \quad V^a_{bc} = Y^{\alpha a}_{bc} = 0; \quad N\ua^a_b = N\ua^A_B f^B_b \delta^a_A.
\end{equation}
Note $K^{\alpha A}_{BC}$ and $K^{\alpha a}_{bc}$ are as yet unassigned to descriptions of different physics.
Nonlinear connections $N\ua^A_B$ ($\alpha  =1, \ldots, r$) are likewise not explicitly assigned in \eqref{eq:connrec}. Once $N\ua^A_B$ are defined, $N\ua^a_b$ in \eqref{eq:connrec} obey \eqref{eq:NrelateC} with \eqref{eq:thetasimp}.  
Conversely, the $N\ua^a_b$ could be defined first and the $N\ua^A_B$ calculated from \eqref{eq:connrec}.
Metric $G_{AB}$ ($g_{ab}$) need not be homogeneous of degree zero with respect to $D_\alpha$ ($d_\alpha$) as in strict Finsler geometry \cite{rund1959,bao2000}, but it can be.

Decomposing $G_{AB}$ into a Riemannian part $\bar{G}_{AC}$ and
 director-dependent part $\hat{G}^C_B$, both assumed invertible with positive determinants, is helpful for analysis of problems in elasticity \cite{claytonTR2016,claytonJGP2017,claytonZAMP2017,claytonIJGMMP2018}:
\begin{equation}
\label{eq:metricdecomp}
\bm{G}  = \bar{\bm{G}} \hat{\bm{G}}; \quad G_{AB}(X,D_\alpha) = \bar{G}_{AC}(X) \hat{G}^C_B (X,D_\alpha); \quad
 \bar{\bm{G}}  = \bar{G}_{AB} \, d X^A \otimes d X^B; \qquad
\hat{\bm{G}} = \hat{G}^A_B \frac{\delta}{\delta X^A} \otimes d X^B.
\end{equation}
Dependence is generally on all fiber families $\alpha = 1, \ldots, r$. More specific forms of \eqref{eq:metricdecomp} are used henceforth for $r \geq 1$, as used previously in Refs.~\cite{claytonMMS2022,claytonSYMM2023,claytonPRE2024} when $r = 1$:
\begin{equation}
\label{eq:mdec}
G_{AB}(X,D_\alpha)  = \bar{G}_{AC}(X) \hat{G}^C_B (D_\alpha(X))
= \hat{G}^C_A (D_\alpha(X))  \bar{G}_{CB}(X) ; \quad
  \bar{G}_{AB}  = \bar{G}_{BA},
\quad \hat{G}^C_A G_{BC} = \hat{G}^C_B G_{CA}.
\end{equation}
Metric $\bar{G}_{AB}$ depicts symmetry of the physical body in the absence of microstructure. Analogously to \eqref{eq:mdec} for spatial metric $g_{ab}(x,d_\alpha)$,
\begin{equation}
\label{eq:mdecc}
g_{ab}(x,d_\alpha)  = \bar{g}_{ac}(x) \hat{g}^c_b (d_\alpha(x))
= \hat{g}^c_a (d_\alpha (x))  \bar{g}_{cb}(x) ; \qquad
 \bar{g}_{ab}  = \bar{g}_{ba}, \quad
\qquad \hat{g}^c_a g_{bc} = \hat{g}^c_b g_{ca}.
\end{equation}
Prescribing $\bar{G}_{AB}(X)$ and $\bar{g}_{ab}(x)$ to be (locally) Euclidean metrics, meaning coordinate transformations
of $\{X^A\}$ and $\{x^a\}$ are available over (regions of) $\mathcal{M}$ and $\mathfrak{m}$ enabling $\bar{G}_{AB} \rightarrow \delta_{AB}$ and $ \bar{g}_{ab} \rightarrow \delta_{ab}$, is useful, though not essential.
In such cases, Riemann-Christoffel curvature tensors from Levi-Civita connections of
$\bar{G}_{AB}(X)$ and $\bar{g}_{ab}(x)$ vanish identically.  The latter are necessary and sufficient conditions for existence of locally Euclidean charts $\{X^A\}$ and $\{x^a\}$ \cite{marsden1983,claytonDGKC2014,sokolnikoff1951}.

A symmetric Lagrangian tensor $\bar{\bm{C}}$ 
and volume ratio $\bar{J} > 0$ exclude director ($D_\alpha$) dependence \cite{claytonZAMP2017}:
\begin{equation}
\label{eq:Cbar}
\bar{\bm{C}}(X) = \bar{C}_{AB}(X) \, d X^A \otimes d X^B, \qquad 
\bar{C}_{AB} = F^a_A \bar{g}_{ab} F^b_B, \quad \bar{C}^A_B = \bar{G}^{AC} \bar{C}_{CB};
\end{equation}
\begin{equation}
\label{eq:Jbar}
\bar{J}(X) = \sqrt{ \det (\bar{C}^A_B (X)) }; \qquad \bar{J}  = J \sqrt{ \hat{G} / \hat{g}},
\quad \hat{G} = \det (\hat{G}^A_B), \quad
\hat{g} = \det (\hat{g}^a_b).
\end{equation}
Constructions \eqref{eq:Cbar} and \eqref{eq:Jbar} can be useful for formulating strain energy potentials \cite{claytonTR2016,claytonZAMP2017,claytonPRE2024}.

\section{Variational principles and governing equations}

Variational principles are used to derive governing equations in the
setting of quasi-statics (i.e., incremental equilibrium states).
Thermal and inertial effects and entropy production
are omitted in this setting.
The present derivations differ from prior work \cite{claytonJGP2017,claytonMMS2022,claytonSYMM2023}
in three primary ways. First, the theory originally developed for the case $r = 1$ is
extended to $\alpha = 1, \ldots, r$ fiber families, where $r \geq 1$.
Second, concepts of mass and mass density are introduced, important for modeling biologic growth
and resorption.
Third, free energy is expressed on a per-unit-mass basis rather than per-unit-volume basis.
Though a volumetric basis can also be acceptable, a mass basis is adopted here for energy density as
it appears more widespread in theories of growth mechanics \cite{lubarda2002,menzel2005,yavari2010,epstein2012} as well as mixture theories \cite{bowen1976,claytonIJES2022,irwin2024,claytonPRE2024}.
In the present setting, the body has but one homogenized constituent, and
interactions with an external chemical or biological environment enable mass and energy exchange with
this environment (i.e., an open-system paradigm is adopted \cite{kuhl2003,volokh2006,epstein2012}). 
Energy exchange can occur even if mass exchange does not, as in remodeling \cite{epstein2007,epstein2012,kumar2023}.

\subsection{Energy, mass, and variations}
Free energy per unit mass of the body $\mathcal M$ is the scalar field $\psi$.
Mass per unit reference volume of $\mathcal M$ is the scalar field $\rho_0$, called referential mass density,
which need not be constant under growth, resorption, or degradation. 
Energy density and mass density are functions of the following arguments, and total energy and mass of the body manifold are
the functionals $\Psi$ and $M$:
\begin{equation}
\label{eq:psivars}
\psi = \psi (\bm{F},\bm{D}_\alpha, \nabla \bm{D}_\alpha, \bm{X} ) = \psi (F^a_A, D^A_\alpha, D^A_{\alpha | B},X^A),
\quad
\rho_0 = \rho_0 (\bm{D}_\alpha, \bm{X}) = \rho_0 ( D^A_{\alpha},X^A),
\end{equation}
\begin{equation}
\label{eq:PsiM}
\Psi[\bm{\varphi} , \bm{D}_\alpha ] = \int_{\mathcal M} \rho_0 (\bm{D}_\alpha,\bm{X}) \psi (\bm{F},\bm{D}_\alpha, \nabla \bm{D}_\alpha, \bm{X} ) d \Omega  (\bm{D}_\alpha, \bm{X} ),
\quad M[\bm{D_\alpha}] = \int_{\mathcal M} \rho_0 (\bm{D}_\alpha, \bm{X} ) d \Omega  (\bm{D}_\alpha, \bm{X} ).
\end{equation}
Deformation gradient $\bm{F}$ is defined in \eqref{eq:defgradC}, and
director gradient $\nabla \bm{D}_\alpha$ is calculated in \eqref{eq:gradD} of Appendix B.
Dependence of $\psi$ on the horizontal part $D^A_{\alpha | B}$, for all $\alpha = 1, \ldots, r$, is sufficient
since the vertical part of $\nabla \bm{D}_\alpha$ contains only a set of constants. 
Heterogeneous material properties are admitted by explicit $\bm{X}$ dependence.
Denote the spatial mass density on $\mathfrak{m}$ by $\rho$,
the spatial mass form by $d m$, and the referential mass form by $dm_0$. Then consistently with
\eqref{eq:voltrans} and \eqref{eq:voltrans2} \cite{marsden1983},
\begin{equation}
\label{eq:massforms}
\rho_0 = \rho J \quad \Leftrightarrow \quad d m_0 = \varphi^* dm ; \qquad M = \int_{\mathcal M} d m_ 0 =  \int_{\mathfrak m} d m .
\end{equation}
Define the following derivatives of the free energy, multiplied by mass density:
\begin{equation}
\label{eq:conforces}
P^A_a = \rho_0 \frac{\partial \psi}{\partial F^a_A}, \qquad
Q^\alpha_A =  \rho_0 \frac{\partial \psi}{\partial D^A_\alpha}, \qquad
Z\ua^B_A =  \rho_0 \frac{\partial \psi}{\partial D^A_{\alpha |B}}.
\end{equation}
Later, $\bm{P} = P^A_a dx^a \otimes \frac{\delta}{\delta X^A}$ is identified
as the first Piola-Kirchhoff stress tensor. 
Conjugate forces to internal state and its gradient are covector 
$\bm{Q}^\alpha = Q^\alpha_A \delta D^A_\alpha $ and mixed tensor
$\bm{Z}\ua = Z\ua^B_A \delta D^A_\alpha \otimes \frac{\delta}{\delta X^B}$.

In forthcoming variational postulates, independently varied parameters are $\delta \varphi^a (X)$ and $\delta D^A_\alpha (X)$ defined in \eqref{eq:varphivarD} of Appendix B.
From \eqref{eq:varderivsimp}, \eqref{eq:varF}, \eqref{eq:vargradD},
and \eqref{eq:omegavar1}, the following variations are noted:
\begin{equation}
\label{eq:variations}
\delta {F^a_A} = (\delta \varphi^a)_{|A}, \qquad
\delta D^A_{\alpha |B} = [\delta(D^A_\alpha)]_{|B} - [ \bar{\partial}^\alpha_C N\ua^A_B - \bar{\partial}^\alpha_C K^{\alpha A}_{BD} D^D_\alpha ] \delta(D^C_\alpha),
\end{equation}
\begin{equation}
\label{eq:varrho}
\delta \rho_0 = \sum_{\alpha = 1}^r \bar{\partial}^\alpha_A \rho_0 \, \delta (D^A_\alpha), \qquad
\delta (d \Omega)  = \sum_{\alpha = 1}^r C^{\alpha B}_{AB} \, \delta (D^A_\alpha) \, d \Omega.
\end{equation}

\subsection{Variational mass balance}
A variational principle for mass conservation in the quasi-static setting (i.e., incremental equilibrium states,
no explicit time dependence) is invoked, similar to Ref.~\cite{eringen1962}.
Denote by $\bm{S}^\alpha = \bm{S}^\alpha(X,D_\beta)$, $\alpha, \beta = 1, \ldots, r$ a field of mass sources or sinks representing growth or resorption from interactions between local microstructure and
the ambient environment in the open-system interpretation \cite{kuhl2003,epstein2012}.
The global mass balance is the variational principle 
\begin{equation}
\label{eq:globmass}
\delta  M = \int_{\mathcal M} \sum_{\alpha = 1}^r \langle \bm{S}^\alpha, \delta (\bm{D}_\alpha) \rangle d \Omega .
\end{equation}
Using \eqref{eq:psivars}--\eqref{eq:massforms} and \eqref{eq:varrho}, \eqref{eq:globmass} becomes, in component form,
\begin{equation}
\label{eq:globmass2}
\delta \int_{\mathcal M} \rho_0 \, d \Omega = 
\int_{\mathcal M} \sum_{\alpha = 1}^r [ \bar{\partial}^\alpha_A \rho_0 
+ \rho_0 C^{\alpha B}_{AB} ] \delta (D^A_\alpha)  d \Omega
=
\int_{\mathcal M} \sum_{\alpha = 1}^r {S}^\alpha_A \delta ({D}^A_\alpha) d \Omega.
\end{equation}
Localizing and assuming \eqref{eq:globmass2} holds for any admissible $\delta (D^A_\alpha)$, the local mass conservation law is
\begin{equation}
\label{eq:locmass}
\bar{\partial}^\alpha_A \rho_0 =  S^\alpha_A - \rho_0 C^{\alpha B}_{AB}
\quad \Rightarrow \quad
\diff \rho_0(X)  = \sum_{\alpha = 1}^r \bar{\partial}^\alpha_A \rho_0 \,\diff (D^A_\alpha) = \sum_{\alpha = 1}^r [S^\alpha_A - \rho_0 C^{\alpha B}_{AB}]  \diff (D^A_\alpha),
\end{equation}
where here $\diff \rho_0$ and $\diff (D^A_\alpha)$ are increments in referential density and
internal state components within an equilibrium deformation history at material point $X$.
The first term on the far right of
\eqref{eq:locmass} is the increase (decrease) in mass density from external sources if $S^\alpha_A$ is positive (negative). The second term is the decrease (increase) in density from the increase (decrease) in local
volume due to the state dependence of the material metric $G_{AB}(X,D_\alpha)$. From \eqref{eq:Jdet}, \eqref{eq:connrec}, \eqref{eq:massforms}, $\partial J / \partial F^a_A = J (F^{-1})^A_a$ \cite{claytonDGKC2014}, and an identity similar
to the first of \eqref{eq:variations}, the spatial form of \eqref{eq:locmass} is
\begin{equation}
\label{eq:locmassspatial}
\frac{\diff \rho}{\rho} = \frac{\diff \rho_0}{\rho_0 } - (F^{-1})^A_a \diff F^a_A - \diff \ln \sqrt{g} + \diff \ln \sqrt {G} =
- (\diff \varphi^a)_{|a} +  
 \frac{1}{\rho } \sum_{\alpha = 1}^r [s^\alpha_a - \rho C^{\alpha b}_{ab} ]  \diff (d^a_\alpha),
\end{equation}
where $s^\alpha_a = J^{-1} \delta^A_a S^\alpha_A$ and state vector increment $\diff (d^a_\alpha) = \delta^a_A \diff (D^A_\alpha)$
via \eqref{eq:thetasimp}.
For closed systems in Euclidean space, \eqref{eq:globmass}, \eqref{eq:locmass},
and \eqref{eq:locmassspatial} reduce to
$\delta M = 0$, $\diff \rho_0 = 0$, and $\diff \rho + \rho (\diff \varphi^a)_{| a} = 0$.

\subsection{Variational energy balance and Euler-Lagrange equations}
A variational principle for energy conservation in quasi-static equilibrium states is posited,
extending Refs.~\cite{claytonJGP2017,claytonSYMM2023} to multiple fiber families and explicit
mass exchange with the environment.  This principle enables derivation of balance laws for stress equilibrium and
internal state equilibrium as well as boundary conditions for mechanical traction and forces conjugate to internal state. 

Let $\bm{b} = b_a dx^a$ with $b_a = b_a(X,D_\alpha)$ be the macroscopic body force field per unit mass.
Denote by $\bm{E}^\alpha = \bm{E}^\alpha(X,D_\beta)$, $\alpha, \beta = 1, \ldots , r$
a field of energy sources or sinks accounting for interactions between local microstructure
and the ambient environment. These energy exchanges are independent of
energy supplies accompanying local mass exchange $\bm{S}^\alpha$ and can
account for remodeling that occurs irrespective of changes in local mass or volume \cite{epstein2007,epstein2012,kumar2023}. Each $\bm{E}^\alpha$ serves the role of a co-vector of
external micro-forces in the context of Refs.~\cite{fried1993,fried1994,gurtin1996}, whereas
each $\bm{S}^\alpha$ serves the role of a co-vector of external mass supplies.
The global energy balance is the variational principle
\begin{equation}
\label{eq:globen}
\begin{split}
\delta \Psi  = \oint_{\partial \! \mathcal{M}} \bigr{(} \langle \bm{p}, 
\delta \pmb{\varphi} \rangle & + \sum_{\alpha = 1}^r \langle \bm{z}^\alpha, \delta \bm{D}_\alpha \rangle \bigr{)} \Omega
+ \int_{\mathcal{M}} \rho_0 \langle \bm{b}, \delta \pmb{\varphi} \rangle \, d \Omega
\\ & + \int_{\mathcal{M}}   \sum_{\alpha = 1}^r  \langle \bm{E}^\alpha, \delta \bm{D}_\alpha \rangle  \, d \Omega
+  \int_{\mathcal{M}}  \psi \sum_{\alpha = 1}^r  \langle \bm{S}^\alpha, \delta \bm{D}_\alpha \rangle  \, d \Omega.
\end{split}
\end{equation}
The mechanical traction co-vector (force per unit reference area) conjugate to particle displacement on
boundary $\partial  \! \mathcal{M}$ is ${\bm p} = p_a dx^a$.
The thermodynamic traction conjugate to changes in internal state for each fiber family
$\alpha$ on $\partial  \! \mathcal{M}$ is
$\bm{z}^\alpha  = z_A^\alpha \delta D^A_\alpha$.
Energetic contributions from the latter are standard in theories of phase-field (e.g, Allen-Cahn or Ginzburg-Landau) 
type \cite{gurtin1996,claytonPHYSD2011,claytonJMPS2021}.
The rightmost term in \eqref{eq:globen} is free energy injected or extracted from growth or resorption \cite{lubarda2002,kuhl2003}.

Subtracting the rightmost integral in \eqref{eq:globen} from the left side of \eqref{eq:globen} and using \eqref{eq:conforces} and \eqref{eq:locmass},
\begin{equation}
\label{eq:globen2}
\begin{split}
 \delta \Psi   & -  \int_{\mathcal{M}}  \psi \sum_{\alpha = 1}^r  {S}^\alpha_A \delta (D^\alpha_a)  \, d \Omega = 
\int_{\mathcal{M}} \rho_0 \delta \psi \, d \Omega + \int_{\mathcal{M}}
\psi \sum_{\alpha = 1}^r [ \bar{\partial}^\alpha_A \rho_0 
+ \rho_0 C^{\alpha B}_{AB} - S^\alpha_A ] \delta (D^A_\alpha)  d \Omega 
\\ & =
\int_{\mathcal{M}} \rho_0 \delta \psi \, d \Omega 
= \int_{\mathcal{M}} \rho_0  \biggr{(} \frac{\partial \psi}{\partial F^a_A} \delta F^a_A 
+ \sum_{\alpha = 1}^r \frac{\partial \psi}{\partial D^A_\alpha} \delta (D^A_\alpha)
+  \sum_{\alpha = 1}^r \frac{\partial \psi}{\partial D^A_{\alpha |B}} \delta D^A_{\alpha |B} \biggr{)} d \Omega 
\\ & =  \int_{\mathcal{M}} \biggr{(} P^A_a \delta F^a_A 
+ \sum_{\alpha = 1}^r  Q^\alpha_A \delta (D^A_\alpha)
+  \sum_{\alpha = 1}^r Z\ua^B_A \delta D^A_{\alpha |B} \biggr{)} d \Omega .
\end{split}
\end{equation}
Using \eqref{eq:stokesRef} or \eqref{eq:stokes} and \eqref{eq:variations}, the first term on the right of \eqref{eq:globen2} is integrated by parts to yield
\begin{equation}
\label{eq:Pint}
\begin{split}
 \int_{\mathcal{M}} & P^A_a \delta F^a_A d \Omega
  = \int_{\mathcal{M}} P^A_a (\delta \varphi^a)_{|A} d \Omega
 = \int_{\mathcal{M}} (P^A_a \delta \varphi^a)_{|A} d \Omega - \int_{\mathcal{M}} P^A_{a|A} \delta \varphi^a d \Omega
 \\ & = \oint_{\partial \! \mathcal{M}} P^A_a \delta \varphi^a N_A \Omega
 - \int_{\mathcal{M}} P^A_{a|A} \delta \varphi^a d \Omega
 -  \int_{\mathcal{M}} \sum_{\alpha = 1}^r
[ P^A_a \delta \varphi^a C^{\alpha C}_{BC} 
 + \bar{\partial}^\alpha_B (P^A_a \delta \varphi^a) ] D^B_{\alpha ; \, A} d \Omega.
 \end{split}
\end{equation}
Because
\eqref{eq:globen} should hold for $\delta (D^A_\alpha) = 0$, 
\begin{equation}
\label{eq:Pint2}
\begin{split}
 \int_{\mathcal{M}} & P^A_a \delta F^a_A d \Omega  = \oint_{\partial \! \mathcal{M}} p_a \delta \varphi^a \Omega
 +  \int_{\mathcal{M}}  \rho_0 b_a \delta \varphi^a d \Omega.
\end{split}
\end{equation}
Recall from Appendix B that $\delta \varphi^a = \delta \varphi^a(X)$ and $\delta(D^C_\alpha) = \delta(D^C_\alpha)(X)$. Substituting from \eqref{eq:Pint}, localizing the result for any admissible $\delta \varphi^a$,
and expanding the covariant derivative of $P^A_a (X,D_\alpha)$
gives the linear momentum balance on $\mathcal M$ and natural boundary conditions for stress on
$\partial \! \mathcal{M}$:
\begin{equation}
\label{eq:linmom}
\partial_A P^A_a + \sum_{\alpha = 1}^r \bar{\partial}^\alpha_B P^A_a \partial_A D^B_\alpha
+ P^B_a \Gamma^A_{AB}
- P^A_c \Gamma^c_{ba} F^b_A  + \rho_0 b_a 
= - P^A_a \sum_{\alpha =1}^r C^{\alpha C}_{BC} (\partial_A D^B_\alpha + N\ua^B_A), 
\end{equation}
\begin{equation}
\label{eq:pa}
p_a = P^A_a N_A.
\end{equation}
From \eqref{eq:Gids2}, the following identity applies, producing an alternative representation of \eqref{eq:linmom}: 
\begin{equation}
\label{eq:Gammagamma}
\Gamma^A_{AB} = \Gamma^A_{BA} = \delta_B (\ln \sqrt{G}) = \partial_B (\ln \sqrt{G})
-  \sum_{\alpha = 1}^r  N\ua_B^A \bar{\partial}^\alpha_A (\ln \sqrt{G})
 = \gamma^A_{AB} - \sum_{\alpha = 1}^r N\ua_B^A C^{\alpha C}_{AC};
\end{equation}
\begin{equation}
\label{eq:linmom2}
\partial_A P^A_a + \sum_{\alpha = 1}^r \bar{\partial}_B^\alpha P^A_a \partial_A D^B_\alpha
+ P^B_a \gamma^A_{AB}
- P^A_c \Gamma^c_{ba} F^b_A + \rho_0 b_a
= - P^A_a \sum_{\alpha =1}^r C^{\alpha C}_{BC} \partial_A D^B_\alpha.
\end{equation}

The third term on the right of \eqref{eq:globen2} is
integrated by parts using \eqref{eq:stokesRef} and \eqref{eq:variations} for $\alpha = 1 ,\dots, r$:
\begin{equation}
\label{eq:zint}
\begin{split}
& \int_{\mathcal{M}}  Z\ua^B_A \delta D^A_{\alpha |B} \, d \Omega = 
\int_{\mathcal{M}}  Z\ua^B_A 
\{ [ \delta(D^A_\alpha)]_{|B} - [ \bar{\partial}^\alpha_C N\ua^A_B - \bar{\partial}^\alpha_C K^{\alpha A}_{BD} D^D_\alpha ] \delta(D^C_\alpha) \} d \Omega
\\
& = \int_{\mathcal{M}}  [ Z\ua^B_A \delta(D^A_\alpha)]_{|B} \, d \Omega 
- \int_{\mathcal{M}}   \{ Z\ua^A_{C|A} 
+  Z\ua^B_A  [ \bar{\partial}^\alpha_C N\ua^A_B - \bar{\partial}^\alpha_C K^{\alpha A}_{BD} D^D_\alpha ] 
\} \delta(D^C_\alpha) d \Omega
\\ & =
\oint_{\partial \! \mathcal{M}}  Z\ua^B_C \delta(D^C_\alpha)  N_B \Omega
- \int_{\mathcal{M}}   \{ Z\ua^A_{C|A} 
+  Z\ua^B_A  [ \bar{\partial}^\alpha_C N\ua^A_B - \bar{\partial}^\alpha_C K^{\alpha A}_{BD} D^D_\alpha ] 
\} \delta(D^C_\alpha) d \Omega
\\ & \qquad \qquad \qquad \qquad \quad \, \, \, - \int_{\mathcal{M}} \sum_{\beta = 1}^r 
\{ Z\ua^B_C \delta(D^C_\alpha) C^{\beta D}_{ED} 
 + \bar{\partial}^\beta_E [Z\ua^B_C \delta(D^C_\alpha) ] \} D^E_{\beta ; \, B} d \Omega.
\end{split}
\end{equation}
Since \eqref{eq:globen} must hold for $\delta \varphi^a = 0$
and independently varied $\delta ( D^A_\alpha)$ for $\alpha = 1 ,\ldots, r$, \eqref{eq:globen2} gives
\begin{equation}
\label{eq:Qint}
\int_{\mathcal M} [Q^\alpha_A  \delta ( D^A_\alpha)
+ Z\ua^B_A \delta D^A_{\alpha | B} ] d \Omega = \oint_{\partial \! \mathcal{M}} z_A^\alpha \delta ( D^A_\alpha) \Omega, \qquad
(\alpha = 1, \ldots, r).
\end{equation}
Substituting from \eqref{eq:zint}, localizing the result for any admissible $\delta (D^C_\alpha)$,
and expanding the covariant derivative of $Z^A_C (X,D_\alpha)$
produces the micro-momentum balance on $\mathcal M$ and natural boundary conditions for conjugate forces to changes in internal state on
$\partial \! \mathcal{M}$:
\begin{equation}
\label{eq:micromom}
\begin{split}
  \partial_A Z\ua^A_C  +   \sum_{\beta = 1}^r & \bar{\partial}^\beta_B Z\ua^A_C \partial_A D^B_\beta  + Z\ua^B_C \Gamma^A_{AB}  - Z\ua^A_B K^{\alpha B}_{AC} -   (Q^\alpha_C - E^\alpha_C)
 \\ &  = 
 - Z\ua^B_A ( \bar{\partial}^\alpha_C N\ua^A_B - \bar{\partial}^\alpha_C K^{\alpha A}_{BD} D^D_\alpha)
    - Z\ua^A_C  \sum_{\beta = 1}^r 
  C^{\beta D}_{BD} 
   (\partial_A D^B_{\beta} + N\ub^B_A),
\end{split} 
\end{equation}
\begin{equation}
\label{eq:zA}
z^\alpha_A = Z\ua^B_A N_B , \qquad (\alpha = 1,\ldots, r).
\end{equation}
Setting $\alpha \rightarrow \beta$ in \eqref{eq:Gammagamma}, an alternative version of \eqref{eq:micromom} is derived:
\begin{equation}
\label{eq:micromom2}
\begin{split}
  \partial_A Z\ua^A_C  +   \sum_{\beta = 1}^r & \bar{\partial}^\beta_B Z\ua^A_C \partial_A D^B_\beta  + Z\ua^B_C \gamma^A_{AB}  - Z\ua^A_B K^{\alpha B}_{AC} -   (Q^\alpha_C - E^\alpha_C)
 \\ &  = 
 - Z\ua^B_A ( \bar{\partial}^\alpha_C N\ua^A_B - \bar{\partial}^\alpha_C K^{\alpha A}_{BD} D^D_\alpha)
    - Z\ua^A_C  \sum_{\beta = 1}^r 
  C^{\beta D}_{BD} 
   \partial_A D^B_{\beta}, \qquad (\alpha = 1,\ldots, r).
\end{split} 
\end{equation}
Given natural boundary conditions \eqref{eq:pa} and \eqref{eq:zA} or essential boundary conditions 
$\bm{\varphi}(X)$ and $\bm{D}_\alpha (X)$ for $X \in \partial \! \mathcal{M}$, mass density field $\rho_0(X,D_\alpha)\, \forall \, X \in \mathcal{M}$, and local micro-forces $\bm{E}^\alpha(X,D_\beta(X))$ $\forall \, X \in \mathcal{M}$, the Euler-Lagrange equations in \eqref{eq:linmom} and \eqref{eq:micromom} comprise $(1+r) \times n$ coupled PDEs in $(1+r) \times n$ degrees-of-freedom
$\varphi^a(X) $ and $D^A_\alpha (X)$ at each $X \in \mathcal{M}$.

\subsection{Spatial invariance and material symmetry}
Changes of observer are described by constant orthonormal transformation
$q^a_b = q\ua^a_b \, \forall \alpha = 1, \ldots, r$ in \eqref{eq:trans1c} and \eqref{eq:subseqc}, whereby 
$\det (q^a_b) = 1$ and $\tilde{q}^a_b = g^{ac} q^d_c g_{bd}$, meaning $\bm{q}^{-1} = \bm{q}^\text{T}$ \cite{marsden1983}.
As $\bm{F} \rightarrow \bm{q} \bm{F}$ under such coordinate changes,
$\psi$ in \eqref{eq:psivars} is restricted. 
To ensure $\bm{F} \rightarrow \bm{q} \bm{F} \Rightarrow \psi \rightarrow \psi$, two options are \cite{claytonSYMM2023}
\begin{equation}
\label{eq:psi2}
\psi = \hat{\psi}[\bm{C}(\bm{F},\bm{g}),\bm{D}_\alpha,\nabla \bm{D}_\alpha,\bm{X}] = \hat{\psi}(C_{AB}, D^A_\alpha, D^A_{\alpha |B},X^A),
\end{equation}
\begin{equation}
\label{eq:psi3}
\psi = \bar{\psi}[\bar{\bm{C}}(\bm{F},\bar{\bm{g}}),\bm{D}_\alpha,\nabla \bm{D}_\alpha,\bm{X}] = \bar{\psi}(\bar{C}_{AB}, D^A_\alpha, D^A_{\alpha|B}, X^A).
\end{equation}
Notice that the first of \eqref{eq:psivars} can be expressed 
from \eqref{eq:varphiC}, \eqref{eq:thetaC}, \eqref{eq:mdec}, and \eqref{eq:mdecc} as follows:
\begin{equation}
\label{eq:psivars2}
 \psi (\bm{F},\bm{D}_\alpha, \nabla \bm{D}_\alpha, \bm{X} ) = \check{\psi} 
 (\bm{F},\bm{D}_\alpha, \bar{\bm{G}}(\bm{X}), \hat{\bm{G}}(\bm{D}_\alpha), 
 \bar{\bm{g}}({\varphi} (\bm{X})), \hat{\bm{g}}({\theta} (\bm{X},\bm{D}_\alpha)), \nabla \bm{D}_\alpha, \bm{X} ).
\end{equation}
From \eqref{eq:lengthC}, \eqref{eq:Cbar}, \eqref{eq:psi2} or \eqref{eq:psi3}, first Piola-Kirchhoff stress $P^A_a$ of \eqref{eq:conforces} is
given by either of
\begin{equation}
\label{eq:Pk1}
P^A_a = \rho_0 \frac{\partial \psi} {\partial F^a_A} = 2 g_{ab} F^b_B \rho_0 \frac{\partial \hat{\psi}}{\partial C_{AB}},
\qquad P^A_a = \rho_0 \frac{\partial \psi} {\partial F^a_A}  = 2 \bar{g}_{ab} F^b_B \rho_0 \frac{\partial \bar{\psi}}{\partial \bar{C}_{AB}}.
\end{equation}
Cauchy stresses $\sigma^{ab}$ or  $\bar{\sigma}^{ab}$ obey symmetry of the classical angular momentum balance \cite{marsden1983,claytonNMC2011}:
\begin{equation}
\label{eq:cauchy}
\sigma^{ab} = \frac{1}{J} g^{ac} P_c^A F^b_A = 
\frac{2}{J} F^a_A F^b_B \rho_0 \frac{\partial \hat{\psi}}{\partial C_{AB}}  = \sigma^{ba}, \quad
\bar{\sigma}^{ab} =
\frac{1}{\bar{J}} \bar{g}^{ac} P_c^A F^b_A =
\frac{2}{\bar{J}} F^a_A F^b_B  \rho_0 \frac{\partial \bar{\psi}}{\partial \bar{C}_{AB}} 
= \bar{\sigma}^{ba}.
\end{equation}

Changes of material coordinates are given by transformation $Q^A_B = Q\ua^A_B$ of
\eqref{eq:trans1} and \eqref{eq:subseq} with inverse $\tilde{Q}^B_A$.
Under affine changes of coordinates $X^A \rightarrow Q^C_A X^A$, it follows that $d X^A \rightarrow  Q^C_A d X^A$, 
$F^a_A \rightarrow \tilde{Q}^A_C F^a_A$, $G_{AB} \rightarrow \tilde{Q}^A_C \tilde{Q}^B_D G_{AB}$,
$C_{AB} \rightarrow \tilde{Q}^A_C \tilde{Q}^B_D C_{AB}$,
$\bar{G}_{AB} \rightarrow \tilde{Q}^A_C \tilde{Q}^B_D \bar{G}_{AB}$, 
$\bar{C}_{AB} \rightarrow \tilde{Q}^A_C \tilde{Q}^B_D \bar{C}_{AB}$,
$D^A_\alpha \rightarrow Q^C_A D^A_\alpha$, $\delta  D^A_\alpha \rightarrow  Q^C_A \delta D^A_\alpha$, and 
$D^A_{\alpha |B} \rightarrow Q^C_A \tilde{Q}^B_D D^A_{\alpha |B}$.
Scalar free energy densities $\psi$, $\hat{\psi}$, and $\bar{\psi}$ should remain
 unchanged for all transformations $\tilde{Q}^A_B$ belonging to the material symmetry group $\mathbb{Q}$ \cite{claytonNMC2011,spencer1971,balzani2006}.
For the case of polynomial invariants with basis $\mathcal{P}$ of invariant functions with respect to $\tilde{\bm{Q}} \in \mathbb{Q}$ \cite{spencer1971,balzani2006} and energy offsets $\hat{\psi}_0 = \text{constant}$,  $\bar{\psi}_0 = \text{constant}$, developments in Ref.~\cite{claytonSYMM2023} extend to 
\begin{equation}
\label{eq:pbasis1}
\hat{\mathcal{P}} = \{ I_1, I_2, \ldots, I_\upsilon \}; \qquad I_\alpha = I_\alpha (\bm{C},\bm{D}_\alpha,\nabla \bm{D}_\alpha), \qquad
\hat{\psi} = \hat{\psi} (I_1, I_2, \ldots, I_\upsilon, \bm{X}) + \hat{\psi}_0;
\end{equation}
\begin{equation}
\label{eq:pbasis2}
\bar{\mathcal{P}} = \{ \bar{I}_1, \bar{I}_2, \ldots, \bar{I}_\zeta \}; \qquad \bar{I}_\alpha = \bar{I}_\alpha (\bar{\bm{C}},\bm{D}_\alpha,\nabla \bm{D}_\alpha), \qquad
\bar{\psi} = \bar{\psi} (\bar{I}_1, \bar{I}_2, \ldots, \bar{I}_\zeta, \bm{X}) + \bar{\psi}_0.
\end{equation}
The total number of applicable invariants is $\upsilon$ or $\zeta$ for \eqref{eq:psi2} or \eqref{eq:psi3}.  Stresses in \eqref{eq:Pk1} are then
\begin{equation}
\label{eq:Pk1b}
P^A_a = 2 g_{ab} F^b_B \rho_0 \sum_{\alpha = 1}^\upsilon \hat{\psi}_\alpha \frac{\partial I_\alpha }{\partial C_{AB}}, \quad \hat{\psi}_\alpha = \frac{ \partial \hat{\psi}} {\partial I_\alpha}; \quad
P^A_a = 2 \bar{g}_{ab} F^b_B \rho_0 \sum_{\alpha = 1}^\zeta \bar{\psi}_\alpha \frac{\partial \bar{I}_\alpha} {\partial \bar{C}_{AB}},
 \quad \bar{\psi}_\alpha = \frac{ \partial \bar{\psi}} {\partial \bar{I}_\alpha}.
\end{equation}

\subsection{Spatial momentum balances}
Spatial versions of \eqref{eq:linmom}, \eqref{eq:linmom2}, \eqref{eq:micromom}, and \eqref{eq:micromom2}
are derived in terms of Cauchy stress ${\bm{\sigma}} = \sigma^b_a dx^a \otimes \frac{\delta}{\delta x^b} $ 
and thermodynamic stresses ${\bm {\zeta}}^\alpha = \zeta\ua^b_A \delta D^A_\alpha \otimes \frac{\delta}{\delta x^b}$, $\alpha = 1, \ldots, r$:
\begin{equation}
\label{eq:spatialstresses}
J \sigma^b_a = F^b_A P^A_a, \qquad J \zeta\ua^b_A = F^b_B Z\ua^B_A.
\end{equation}
Let $N_A \diff S$ and $n_a \diff s$ be oriented area elements of $\partial \! \mathcal{M}$ and $\partial \mathfrak{m}$, and let $t_a$ and $\varsigma^\alpha_A$ be spatial tractions associated with $\bm \sigma$ and $\bm{\zeta}^\alpha$. Then from \eqref{eq:pa}, \eqref{eq:zA}, \eqref{eq:spatialstresses}, and Nanson's formula \cite{claytonDGKC2014,claytonIJF2017},
\begin{equation}
\label{eq:spatialtractions}
J N_A \diff S = F^a_A n_a \diff s ; \qquad
p_a \diff S = t_a \diff s, \quad z^\alpha_A \diff S = \varsigma^\alpha_A \diff s ;
\qquad
t_a = \sigma^b_a n_b, \quad \varsigma^\alpha_A = \zeta\ua^b_A n_b.
\end{equation}
Since $F^a_A = F^a_A(X) \Rightarrow \delta_B F^a_A = \partial_B F^a_A = \partial_B \partial_A \varphi^a = 
\partial_A \partial_B \varphi^a= \partial_A F^a_B = \delta_A F^a_B$, applying \eqref{eq:horiztranspf} and \eqref{eq:connrec} gives the horizontal covariant derivative of $\bm F$, recalling $\Gamma^A_{BC} = \Gamma^A_{CB}$ and $\Gamma^a_{bc} = \Gamma^a_{cb}$:
\begin{equation}
\label{eq:Fderiv}
F^a_{A|B} = \partial_B F^a_A - \Gamma^C_{BA} F^a_C + \Gamma^a_{be} F^e_A F^b_B = F^a_{B|A}.
\end{equation}
It follows as $j = J^{-1}$, $f^A_a = (F^{-1})^A_a$, and $(\sqrt{g/G})_{|A} = 0$, from \eqref{eq:Gids2}, \eqref{eq:Gids2c}, \eqref{eq:horiztranspf} and \eqref{eq:Fderiv}, 
\begin{equation}
\label{eq:piola1}
(j F^a_A)_{|a} =  (j F^a_A)_{|B} f^B_a = j f^B_a F^a_{A|B} + F^a_A f^B_a j_{|B} = 
 j f^B_a (F^a_{A|B} - F^a_{B|A}) = 0,  \quad (J f^A_a)_{|A} = 0.
\end{equation}
From \eqref{eq:thetasimp}, \eqref{eq:spatialstresses}, and \eqref{eq:piola1},
and defining spatial internal and external micro-forces ${\bm q}^\alpha$ and $\bm{e}^\alpha$:
\begin{equation}
\label{eq:spatialids}
P^A_{a|A} = J \sigma^b_{a|b}, \quad Z\ua^A_{C|A} = J \zeta\ua^b_{C|b}; \qquad
\bar{\partial}^\alpha_B (\cdot) = \delta_B^a \bar{\partial}^\alpha_a (\cdot);
\qquad J q^\alpha_C = Q^\alpha_C, \quad J e^\alpha_C = E^\alpha_C.
\end{equation}
Dividing \eqref{eq:linmom} and \eqref{eq:micromom} by $J$ and using \eqref{eq:connrec} and \eqref{eq:spatialids}
gives the spatial momentum balances:
\begin{equation}
\label{eq:linmoms}
\partial_b \sigma^b_a + \sum_{\alpha = 1}^r \bar{\partial}^\alpha_c \sigma^b_a \partial_b d^c_\alpha
+ \sigma^c_a \Gamma^b_{bc}
- \sigma^b_d \Gamma^d_{ba} + \rho b_a
= - \sigma^b_a \sum_{\alpha =1}^r C^{\alpha c}_{ec} (\partial_b d^e_\alpha + N\ua^e_b), 
\end{equation}
\begin{equation}
\label{eq:micromoms}
\begin{split}
  \partial_b \zeta\ua^b_C  +   \sum_{\beta = 1}^r & \bar{\partial}^\beta_e \zeta\ua^b_C \partial_b d^e_\beta  + \zeta\ua^b_C \Gamma^a_{ab}  - \zeta\ua^b_B K^{\alpha B}_{AC} f^A_b -   (q^\alpha_C - e^\alpha_C)
 \\ &  = 
 - \zeta\ua^b_A \delta_C^e ( \delta_d^A \bar{\partial}^\alpha_e N\ua^d_b 
 - \delta^D_a f_b^B \bar{\partial}^\alpha_e K^{\alpha A}_{BD} d^a_\alpha)
    - \zeta\ua^b_C  \sum_{\beta = 1}^r 
  C^{\beta d}_{ed} 
   (\partial_b d^e_{\beta} + N\ub^e_b).
\end{split} 
\end{equation}
Then from \eqref{eq:Gids2c}, balances \eqref{eq:linmoms} and \eqref{eq:micromoms} can be recast
to produce analogs of \eqref{eq:linmom2} and \eqref{eq:micromom2}: 
\begin{equation}
\label{eq:Gammagammas}
\Gamma^a_{ab} = \Gamma^a_{ba} = \delta_b (\ln \sqrt{g}) = \partial_b (\ln \sqrt{g})
-  \sum_{\alpha = 1}^r  N\ua_b^a \bar{\partial}^\alpha_a (\ln \sqrt{g})
 = \gamma^a_{ab} - \sum_{\alpha = 1}^r N\ua_b^a C^{\alpha d}_{ad} ;
\end{equation}
\begin{equation}
\label{eq:linmom2s}
\partial_b \sigma^b_a + \sum_{\alpha = 1}^r \bar{\partial}^\alpha_c \sigma^b_a \partial_b d^c_\alpha
+ \sigma^c_a \gamma^b_{bc}
- \sigma^b_d \Gamma^d_{ba} + \rho b_a
= - \sigma^b_a \sum_{\alpha =1}^r C^{\alpha c}_{ec} \partial_b d^e_\alpha , 
\end{equation}
\begin{equation}
\label{eq:micromom2s}
\begin{split}
  \partial_b \zeta\ua^b_C  +   \sum_{\beta = 1}^r & \bar{\partial}^\beta_e \zeta\ua^b_C \partial_b d^e_\beta  + \zeta\ua^b_C \gamma^a_{ab}  - \zeta\ua^b_B K^{\alpha B}_{AC} f^A_b -   (q^\alpha_C - e^\alpha_C)
 \\ &  = 
 - \zeta\ua^b_A \delta_C^e ( \delta_d^A \bar{\partial}^\alpha_e N\ua^d_b 
 - \delta^D_a f_b^B \bar{\partial}^\alpha_e K^{\alpha A}_{BD} d^a_\alpha)
    - \zeta\ua^b_C  \sum_{\beta = 1}^r 
  C^{\beta d}_{ed} \partial_b d^e_{\beta} .
\end{split}
 \end{equation}

\section{Interpretation and comparison with Riemannian geometry}
If the internal state or director vectors ($D_\alpha,d_\alpha$) are known as explicit, one-to-one and continuously differentiable functions
of (reference, spatial) position ($X,x$) over some or all of the base manifold ($\mathcal{M},\mathfrak{m}$), then horizontal Finslerian metrics 
($\bm{G},\bm{g}$) entering \eqref{eq:Sasaki} and \eqref{eq:Sasakic} can be recast as locally osculating Riemannian metrics.
In this setting, balance laws for mass, linear momentum, and micro-momentum can be rewritten in
forms more similar to those of nonlinear elasticity and phase-field mechanics.
Under certain simplifying conditions, the former reduce to the latter (i.e., classical) forms.
The osculating Riemannian setting is useful for analyzing residual stress and strain, as in theories
for nonlinear elastic biologic continua with Riemannian metrics \cite{taka1990,taka2013,klarbring2007,yavari2010}.

\subsection{Osculating Riemannian metrics}
Recall \eqref{eq:funcC}: $D^A_\alpha = D^A_\alpha(X)$ and $d^a = d^a_\alpha(x)$, one-to-one 
for $\alpha = 1, \ldots, r$ and at least $C^2$ differentiable 
over coordinate patches parameterized by charts $\{X^A\}$ and $\{x^a\}$ with $a,A = 1, \ldots, n$ \cite{rund1959,amari1962,rund1975,claytonJGP2017}. Define the osculating Riemannian metric tensors and their determinants as in Corollary A.1 of Appendix A \cite{claytonSYMM2023}:
\begin{equation}
\label{eq:oscGg}
\tilde{G}_{AB} (X) = G_{AB}(X,D_\alpha(X)), \quad \tilde{G} = \det (\tilde{G}_{AB}) ;
\quad
\tilde{g}_{ab}(x) = g_{ab}(x,d_\alpha(x)), \quad \tilde{g} = \det (\tilde{g}_{ab}).
\end{equation}
Tensors are $\tilde{\bm{G}} = \tilde{G}_{AB} dX^A \otimes dX^B$ and
 $\tilde{\bm{g}} = \tilde{g}_{ab} dx^a \otimes dx^b$.
 Inverse components are $\tilde{G}^{AB}(X)$ and $\tilde{g}^{ab}(x)$.
Osculating Levi-Civita connection $\tilde{\gamma}^A_{BC}(X)$ is defined in terms of
$\tilde{G}_{AB}$
in \eqref{eq:LC1o}. Spatial coefficients $\tilde{\gamma}^a_{bc}(x)$ are
 \begin{equation}
\label{eq:LC1oc}
\tilde{\gamma}^a_{bc}={\textstyle{\frac{1}{2}}} \tilde{g}^{ad} (\partial_c \tilde{g}_{bd} + \partial_b \tilde{g}_{cd} - \partial_d \tilde{g}_{bc})
=\tilde{g}^{ad} \tilde{\gamma}_{bcd} = \tilde{\gamma}^a_{cb}.
\end{equation}
Per usual identities of Riemannian geometry \cite{claytonDGKC2014}, where $\partial_A(\cdot) = \partial(\cdot)/\partial X^A$ and $\partial_a(\cdot) = \partial(\cdot)/\partial x^a$,
\begin{equation}
\label{eq:Gidsosc}
\partial_A (\ln \sqrt{\tilde{G}}) = \tilde{\gamma}^B_{AB} = \tilde{\gamma}^B_{BA}, \qquad
\partial_a (\ln \sqrt{\tilde{g}}) = \tilde{\gamma}^b_{ab} =  \tilde{\gamma}^b_{ba}.
\end{equation}
From \eqref{eq:Gids2}, \eqref{eq:Gids2c}, \eqref{eq:oscGg}, and \eqref{eq:Gidsosc},
\begin{equation}
\label{eq:osc1}
\tilde{\gamma}^B_{AB} =  \partial_A (\ln \sqrt{\tilde{G}}) = \partial_A (\ln \sqrt{G}) + \sum_{\alpha = 1}^r \bar{\partial}^\alpha_B (\ln \sqrt{G})
\partial_A D^B_\alpha 
= \gamma^B_{AB} +  \sum_{\alpha = 1}^r C^{\alpha E}_{BE}  \partial_A D^B_\alpha,
\end{equation}
\begin{equation}
\label{eq:osc1c}
\tilde{\gamma}^b_{ab} = \partial_a (\ln \sqrt{\tilde{g}}) = \partial_a (\ln \sqrt{g}) + \sum_{\alpha = 1}^r \bar{\partial}^\alpha_b (\ln \sqrt{g})
\partial_a d^b_\alpha 
= \gamma^b_{ab} +  \sum_{\alpha = 1}^r C^{\alpha e}_{be}  \partial_a d^b_\alpha.
\end{equation}

\subsection{Local equilibrium equations}
Similar to \eqref{eq:oscGg} and appealing to \eqref{eq:funcC}, define the following functional forms for 
mass densities, mechanical stresses, and thermodynamic micro-stresses:
\begin{equation}
\label{eq:defosc}
\tilde{\rho}_0 (X) = \rho_0(X,D_\alpha(X)), \quad \tilde{\rho}(x) = \rho(x,d_\alpha(x)); \quad
\tilde{P}^A_a (X) = P^A_a (X,D_\alpha(X)), \quad \tilde{\sigma}^b_a (x) = \sigma^b_a (x,d_\alpha(x));
\end{equation}
\begin{equation}
\label{eq:defosc2}
\tilde{Z}\ua^A_B (X) = Z\ua^A_B (X,D_\alpha(X)), \qquad \tilde{\zeta}\ua^a_B (x) = \zeta\ua^a_B (x,d_\alpha(x)).
\end{equation}

First consider the local balance of mass. Define the cumulative mass source field for all fiber families $\tilde{S}^\alpha(X) = \sum_{\alpha = 1}^r S^\alpha_A \delta (D^A_\alpha)$. Then localizing \eqref{eq:globmass} with \eqref{eq:oscGg} and \eqref{eq:defosc} produces the following local variational mass conservation law and the analogs of
incremental equations \eqref{eq:locmass} and \eqref{eq:locmassspatial} in the osculating Riemannian setting, at each material point $X \in \mathcal{M}$ where \eqref{eq:oscGg} and \eqref{eq:defosc}  hold:
\begin{equation}
\label{eq:locmassosc}
\delta \tilde{\rho}_0 = \tilde{S} - \tilde{\rho}_0 \delta ( \ln \sqrt{\tilde{G}}); \qquad
\frac{\diff \tilde{\rho}_0}{\tilde{\rho}_0} = \frac{ \tilde{S}}{\tilde{\rho}_0} - \diff  \ln \sqrt{\tilde{G}},
\quad
\frac{\diff \tilde{\rho}}{\tilde{\rho}} = - \tilde{\nabla}_a (\diff \varphi^a) + \frac{ \tilde{s}}{\tilde{\rho}} - \diff  \ln \sqrt{\tilde{g}}.
\end{equation}
Here $\tilde{\nabla}_a (\diff \varphi^a) = \partial_a (\diff \varphi^a) + \tilde{\gamma}^a_{ab} \diff \varphi^b$ and $ \tilde{s} = j \tilde{S}$. Balances similar to \eqref{eq:locmassosc} are given in Refs.~\cite{yavari2010,sadik2016}.

Next consider the local balance of linear momentum. Differentiating the mechanical stress measures in \eqref{eq:defosc} produces
\begin{equation}
\label{eq:diffstressosc}
\partial_A \tilde{P}^A_a = \partial_A P^A_a + \sum_{\alpha = 1}^r \bar{\partial}^\alpha_B P^A_a \partial_A D^B_\alpha, 
\qquad
\partial_b \tilde{\sigma}^b_a = \partial_b \sigma^b_a + \sum_{\alpha = 1}^r \bar{\partial}^\alpha_c \sigma^b_a \partial_b d^c_\alpha.
\end{equation}
Using \eqref{eq:diffstressosc}, substituting \eqref{eq:osc1} into \eqref{eq:linmom2}, and substituting \eqref{eq:osc1c}
into \eqref{eq:linmom2s} gives the referential and spatial static equilibrium equations for linear momentum in
the osculating Riemannian setting:
\begin{equation}
\label{eq:linmom2osc}
\partial_A \tilde{P}^A_a + \tilde{P}^B_a \tilde{\gamma}^A_{AB}
- \tilde{P}^A_c \tilde{\gamma}^c_{ba} F^b_A  + \rho_0 b_a
= \tilde{P}^A_c (\Gamma^c_{ba} -  \tilde{\gamma}^c_{ba}) F^b_A,
\end{equation}
\begin{equation}
\label{eq:linmom2oscs}
\partial_b \tilde{\sigma}^b_a + \tilde{\sigma}^c_a \tilde{\gamma}^b_{bc}
- \tilde{\sigma}^b_d \tilde{\gamma}^d_{ba}  + \rho b_a
=  \tilde{\sigma}^b_d (\Gamma^d_{ba} -  \tilde{\gamma}^d_{ba} ). 
\end{equation}
When $\Gamma^a_{bc} = \tilde{\gamma}^a_{bc}$, right sides vanish, and \eqref{eq:linmom2osc}
and \eqref{eq:linmom2oscs} match static equilibrium equations in Refs.~\cite{taka1990,klarbring2007,taka2013,yavari2010} dealing with growth mechanics and residual stresses.
Note that when $\mathfrak{m}$ is assigned a fully Euclidean metric tensor
with components $g_{ab} = g_{ab}(x)$, then $\Gamma^a_{bc} = \gamma^a_{bc}$,
$C^{\alpha a}_{bc} = 0$, and thus $\Gamma^a_{bc} = \tilde{\gamma}^a_{bc}$.
In this special case, if decomposition \eqref{eq:mdecc} is used, then $\hat{g}^a_b = \delta^a_b$ and
$g_{ab}(x) = \bar{g}_{ab}(x)$.
When both $\mathfrak{m}$ and $\mathcal{M}$ are assigned fully Euclidean metrics, the latter
gives  $G_{AB} = G_{AB}(X)$, $\Gamma^A_{BC} = \gamma^A_{BC} = \tilde{\gamma}^A_{BC}$,
$C^{\alpha A}_{BC} = 0$, 
and if \eqref{eq:mdec} is used, $\hat{G}^A_B = \delta^A_B$ and
$G_{AB}(X) = \bar{G}_{AB}(X)$. In this special case, \eqref{eq:linmom2osc}
and \eqref{eq:linmom2oscs} reduce to equilibrium equations of classical nonlinear elasticity \cite{claytonNMC2011,marsden1983}.

Finally consider the local balance of micro-momentum associated with internal state equilibrium.
Differentiating \eqref{eq:defosc2} gives
\begin{equation}
\label{eq:diffzsosc}
\partial_A \tilde{Z}\ua^A_C = \partial_A Z\ua^A_C + \sum_{\alpha = 1}^r \bar{\partial}^\alpha_B Z\ua^A_C \partial_A D^B_\alpha, 
\qquad
\partial_b \tilde{\zeta}\ua^b_c = \partial_b {\zeta}\ua^b_c + \sum_{\alpha = 1}^r \bar{\partial}^\alpha_e {\zeta}\ua^b_c \partial_b d^e_\alpha.
\end{equation}
Using \eqref{eq:diffzsosc}, substituting \eqref{eq:osc1} into \eqref{eq:micromom2}, and substituting \eqref{eq:osc1c}
into \eqref{eq:micromom2s} gives the referential and spatial static equilibrium equations for linear momentum in
the osculating Riemannian setting:
\begin{equation}
\label{eq:microsc}
\begin{split}
  \partial_A \tilde{Z}\ua^A_C  +  \tilde{Z}\ua^B_C \tilde{\gamma}^A_{AB}  -  \tilde{Z}\ua^A_B \tilde{\gamma}^{B}_{AC}
   -   (Q^\alpha_C - E^\alpha_C)
  = 
\tilde{Z}\ua^A_B [(K^{\alpha B}_{AC} - \tilde{\gamma}^{B}_{AC}) - ( \bar{\partial}^\alpha_C N\ua^B_A - \bar{\partial}^\alpha_C K^{\alpha B}_{AD} D^D_\alpha)],
\end{split} 
\end{equation}
\begin{equation}
\label{eq:microscs}
\begin{split}
  \partial_b \tilde{\zeta}\ua^b_C   + \tilde{\zeta}\ua^b_C \tilde{\gamma}^a_{ab}  & - \tilde\zeta\ua^b_B \tilde{\gamma}^{B}_{AC} f^A_b -   (q^\alpha_C - e^\alpha_C)
 \\ & \qquad  = \tilde\zeta\ua^b_B [(K^{\alpha B}_{AC}-\tilde{\gamma}^{B}_{AC} ) f^A_b 
 - \delta_C^e ( \delta_d^B \bar{\partial}^\alpha_e N\ua^d_b 
 - \delta^D_a f_b^A \bar{\partial}^\alpha_e K^{\alpha B}_{AD} d^a_\alpha)].
\end{split}
 \end{equation}
 Consider first the special case when $N\ua^A_B = N\ua^A_B(X)$, 
 $N\ua^a_b = N\ua^a_b(x)$, and $K^{\alpha A}_{BC} = K^{\alpha A}_{BC}(X)$.
 Then the rightmost two terms in \eqref{eq:microsc} and \eqref{eq:microscs} vanish.
 Now further choosing $K^{\alpha A}_{BC}(X) = \tilde{\gamma}^A_{BC}(X) \, \forall \, \alpha = 1, \ldots, r$,
 the right sides of \eqref{eq:microsc} and \eqref{eq:microscs} vanish entirely.
 Then if $\mathcal{M}$ is assigned a Euclidean metric $G_{AB} = G_{AB}(X)$, $\tilde{\gamma}^A_{BC} = \gamma^A_{BC}$ and \eqref{eq:microsc} reduces to $r$ Euler-Lagrange equations
 matching those of non-dissipative phase-field theories of Allen-Cahn or Ginzburg-Landau type \cite{gurtin1996,claytonPHYSD2011,claytonJMPS2021} for $r$ vector-valued order parameters ${\bm D}_\alpha$.
 The spatial analog follows from \eqref{eq:microscs} if $\mathfrak{m}$ is also assigned Euclidean metric $g_{ab} = g_{ab}(x)$ giving $\tilde{\gamma}^a_{bc} = \gamma^a_{bc}$.
 In the fully Euclidean setting, $\hat{G}^A_B = \delta^A_B$ and $\hat{g}^a_b = \delta^a_b$.

\subsection{Curvature and residual stress}
Components of Riemann-Christoffel curvature tensors $\tilde{\mathcal{R}}^A_{BCD}(X)$ and $\tilde{\mathcal{R}}^a_{bcd}(x)$ on $\mathcal{M}$ on $\mathfrak{m}$ are \cite{claytonDGKC2014}
\begin{equation}
\label{eq:RCurv}
\tilde{\mathcal{R}}^A_{BCD} = 
\partial_B \tilde{\gamma}^A_{CD} - \partial_C \tilde{\gamma}^A_{BD}
+ \tilde{\gamma}^A_{BE}  \tilde{\gamma}^E_{CD}
- \tilde{\gamma}^A_{CE}  \tilde{\gamma}^E_{BD},
\quad
\tilde{\mathcal{R}}^a_{bcd} = 
\partial_b \tilde{\gamma}^a_{cd} - \partial_c \tilde{\gamma}^a_{bd}
+ \tilde{\gamma}^a_{be}  \tilde{\gamma}^e_{cd}
- \tilde{\gamma}^a_{ce}  \tilde{\gamma}^e_{bd}.
\end{equation}
With $\tilde{\gamma}^A_{BC}$ and $\tilde{\gamma}^a_{bc}$ symmetric Levi-Civita connections
of $\tilde{G}_{AB}$ and $\tilde{g}_{ab}$,
substituting \eqref{eq:LC1o} and \eqref{eq:LC1oc} into \eqref{eq:RCurv} produces 
covariant versions $\tilde{\mathcal{R}}_{ABCD} =  \tilde{\mathcal{R}}^E_{ABC} \tilde{G}_{DE}$
and $\tilde{\mathcal{R}}_{abcd} =  \tilde{\mathcal{R}}^e_{abc} \tilde{g}_{de} $
\cite{claytonDGKC2014}:
\begin{equation}
\label{eq:RCurv1}
\tilde{\mathcal{R}}_{ABCD} =  
{\textstyle{\frac{1}{2}}} (\partial_A \partial_C \tilde{G}_{BD} + \partial_B \partial_D \tilde{G}_{AC}
- \partial_A \partial_D \tilde{G}_{BC} - \partial_B \partial_C \tilde{G}_{AD})
+\tilde{G}^{EF}(\tilde{\gamma}_{ACE} \tilde{\gamma}_{BDF} - \tilde{\gamma}_{BCE} \tilde{\gamma}_{ADF}),
\end{equation}
\begin{equation}
\label{eq:RCurv2}
\tilde{\mathcal{R}}_{abcd} =  
{\textstyle{\frac{1}{2}}} (\partial_a \partial_c \tilde{g}_{bd} + \partial_b \partial_d \tilde{g}_{ac}
- \partial_a \partial_d \tilde{g}_{bc} - \partial_b \partial_c \tilde{g}_{ad})
+\tilde{g}^{ef}(\tilde{\gamma}_{ace} \tilde{\gamma}_{bdf} - \tilde{\gamma}_{bce} \tilde{\gamma}_{adf}).
\end{equation}

Now invoke \eqref{eq:oscGg} and linear momentum balances \eqref{eq:linmom2osc} and \eqref{eq:linmom2oscs}.
In the latter two, assume $\Gamma^a_{bc} = \tilde{\gamma}^a_{bc}$ so that right sides vanish indentically.
Further invoke the free energy density $\psi = \hat{\psi}(\bm{C}, \bm{D}_\alpha, \nabla \bm{D}_\alpha, \bm{X})$
of \eqref{eq:psi2}. Assume, per convention in (compressible\footnote{
If the material is truly incompressible, the variational framework can include a Lagrange multiplier
contributing to local pressure along with constraint $J(X,D_\alpha) = 1$ \cite{kumar2023,marsden1983}.
Local stress depends on boundary conditions in addition to local strain.
However, per convention, stress should vanish if
boundaries are traction-free and elastic strain vanishes. 
}) nonlinear elasticity, that
$\hat{\psi}$ depends on ${\bm{C}}$ in such a way that at any point $X \in \mathcal M$,
$C^A_B = \delta^A_B \Rightarrow C_{AB} = G_{AB} \Rightarrow 
\epsilon^e_{AB} = \frac{1}{2} (C_{AB} - G_{AB}) = 0 \Rightarrow P^A_a = 0 \Rightarrow \sigma^b_a = 0$.
In other words, local vanishing of ``elastic strain'' $\epsilon^e_{AB}$ gives locally vanishing stress.
Define $\check{\mathcal{R}}^A_{BCD}(X)$ as components of the Riemann-Christoffel curvature tensor
formed from the Levi-Civita connection with $\tilde{C}_{AB}(X) = C_{AB} (X,D_\alpha(X))$ as the metric.
Then since $F^a_A(X) = \partial_A \varphi^a (X) $ is integrable, pull-back relations apply: $\tilde{C}_{AB} = \varphi^* (\tilde{g}_{ab}) =
F^a_A \tilde{g}_{ab} F^b_B$
and  $\check{\mathcal{R}}^A_{BCD} = \varphi^* ( \tilde{\mathcal{R}}^a_{bcd})$ \cite{marsden1983,claytonDGKC2014}.
It follows that $\tilde{\mathcal{R}}^a_{bcd} (\varphi(X)) = 0 \Leftrightarrow \check{\mathcal{R}}^A_{BCD} (X)= 0$.
Four cases are conceivable for curvatures in \eqref{eq:RCurv}--\eqref{eq:RCurv2} over (regions of) $\mathcal{M}$ and $\mathfrak{m}$:
\begin{enumerate}
\item 
$\tilde{\mathcal{R}}^a_{bcd} = 0$ and $\tilde{\mathcal{R}}^A_{BCD} = 0$.
Both $\mathcal{M}$ and $\mathfrak{m}$ admit Euclidean coordinate charts $\{ X^A \}$ and $\{x ^a \}$ and Euclidean metric tensors $\tilde{G}_{AB}$ and $\tilde{g}_{ab}$.
It is possible to prescribe $\tilde{C}_{AB} = \tilde{G}_{AB} = \delta_{AB}$ over corresponding regions of $\mathcal{M}$.
From constitutive assumptions above, stresses $P^A_a$ and $\sigma^b_a$ must vanish for this prescription, so tractions $p_a = P^A_a N_A = 0$ on $\partial \! \mathcal{M}$  and $t_a = \sigma^b_a n_b = 0$ on $\partial \mathfrak{m}$. The body is free of residual stress.
This is the case of classical nonlinear elasticity \cite{claytonNMC2011,marsden1983}.
\item 
$\tilde{\mathcal{R}}^a_{bcd} = 0$ and $\tilde{\mathcal{R}}^A_{BCD} \neq 0$.
Here $\mathfrak{m}$ is Euclidean but $\mathcal{M}$ is not.  Since $\check{\mathcal{R}}^A_{BCD} = 0$,
it is not possible to prescribe $\tilde{C}_{AB} = \tilde{G}_{AB} \, \forall \, X \in \mathcal{M}$.
Therefore, even if $p_a = 0 \, \forall \, X \in \partial \! \mathcal{M}$, $\epsilon^e_{AB}(X) \neq 0$ for
some $X \in \mathcal{M}$ (i.e., in the interior of the body), meaning the body is residually stressed.
Physically, internal stresses are needed to fit a curved material manifold into a flat spatial domain.
This is the case modeled in Refs.~\cite{taka1990,taka2013,yavari2010}.
\item 
$\tilde{\mathcal{R}}^a_{bcd} \neq 0$ and $\tilde{\mathcal{R}}^A_{BCD} = 0$.
Now $\mathcal{M}$ is Euclidean but $\mathfrak{m}$ is not.  Since $\check{\mathcal{R}}^A_{BCD} \neq 0$
but $\tilde{G}_{AB}$ is Euclidean,
it is not possible to prescribe $\tilde{C}_{AB} = \tilde{G}_{AB} \, \forall \, X \in \mathcal{M}$.
As such, even if $p_a = 0 \, \forall \, X \in \partial \! \mathcal{M}$, $\epsilon^e_{AB}(X) \neq 0$ for
some $X \in \mathcal{M}$, so the body is residually stressed.
Physically, internal stresses are needed to instill curvature into a flat material manifold.
\item $\tilde{\mathcal{R}}^a_{bcd} \neq 0$ and $\tilde{\mathcal{R}}^A_{BCD} \neq 0$.
Both $\mathcal{M}$ and $\mathfrak{m}$ are non-Euclidean.
It is possible, but not necessary, that $\epsilon^e_{AB}(X) = 0 \, \forall \, X \in \mathcal{M}$.
No definitive conclusions arise on existence of residual stresses.
\end{enumerate}
Further consider the metric decompositions \eqref{eq:mdec} and \eqref{eq:mdecc}.
Let $\bar{G}_{AB}$ and $\bar{g}_{ab}$ be Euclidean, and choose preferred coordinate systems
wherein $\bar{G}_{AB}  = \delta_{AB} $ and $\bar{g}_{ab} = \delta_{ab}$.
It follows that $\tilde{G}_{AB}(X) = \delta_{AC} \hat{G}^C_B (D_\alpha(X)) = \hat{G}_{AB}(X)$ and
$\tilde{g}_{ab}(x) = \delta_{ac} \hat{g}^c_b (d_\alpha(x)) = \hat{g}_{ab}(x)$.
Furthermore, $\tilde{\mathcal{R}}^A_{BCD} = \hat{\mathcal{R}}^A_{BCD}$
and $\tilde{\mathcal{R}}^a_{bcd} = \hat{\mathcal{R}}^a_{bcd}$, where
$\hat{\mathcal{R}}^A_{BCD}$ and $ \hat{\mathcal{R}}^a_{bcd}$ are curvature tensors of $\hat{G}_{AB}(X)$ and
$\hat{g}_{ab}(x)$. In the former,
\begin{equation}
\label{eq:ghatderiv}
\begin{split}
\partial_A \hat{G}_{CD} & = \delta_{CG} \sum_{\alpha = 1}^r (\bar{\partial}^\alpha_E \hat{G}^G_D) \partial_A D^E_\alpha, \\
\partial_A \partial_B \hat{G}_{CD} & = \delta_{CG} \bigr{[} \sum_{\alpha = 1}^r
\sum_{\beta = 1}^r 
( \bar{\partial}^\alpha_E  \bar{\partial}^\beta_F \hat{G}^G_D)   \partial_B D^F_\beta   \partial_A D^E_\alpha
+ \sum_{\alpha = 1}^r (\bar{\partial}^\alpha_E \hat{G}^G_D) \partial_B \partial_A D^E_\alpha \bigr{]}.
\end{split}
\end{equation}
Analogous expressions hold for $\partial_a \hat{g}_{cd}$ and $\partial_a \partial_b \hat{g}_{cd}$.
In the current setting, necessary and sufficient conditions for $\hat{G}^A_B(D_\alpha(X))$$=\text{constant}$
are $C^{\alpha A}_{BC} = 0$ \cite{sokolnikoff1951}
and similarly for $\hat{g}^a_b(d_\alpha(x))$$=\text{constant}$, $C^{\alpha a}_{bc} = 0$.
From this statement and \eqref{eq:ghatderiv}, $\hat{\mathcal{R}}^A_{BCD} = 0$
holds if one or more of the following conditions applies for all $\alpha = 1, \ldots, r$:
$C^{\alpha A}_{BC} = 0$, $\bar{\partial}^\alpha_E \hat{G}^G_D = 0$, or
$\partial_A D^E_\alpha = 0$.
Similarly, $\hat{\mathcal{R}}^a_{bcd} = 0$
holds if any or all of the following apply for all $\alpha = 1, \ldots, r$:
$C^{\alpha a}_{bc} = 0$, $\bar{\partial}^\alpha_e \hat{g}^c_d = 0$, or
$\partial_a d^e_\alpha = 0$.
Residual stresses thus originate from non-vanishing Cartan tensors and material or spatial gradients of
internal state.
It is shown by example in \S5 that $\tilde{\mathcal{R}}^A_{BCD}$ ($\tilde{\mathcal{R}}^a_{bcd}$) can
be nonzero even if $\partial_A D^B_\alpha = 0 $ ($\partial_a d^b_\alpha = 0$) when $\bar{G}_{AB} \neq \delta_{AB}$
($\bar{g}_{ab} \neq \delta_{ab}$). In such general settings, uniform internal states can induce residual stress.

Next, instead of $\hat{\psi}$, prescribe $\psi  
 = \bar{\psi}(\bar{\bm{C}}, \bm{D}_\alpha, \nabla \bm{D}_\alpha, \bm{X})$
of \eqref{eq:psi3}. Let 
$\bar{\psi}$ depend on $\bar{\bm{C}}$ in a manner that any point $X \in \mathcal M$,
$\bar{C}^A_B = \delta^A_B \Rightarrow \bar{C}_{AB} = \bar{G}_{AB} \Rightarrow 
\bar{\epsilon}^e_{AB} = \frac{1}{2} (\bar{C}_{AB} - \bar{G}_{AB}) = 0 \Rightarrow P^A_a = 0 \Rightarrow \bar{\sigma}^b_a = 0$.
Again take $\Gamma^a_{bc} = \tilde{\gamma}^a_{bc}$ so the right side of \eqref{eq:linmom2osc} vanishes.
Define $\bar{\mathcal{R}}^A_{BCD}$ and $\bar{\mathcal{R}}^a_{bcd}$ as curvature tensors of $\bar{G}_{AB}(X)$ and $\bar{g}_{ab}(x)$ in respective \eqref{eq:mdec} and \eqref{eq:mdecc}, and define
$\underline{\mathcal{R}}^A_{BCD}$ as the curvature tensor of $\bar{C}_{AB}(X)$.
Then pull-back relations are $\bar{C}_{AB} = \varphi^*(\bar{g}_{ab})$ and $\underline{\mathcal{R}}^A_{BCD}
=  \varphi^*(\bar{\mathcal{R}}^a_{bcd})$.
Assume again that $\bar{G}_{AB}$ and $\bar{g}_{ab}$ are Euclidean. Then
all three curvatures vanish: $\bar{\mathcal{R}}^A_{BCD} = 0$ and $\bar{\mathcal{R}}^a_{bcd} =0 \Leftrightarrow \underline{\mathcal{R}}^A_{BCD} = 0$. The body is free of residual stress, irrespective of Cartan tensors or gradients of internal state, because strain energy depends on $\bar{\bm C}$ rather than $\bm C$.
This is one model used in Ref.~\cite{claytonSYMM2023}.

Now consider more general situations where $\Gamma^a_{bc} \neq \tilde{\gamma}^a_{bc}$.
In such instances, right sides of \eqref{eq:linmom2osc} and \eqref{eq:linmom2oscs}
produce effective body forces, linearly proportional to stress components, that can affect residual
stress distributions. However, aforementioned remarks on vanishing of residual stresses still hold, since
right sides of  \eqref{eq:linmom2osc} and \eqref{eq:linmom2oscs} are zero when (residual) stresses
are zero, and static equilibrium is maintained.  Sufficient conditions for $\Gamma^a_{bc} = \tilde{\gamma}^a_{bc}$
are $N\ua^b_a = 0$ along with $C^{\alpha a}_{bc} = 0$ and/or $\partial_a d^b_\alpha = 0 $,
for all $\alpha = 1, \ldots, r$.  These sufficiency conditions render $\hat{\mathcal{R}}^a_{bcd} = 0$, but it is not necessary that residual stresses always vanish when $\Gamma^a_{bc} = \tilde{\gamma}^a_{bc}$.

In summary, the explicit role of Finslerian metric dependence of internal state vectors $(D^A_\alpha, d^a_\alpha)$ and their gradients $(\partial_B D^A_\alpha, \partial_b d^a_\alpha)$ on residual stresses has been identified.
Although the osculating Riemannian setting, with  momentum conservation equations of the form \eqref{eq:linmom2osc} and \eqref{eq:linmom2oscs}, has been proven useful for understanding origins of residual stress, the proposed Finslerian theory potentially enables description of broader kinds of physics than
Riemannian theory \cite{taka1990,taka2013,yavari2010}.
In contrast to the latter, the Finslerian theory admits non-vanishing right sides of \eqref{eq:linmom2osc}, \eqref{eq:linmom2oscs},  \eqref{eq:microsc}, and \eqref{eq:microscs}.
Effects of uniquely Finslerian terms, including nonlinear connection coefficients $N\ua^A_B$ and
$N\ua^a_b$, have been quantified elsewhere in the context of modeling of skin tissue \cite{claytonSYMM2023} with $r =1$.

\section{Left ventricle with residual stress, remodeling, and damage}
Following Refs.~\cite{janz1976,taka1990}, the left ventricle is idealized as a hollow (thick-walled) sphere
subjected to internal pressurization. However, the present analysis
invokes a generalized Finsler theory of \S3 rather than Riemannian geometry of Ref.~\cite{taka1990}, and 
a different strain energy function and elastic stress-strain law are used for nonlinear elasticity than those of Refs.~\cite{janz1976,taka1990}.
Second, in some calculations, internal state is allowed to evolve with loading, leading to redistribution of
internal (i.e., residual or remnant) strains and stresses.
Finally, degradation (i.e., damage) of cardiac tissue is included, as might be induced by very large
internal (tension) or external (compression) pressure.

\subsection{Geometry and kinematics}
In a reference configuration parameterized by spherical coordinates $\{X^A\}$, the base manifold is
the simply connected domain
$\{ \mathcal{M}: X^1 = R \in [R_i, R_o], X^2 = \Theta \in [0,\pi], X^3 = \Phi \in (-\pi,\pi] \}$.
The inner and outer radii are the constants $R_i$ and $R_o$.
Two fiber families with internal state coordinates $\{ D^A_\alpha \}$ are considered ($r = 2$).
All $D^A_\alpha$ components are
defined as physically dimensionless. 
The first family ($\alpha = 1$) measures effects of growth or remodeling in $(R,\Theta,\Phi)$ fiber directions,
that is, $D^1_1 = D_R$, $D^2_1 = D_\Theta$, $D^3_1 = D_\Phi$, where all three components are unrestricted in sign and magnitude.  
The second measures isotropic damage so requires only one nonzero component $\eta$ with physical range $\eta \in [0,1]$.  Per usual theories of continuum damage mechanics \cite{claytonNMC2011} and phase-field fracture \cite{claytonIJF2014,claytonIJF2017,gultekin2019,claytonPRE2024}, degradation increases with increasing $\eta$, with $\eta(X) = 0$ the pristine state at $X$ and $\eta(X) = 1$ a locally ``failed'' or fluidic state with no material shear strength at location $X$. 
Without loss of generality, $D^1_2 = \eta$, $D^2_2 = D^3_2 = 0$. The choice of nonzero first slot for $\{ D^A_2 \}$ rather than second or third is arbitrary and does not imply any (e.g., purely radial) directionality of damage.

In a spatial configuration, the base manifold is $\{ \mathfrak{m}: x^1 = r \in [r_i, r_o], x^2 = \theta \in [0,\pi], x^3 = \phi \in (-\pi,\pi] \}$ with inner and outer radii $r_i$ and $r_o$.
From \eqref{eq:thetasimp}, $d^a_\alpha (\varphi(X)) = \delta^a_A D^A_\alpha (X)$, so, with arguments suppressed, 
 $d^1_1 = D_R$, $d^2_1 = D_\Theta$, $d^3_1 = D_\Phi$,
  $d^1_2 = \eta$, and $d^2_2 = d^3_2 = 0$.
  Summarizing,
  \begin{equation}
  \label{eq:spherecoord}
  \begin{split}
  \{X^1 ,X^2 ,X^3 \} & = \{R,\Theta,\Phi \},  \qquad \quad \quad \, \, \, \, \{x^1,x^2,x^3 \}  = \{ r,\theta, \phi \}, \
  \\ \{ D^1_1,D^2_1,D^3_1 \} &= \{ D_R, D_\Theta,D_\Phi \}, \qquad \{ D^1_2,D^2_2,D^3_2 \} = \{ \eta, 0, 0 \}.
  \end{split}
  \end{equation}
  The reference configuration is assigned a Finslerian metric $G_{AB}(X,D_\alpha)$, whereas
  the current configuration is assigned a Riemannian, and more specifically, Euclidean metric $g_{ab}(x)$
  of the usual form for spherical coordinates.  Decompositions in \eqref{eq:mdec} and \eqref{eq:mdecc}
  are used, where in matrix form,
  \begin{equation}
  \label{eq:spheremetR}
  [G_{AB}] = [\bar{G}_{AC}] [\hat{G}^C_B] = 
  \begin{bmatrix}
  \bar{G}_{RR} &  0 &  0 \\
  0 & \bar{G}_{\Theta \Theta} & 0 \\
  0 & 0 & \bar{G}_{\Phi \Phi} 
    \end{bmatrix}
    \begin{bmatrix}
  G_\eta^{1/3} \hat{G}_R &  0 &  0 \\
  0 &  G_\eta^{1/3} \hat{G}_{\Theta} & 0 \\
  0 & 0 &  G_\eta^{1/3} \hat{G}_{\Phi} 
    \end{bmatrix},
  \end{equation}
  \begin{equation}
  \label{eq:Gsphere}
  \begin{split}
 &  \bar{G}_{RR} = 1, \quad \bar{G}_{\Theta \Theta} =  R^2, \quad \bar{G}_{\Phi \Phi} = R^2 \sin^2 \Theta; 
  \qquad G_\eta = \exp ( 2 {k_\eta \eta^{\hat{m}}}/{\hat{m}} );
 \\ &  \hat{G}_R = \frac{1}{ \Lambda_R^2} 
 = \exp \biggr{[} \frac{ 2 k_D}{\hat{n}} \{ D^{\hat{n}}_R - {\textstyle{\frac{1}{2}}}( D^{\hat{n}}_\Theta + D^{\hat{n}}_\Phi ) \} \biggr{]},
 \quad 
  \hat{G}_\Theta = \frac{1}{ \Lambda_\Theta^2} 
 = \exp \biggr{[} \frac{ 2 k_D}{\hat{n}} \{ D^{\hat{n}}_\Theta - {\textstyle{\frac{1}{2}}}( D^{\hat{n}}_\Phi + D^{\hat{n}}_R ) \} \biggr{]},
 \\
 & 
 \hat{G}_\Phi = \frac{1}{ \Lambda_\Phi^2} 
 = \exp \biggr{[} \frac{ 2 k_D}{\hat{n}} \{ D^{\hat{n}}_\Phi - {\textstyle{\frac{1}{2}}}( D^{\hat{n}}_R+ D^{\hat{n}}_\Theta ) \} \biggr{]}; \qquad
 G = \det [G_{AB}] =  \bar{G} \hat{G} = G_\eta R^4 \sin^2 \Theta;
  \end{split}
  \end{equation}
  \begin{equation}
   \label{eq:spheremetr}
  [g_{AB}] = [\bar{g}_{ab}] =  
  \begin{bmatrix}
  \bar{g}_{rr} &  0 &  0 \\
  0 & \bar{g}_{\theta \theta} & 0 \\
  0 & 0 & \bar{g}_{\phi \phi} 
    \end{bmatrix} =
    \begin{bmatrix}
 1&  0 &  0 \\
  0 &  r^2  \\
  0 & 0 &  r^2 \sin^2 \theta
    \end{bmatrix};
    \qquad g =\bar{g} = \det [g_{ab}] = r^4 \sin^2 \theta.
      \end{equation}
   Exponential forms as in \eqref{eq:Gsphere} are routinely used
   since derivatives and inverses are easily obtained, facilitating analytical solutions
   \cite{yavari2010,claytonSYMM2023,claytonMMS2022}.
   Material constants are $k_\eta$, $\hat{m} > 0$, $k_D$, and $\hat{n} > 0$.
   As in Refs.~\cite{taka1990,taka2013}, internal stretches from growth and remodeling are isochoric
   and non-negative in \eqref{eq:Gsphere}:
   \begin{equation}
   \label{eq:isoGhat}
   \hat{G}_R \hat{G}_\Theta \hat{G}_\Phi = 1 \quad \Leftrightarrow \quad \Lambda_R \Lambda_\Theta \Lambda_\Phi = 1, \qquad (\Lambda_R > 0, \, \Lambda_\Theta > 0, \, \Lambda_\Phi > 0).
   \end{equation}
     Any possible internal strain from isotropic damage is spherical, measured by $G_\eta > 0$.
   Regarding connections, \eqref{eq:connrec} is invoked in conjunction with the simplest assumptions
   \begin{equation}
   \label{eq:consphere}
   K^{\alpha A}_{BC} = 0, \quad K^{\alpha a}_{bc} = 0; \qquad N\ua^A_B = 0 \, \Rightarrow \, N\ua^a_b = 0, \quad (\alpha = 1,2).
   \end{equation}
   Therefore $\delta_A(\cdot) = \partial_A (\cdot) = F^a_A \delta_a (\cdot) = F^a_A \partial_a (\cdot)$,
  and horizontal covariant derivatives of directors are
   \begin{equation}
   \label{eq:sphereDd}
   (D^1_{1|A},D^2_{1|A},D^3_{1|A}) = (\partial_A D_R, \partial_A D_\Theta, \partial_A D_\Phi), \qquad
   D^1_{2|A} = \partial_A \eta.
   \end{equation}
   From \eqref{eq:spheremetR}--\eqref{eq:spheremetr} and \eqref{eq:consphere},
     $\Gamma^A_{BC} = \gamma^A_{BC}$, $\Gamma^a_{bc} = \gamma^a_{bc}$ and
     pertinent nonzero connection coefficients are
   \begin{equation}
   \label{eq:LCsphere}
   \Gamma^B_{AB} =  \partial_A \ln \sqrt{G} = \partial_A \sqrt{ \ln (\bar{G} \hat{G}) } =
   \partial_A \sqrt{ \ln \bar{G}}; \quad \Gamma^B_{BR} = 2/R, \quad \Gamma^B_{B \Theta} = \cot \Theta;
      \end{equation}
   \begin{equation}
   \label{eq:LCspherec}
   \Gamma^\theta_{r \theta} =   \Gamma^\phi_{r \phi} = 1/r, \quad \Gamma^r_{\theta \theta} = -r,
   \quad \Gamma^r_{\phi \phi} = -r \sin^2 \theta, \quad  \gamma^\theta_{\phi \phi} = - \sin \theta \cos \theta,
   \quad \gamma^\phi_{\theta \phi} = \cot \theta;
   \end{equation}
   \begin{equation}
   \label{eq:Cartansphere}
   C^{1 B}_{AB} = \bar{\partial}^1_A \ln \sqrt{G(R,\Theta,\eta)} = 0, \quad 
   C^{2 B}_{AB} = \bar{\partial}^2_A \ln \sqrt{G} \rightarrow \partial \ln \sqrt{G} / \partial \eta = k_\eta \eta^{\hat{m} - 1};
   \quad C^{\alpha a}_{bc} = 0.
   \end{equation}
   
   Refer to the osculating Riemannian analysis of \S4.3. In the spatial configuration,
   $g_{ab} = g_{ab}(x) = \tilde{g}_{ab}(x) = \bar{g}_{ab}(x)$ is a Euclidean metric, so
   $\tilde{\mathcal{R}}^a_{bcd} = 0$ and $\mathfrak{m}$ is flat.
   In contrast, $G_{AB} = G_{AB}(X,D_\alpha(X))$ $= \bar{G}_{AC}(X) \hat{G}^C_B (D_\alpha(X))= \tilde{G}_{AB}(X)$ is generally non-Euclidean, and $\bar{G}_{AC}$ is Euclidean but not Cartesian: $\bar{G}_{AC} \neq \delta_{AC}$.
    As a simple example, temporarily assume each $D^\alpha_A$ is a constant over $\mathcal{M}$, so
 $\partial_B D^\alpha_A = 0$.  From \eqref{eq:spheremetR} and 
 \eqref{eq:Gsphere}, $G_\eta$, $\Lambda_R$, $\Lambda_\Theta$, and $\Lambda_\Phi$ are constants;
nonzero entries of \eqref{eq:LC1o} are 
\begin{equation}
\label{eq:LCtildesphere}
\begin{split}
\tilde{\gamma}^\Theta_{R \Theta} & = \tilde{\gamma}^\Phi_{R \Phi} = {1}/{R}, \qquad
\tilde{\gamma}^R_{\Theta \Theta} = - ({\Lambda_R^2}/{\Lambda_\Theta^2}) R, \qquad 
\tilde{\gamma}^R_{\Phi \Phi} =  - ({\Lambda_R^2}/{\Lambda_\Phi^2}) R \sin^2 \Theta, \\
\tilde{\gamma}^\Theta_{\Phi \Phi} & = - ({\Lambda_\Theta^2}/{\Lambda_\Phi^2}) \sin \Theta \cos \Theta, \qquad
\tilde{\gamma}^\Phi_{\Theta \Phi} = \cot \Theta.
\end{split}
\end{equation}
Note $\tilde{G}^{A}_{BC}$ is independent of $\eta$ when $\eta = \text{constant}$, as $\eta$ contributes to \eqref{eq:spheremetR} isotropically via $G_\eta$.
All entries of $\tilde{\mathcal{R}}^A_{BCD}$ or $\tilde{\mathcal{R}}_{ABCD}$ need not vanish, implying residual stresses. For example, from \eqref{eq:RCurv}, \eqref{eq:RCurv1}, and \eqref{eq:LCtildesphere}, one of the six independent components of curvature in Riemannian 3-space is
\begin{equation}
\label{eq:RCsphere}
\tilde{\mathcal{R}}^\Phi_{\Theta \Phi \Theta} = \Lambda_R^2/ \Lambda_\Theta^2  - 1,
\qquad
\tilde{\mathcal{R}}_{\Theta \Phi \Theta \Phi} = \tilde{G}_{\Phi \Phi} \tilde{\mathcal{R}}^\Phi_{\Theta \Phi \Theta}
=(  G_\eta^{1/3} R^2 \sin^2 \Theta / \Lambda_\Phi^2)(\Lambda_R^2/ \Lambda_\Theta^2  - 1).
\end{equation}

Resume now the general conditions analyzed in \S5, whereby none of the $D^A_\alpha$ components need be constants over the thick-walled spherical body $\mathcal{M}$. Motions $x =\varphi(X)$ of \eqref{eq:varphiC} and internal state functions $D_\alpha = D_\alpha(X)$ of \eqref{eq:funcC} are of the 
following functional forms for spherical symmetry:
\begin{equation}
\label{eq:sphmotion}
r = r(R), \quad \theta = \Theta, \quad \phi = \Phi; \qquad D_R = D_R(R), \quad D_\Theta = D_\Theta(R),
\quad D_\Phi = D_\Phi(R), \quad \eta = \eta(R).
\end{equation}
Nonzero components of deformation gradient \eqref{eq:defgradC} and Jacobian determinants \eqref{eq:Jdet} and \eqref{eq:Jbar} are
\begin{equation}
\label{eq:Jdetsph}
F^r_R  = \diff r / \diff R, \qquad F^\theta_\Theta = F^\phi_\Phi = 1; \qquad
J =  (\diff r / \diff R) (r^2/ R^2) G^{-1/2}_\eta = \bar{J} G^{-1/2}_\eta.
\end{equation}
Furthermore, as in Ref.~\cite{taka1990}, and with \eqref{eq:isoGhat},
\begin{equation}
\label{eq:Phisymm}
D_\Phi = D_\Theta  \quad \Rightarrow \quad \Lambda_\Phi = \Lambda_\Theta \quad \Rightarrow \quad  \hat{G}_R \hat{G}_\Theta^2 = \Lambda_R \Lambda_\Theta^2 = 1 \quad \Rightarrow \quad \Lambda_R^2 = 1 / \Lambda_\Theta^4.
\end{equation}
Metric $G_{AB}$ depends only on the difference $D_R^{\hat{n}} - D_\Theta^{\hat{n}}$ and not $D_R$ and $D_\Theta$ individually.
When $\eta$ and $D_\Theta$ are constants, $D_R$ is constant, and \eqref{eq:RCsphere} gives
$\tilde{R}^\Phi_{\Theta \Phi \Theta} = 1/\Lambda_\Theta^6 - 1$.

In some forthcoming calculations, it is prudent and physically realistic to assume the material is nearly incompressible.  True elastic incompressibility implies $\bar{J} = 1$, and true anelastic incompressibility implies $G_\eta = 1$, the latter
necessitating $k_\eta = 0$ for nonzero $\eta$. For nearly total (i.e., approximate) paired elastic and anelastic incompressibility, the following approximations apply:
\begin{equation}
\label{eq:incomp1}
k_\eta \approx 0 \, \Rightarrow \, C^{2 B}_{AB} \approx 0; \qquad
\diff  r/ \diff R \approx R^2 / r^2 \quad \Rightarrow \quad  C_i = r_i^3 - R^3_i \approx r^3 - R^3,
\end{equation}
with $C_i$ defined by inner radius $R_i$ in a reference state and any 
spatial state $r_i$ via \eqref{eq:sphmotion}: $r_i = r(R_i)$.

\subsection{Free energy and balance laws}
Mass density is assumed homogeneous when internal state is spatially constant. Define the mass density at any reference state  and at a state where all internal state functions vanish as
\begin{equation}
\label{eq:massref}
\rho_0(\{ D_\alpha \} ) = \rho_0(D_R,D_\Theta, D_\Phi, \eta), \qquad \bar{\rho}_0 = \rho_0(0,0,0,0) = \text{constant}.
\end{equation}
Assume for simplicity that all source terms related to growth or resorption vanish over the load history, meaning
$S^\alpha_A = 0$ for $\alpha = 1,2$ in \eqref{eq:locmass}. Then \eqref{eq:locmass} and \eqref{eq:Cartansphere} give mass conservation law
\begin{equation}
\label{eq:massconsphere}
\rho_0 = \rho_0(\eta), \quad \diff \rho_0 / \rho_0 = - k_\eta \eta^{ \hat{m} - 1} \diff \eta \quad \Rightarrow \quad
\rho_0(\eta) = \bar{\rho}_0  \exp [ - k_\eta \eta^{\hat{m}} / \hat{m} ] = \bar{\rho}_0 \bar{\alpha}(\eta).
\end{equation}

Free energy per unit mass $\hat{\psi}$ of \eqref{eq:psi2} consists of
volumetric strain energy $\psi_v$, deviatoric (i.e., shear) strain energy $\psi_s$, 
cohesive energy $\psi_c$, and internal surface energy $\psi_\nabla$:
\begin{equation}
\label{eq:psis1}
\begin{split}
\psi = \hat{\psi}(C^A_B,D_R,D_\Theta,D_\Phi,\eta,\partial_A \eta)  = 
\omega_v(\eta) & \psi_v (J (C^A_B)) +
\omega_s (\eta) \psi_s( {C}_{AB}, D_R,D_\Theta,D_\Phi, \eta) \\  + & 
\psi_c(\eta) + \psi_\nabla (D_R,D_\Theta,D_\Phi,\eta,\partial_A \eta).
\end{split}
\end{equation}
Degradation functions $\omega_v \in [0,1]$ and $\omega_s \in [0,1]$ obey
$\omega_v(0) = \omega_s(0) = 1$. Cohesive energy and surface energy of fracture are defined
as in isotropic phase-field fracture mechanics \cite{claytonIJF2014,claytonJMPS2021,claytonPRE2024}:
\begin{equation}
\label{eq:cohsphere}
\bar{\rho}_0  \psi_c   = \hat{e}_c \eta^2, \qquad
\bar{\rho}_0  \psi_\nabla = \hat{\Upsilon} l_0 G^{AB} \partial_A \eta  \partial_B \eta, \qquad G^{AB} = G^{AB}(R, \Theta; D_R,D_\Theta,D_\Phi,\eta).
\end{equation}
Cohesive energy per unit volume, surface energy per unit area, and regularization length
are constants $\hat{e}_c \geq 0$, $\hat{\Upsilon} \geq 0$, and $ l_0 \geq 0$.
Volumetric strain energy and resulting pressure $\hat{p}$ are \cite{claytonMOSM2020,claytonBM2020,claytonPRE2024}
\begin{equation}
\label{eq:psiv}
\bar{\rho}_0 \psi_v = \frac{B_0}{2 k_v}\{  \exp [k_v ( \ln J)^2]  - 1 \}, 
\qquad \hat{p} = - \bar{\rho}_0 \frac{\partial \psi_v}{ \partial J} = - \frac{ B_0 \ln J}{J}   \exp [k_v ( \ln J)^2].
\end{equation}
The reference bulk modulus is the constant $B_0 > 0$, and constant $k_v$ provides exponential stiffening (softening) if positive (negative). Deviatoric strain energy and resulting Cauchy stress contribution are, similar to isotropic contributions to cardiac elasticity in Refs.~\cite{holz2009,gultekin2016} when $G_{AB}$ is Euclidean,
\begin{equation}
\label{eq:psis}
\bar{\rho}_0 \psi_s = \frac{\mu_0}{2 k_s} \{  \exp [k_s (\hat{I}_1 -3)]  - 1 \},
\quad \hat{\sigma}^{ab} = \frac{2}{J}   F^a_A \bar{\rho}_0 \frac{\partial \psi_s}{\partial C_{AB}} F^b_B
= \frac{\mu_0}{J}   \exp [k_s (\hat{I}_1 -3)] \hat{B}^{ab},
\end{equation}
\begin{equation}
\label{eq:BI1}
\hat{I}_1 = J^{-2/3} C^A_A = J^{-2/3} F^a_A F^b_B g_{ab} G^{AB},
\qquad
\hat{B}^{ab} = J^{-2/3} (F^a_A F^b_B G^{AB} - {\textstyle{\frac{1}{3}}} C^A_A g^{ab}).
\end{equation}
A shear modulus and exponential stiffening (if positive) term are constants $\mu_0 > 0$ and $k_s$.
Merging \eqref{eq:massconsphere}, \eqref{eq:psis1}, \eqref{eq:psiv}, and \eqref{eq:psis}, total Cauchy stress in \eqref{eq:cauchy} is, with Cauchy pressure $p$, 
\begin{equation}
\label{eq:cauchysph1}
\begin{split}
\sigma^{ab} & = \bar{\alpha} \omega_v  \frac{ B_0 \ln J}{J}   \exp [k_v ( \ln J)^2] g^{ab} +
\bar{\alpha} \omega_s \frac{\mu_0}{J}   \exp [k_s (\hat{I}_1 -3)] \hat{B}^{ab}, \\ 
p & = -{\textstyle{\frac{1}{3}}} \sigma^a_a = - \bar{\alpha} \omega_v  B_0 (\ln J /{J} )  \exp [k_v ( \ln J)^2].
\end{split}
\end{equation}
For spherically symmetric deformation of the forms in \eqref{eq:sphmotion} and \eqref{eq:Jdetsph},
\begin{equation}
\label{eq:cauchsph2}
\hat{I}_1 = \lambda_r^2 + \lambda_\theta^2 + \lambda_\phi^2, 
\quad [\hat{B}^b_a] = {\text {diag}}(\lambda_r^2 - {\textstyle{\frac{1}{3}}} \hat{I}_1,
\lambda_\theta^2 - {\textstyle{\frac{1}{3}}} \hat{I}_1,
\lambda_\phi^2 - {\textstyle{\frac{1}{3}}} \hat{I}_1),
\end{equation}
\begin{equation}
\label{eq:lambdas}
\lambda_r = \bar{J}^{-1/3} (\diff r/ \diff R) \Lambda_R, \quad
\lambda_\theta = \bar{J}^{-1/3} (r/R) \Lambda_\Theta, \quad
\lambda_\phi = \bar{J}^{-1/3} (r/R) \Lambda_\Phi;
\end{equation}
\begin{equation}
\label{eq:sigc2}
\begin{split}
\sigma_r^r  & = - p + \bar{\alpha} \omega_s \frac{\mu_0}{J}   \exp [k_s (\hat{I}_1 -3)] (\lambda_r^2 - {\textstyle{\frac{1}{3}}} \hat{I}_1),
\quad \sigma_\theta^\theta  = - p + \bar{\alpha} \omega_s \frac{\mu_0}{J}   \exp [k_s (\hat{I}_1 -3)] (\lambda_\theta^2 - {\textstyle{\frac{1}{3}}} \hat{I}_1), \\
\sigma_\phi^\phi & = - p + \bar{\alpha} \omega_s \frac{\mu_0}{J}   \exp [k_s (\hat{I}_1 -3)] (\lambda_\phi^2 - {\textstyle{\frac{1}{3}}} \hat{I}_1), \quad \sigma^r_\theta = \sigma^\theta_r = \sigma^r_\phi = \sigma^\phi_r = \sigma^\theta_\phi = \sigma^\phi_\theta = 0.
\end{split}
\end{equation}
When \eqref{eq:Phisymm} applies,
\begin{equation}
\label{eq:phisymm}
D_\Phi = D_\Theta  \quad \Rightarrow \quad \lambda_\phi = \lambda_\theta \quad \Rightarrow \quad \sigma^\phi_\phi = \sigma^\theta_\theta.
\end{equation}

In muscle tissue, including that of the heart, $B_0$ is on the order of GPa whereas $\mu_0$ is
on the order of kPa \cite{taka1990,holz2009,claytonPRE2024}.  In the near incompressible limit,
it is assumed that $k_\eta \rightarrow 0$ and $|\ln J| \ll 1 \Rightarrow J \rightarrow \bar{J} \rightarrow 1$ in \eqref{eq:Jdetsph} and \eqref{eq:lambdas}, but since $B_0$ is large,
the product $\bar{\alpha} \omega_v B_0 \ln J$ is regarded as finite so $p$ can be nonzero in \eqref{eq:cauchysph1}.
Note $k_\eta \rightarrow 0 \Rightarrow \bar{\alpha} \rightarrow 1$.
Assuming $k_v \geq 0$ \cite{claytonPRE2024}, the near incompressibility assumption is valid when $|p|/(\omega_v B_0) \ll 1$.
For minimum $B_0 \approx 2$ GPa (the order of water), $\omega_v = 1$, and maximum
$|p| \approx 100$ kPa, $| \ln J | \approx 5 \times 10^{-5}$, so $J \approx 1$ is justified. In that case,
with \eqref{eq:Phisymm} and \eqref{eq:phisymm},
\begin{equation}
\label{eq:incomplam}
\lambda_r \lambda_\theta \lambda_\phi \approx 1, \quad \lambda_\phi = \lambda_\theta \approx (r/R) \Lambda_\Theta, \quad \lambda_r = R^2 / (r^2 \Lambda_\Theta^2), \quad
\hat{I}_1 \approx R^4 / (r^4 \Lambda_\Theta^4) + 2 (r/R)^2 \Lambda_\Theta^2.
\end{equation}

Returning to the general case, the balance of linear momentum in spatial form \eqref{eq:linmom2s}
becomes, in the absence of body force (i.e., $b_a = 0$), with $\Gamma^a_{bc} = \gamma^a_{bc}$
from \eqref{eq:consphere} and the last of \eqref{eq:Cartansphere},
\begin{equation}
\label{eq:linmomsph}
\partial_b \sigma^b_a + \sum_{\alpha = 1}^2 \bar{\partial}^\alpha_C \sigma^b_a \partial_b D^C_\alpha
+ \sigma^c_a \gamma^b_{bc}
- \sigma^b_d \gamma^d_{ba} 
= 0. 
\end{equation}
From \eqref{eq:LCspherec}, \eqref{eq:linmomsph} is satisfied automatically for $a = \theta$ and $a = \phi$
by the spherical symmetry conditions in \eqref{eq:phisymm}, used henceforth.
This leaves the radial momentum equation
\begin{equation}
\label{eq:linmomrad}
\frac{ \diff \sigma^r_r}{\diff r} = \frac{2}{r} (\sigma^\theta_\theta - \sigma^r_r), \qquad
\frac{ \diff \sigma^r_r}{\diff r} = \frac{ \partial \sigma^r_r}{\partial r} + \sum_{\alpha =1}^2
\frac{ \partial \sigma^r_r}{\partial D^C_\alpha} \frac{ \diff D^C_\alpha}{\diff r}
= \frac{ \partial \sigma^r_r}{\partial r} + \sum_{\alpha =1}^2
\frac{ \partial \sigma^r_r}{\partial D^C_\alpha} \frac{ \diff D^C_\alpha}{\diff R} \biggr{(} \frac{\diff r}{ \diff R} \biggr{)}^{-1}.
 \end{equation}
 
 The internal state equilibrium equations are most easily analyzed in material form of
 \eqref{eq:micromom2}. These become here, with \eqref{eq:consphere} and \eqref{eq:Cartansphere},
\begin{equation}
\label{eq:micromomsph}
\begin{split}
  \partial_A Z\ua^A_C  +   \sum_{\beta = 1}^2 & \bar{\partial}^\beta_B Z\ua^A_C \partial_A D^B_\beta  + Z\ua^B_C \gamma^A_{AB}   -   (Q^\alpha_C - E^\alpha_C)
 =      - Z\ua^A_C ( k_\eta \eta^{\hat{m} - 1} \partial_A \eta ).
\end{split} 
\end{equation}
For $\alpha = 2$, $D^A_\alpha \rightarrow \eta$, and per standard phase-field fracture \cite{bourdin2008,claytonIJF2014,claytonJMPS2021},
external micro-force $E^\alpha_C \rightarrow E_\eta = 0$. Conjugate force to the gradient of $\eta$ is
$Z\ua^A_C \rightarrow Z^ A_\eta$; from \eqref{eq:sphereDd}, \eqref{eq:Phisymm}, and \eqref{eq:cohsphere},
\begin{equation}
\label{eq:Zxi}
Z^A_\eta = \rho_0 \frac{\partial \psi_\nabla}{\partial (\partial_A \eta)} = 
2 \bar{\alpha} \hat{\Upsilon} l_0 G^{AB} \partial_B \eta \quad
\Rightarrow \quad Z^R_\eta =  \frac{2 \bar{\alpha} \hat{\Upsilon} l_0}{\Lambda_\Theta^4}  \exp \biggr{(} \frac{-2k_\eta \eta^{\hat{m}}}{3\hat{m}} \biggr{)} \frac{ \diff \eta}{\diff R}.
\end{equation}
The conjugate force to $\eta$ itself is $Q^\alpha_C \rightarrow Q_\eta$, where from \eqref{eq:psis1} and \eqref{eq:cohsphere}, noting that $J$ in $\psi_v$ depends on $\eta$ through $G = G_\eta \bar{G}$ and that $\hat{I}_1$ in $\psi_s$ of \eqref{eq:psis} is independent of $G_\eta$ and $\eta$,
\begin{equation}
\label{eq:Qxi}
\begin{split}
Q_\eta  = \rho_0 \frac{\partial \psi}{\partial \eta} =  \frac{{\rho}_0}{\bar {\rho}_0} \frac{\partial (\bar{\rho}_0 \psi)}{\partial \eta}  = \bar{\alpha} \frac{ \diff \omega_v}{\diff \eta} \psi_v
& + \bar{\alpha} \frac{ \diff \omega_s}{\diff \eta} \psi_s
+ {p} J k_\eta \eta^{\hat{m} - 1}
+ 2 \bar{\alpha} \hat{e}_c \eta \\ & - \frac{ 2 \bar{\alpha} \hat{\Upsilon} l_0 k_\eta \eta^{\hat{m}-1} }{3 \Lambda_\Theta^4}
 \exp \biggr{(} \frac{-2k_\eta \eta^{\hat{m}}}{3\hat{m}} \biggr{)} \left(\frac{ \diff \eta}{\diff R} \right)^2.
\end{split}
\end{equation}
For $\alpha = 2$, \eqref{eq:micromomsph} is, with $E_\eta =0$, $Z^R_\eta =Z^R_\eta (R)$ in \eqref{eq:Zxi}, and $Q_\eta = Q_\eta(R)$ in \eqref{eq:Qxi},
\begin{equation}
\label{eq:microZ1}
Q_\eta = 2  \hat{\Upsilon} l_0 \frac{ \diff} {\diff R} \biggr{[} \frac{ \bar{\alpha}}{\Lambda_\Theta^4}  \exp \biggr{(} \frac{-2k_\eta \eta^{\hat{m}}}{3\hat{m}} \biggr{)} \frac{ \diff \eta}{\diff R} \biggr{]} 
+ \frac{2 \bar{\alpha} \hat{\Upsilon} l_0}{ \Lambda_\Theta^4}  \exp \biggr{(} \frac{-2k_\eta \eta^{\hat{m}}}{3\hat{m}} \biggr{)}  \biggr{(} \frac{2}{R} + k_\eta \eta^{\hat{m}-1} \frac{ \diff \eta}{\diff R} \biggr{)} \frac{ \diff \eta}{\diff R}.
\end{equation}
For near incompressibility ($k_\eta \approx 0 \Rightarrow \bar{\alpha} \approx 1$, $\omega_v \approx 1$),
\eqref{eq:psis}, and the phase-field quadratic 
form of $\omega_s$ \cite{claytonIJF2014}, \eqref{eq:Qxi} and \eqref{eq:microZ1} are, in combination,
\begin{equation}
\label{eq:Qxinc}
Q_\eta \approx -  [\mu_0 (1-\eta)  /k_s ] \{  \exp [k_s (\hat{I}_1 -3)]  - 1 \} + 2 \hat{e}_c \eta, \qquad
\omega_s = (1-\eta)^2,
\end{equation}
\begin{equation}
\label{eq:Zxinc}
 \hat{e}_c \eta -  \frac{\mu_0 (1-\eta) }{2 k_s } \{  \exp [k_s (\hat{I}_1 -3)]  - 1 \} 
 \approx
   \hat{\Upsilon} l_0 \frac{ \diff} {\diff R} \biggr{(} \frac{1}{\Lambda_\Theta^4}  \frac{ \diff \eta}{\diff R} \biggr{)} 
+ \frac{2}{R} \frac{ \hat{\Upsilon} l_0}{  \Lambda_\Theta^4}  \frac{ \diff \eta}{\diff R}.
\end{equation}

For $\alpha = 1$, $(D^1_\alpha,D^2_\alpha,D^3_\alpha) \rightarrow (D_R,D_\Theta,D_\Phi)$. Generally
any of $E^\alpha_C(R) \rightarrow (E_R(R), E_\Theta(R), E_\Phi(R))$ can be nonzero fields of radial coordinate $R$ on $\mathcal{M}$. Since $\hat{\psi}$ of \eqref{eq:psis1} does not depend on state vector gradients
$(\partial_A D_R, \partial_A D_\Theta, \partial_A D_\Phi)$ and $K^{\alpha A}_{BC} = 0$, $Z\ua^A_B = \rho_0 \partial \psi / \partial (\partial_A D^B_\alpha) = 0$ for $\alpha = 1$.
Since $G$ and $J$ are independent of $(D_R,D_\Theta,D_\Phi)$, conjugate force components $Q^\alpha_C$ for $\alpha =1 $ arise from $\psi_s$ and $\psi_\nabla$ in \eqref{eq:psis1}, \eqref{eq:cohsphere}, and \eqref{eq:psis}:
\begin{equation}
\label{eq:QDR}
Q_R = Q^1_1 = \rho_0 \frac{\partial \psi}{\partial D_R} = \frac{\bar{\alpha} \omega_s
\mu_0}{2}  \exp [k_s (\hat{I}_1 -3)] \frac{ \partial \hat{I}_1} {\partial D_R}
+2  \bar{\alpha} \hat{\Upsilon} l_0 \biggr{(} \frac{ \diff \eta}{\diff R} \biggr{)}^2 G_\eta^{-1/3} \Lambda_R \frac{\partial \Lambda_R}{\partial D_R},
\end{equation}
\begin{equation}
\label{eq:QDTheta}
Q_\Theta = Q^1_2 = \rho_0 \frac{\partial \psi}{\partial D_\Theta} = \frac{\bar{\alpha} \omega_s
\mu_0}{2}  \exp [k_s (\hat{I}_1 -3)] \frac{ \partial \hat{I}_1} {\partial D_\Theta}
+2  \bar{\alpha} \hat{\Upsilon} l_0 \biggr{(} \frac{ \diff \eta}{\diff R} \biggr{)}^2 G_\eta^{-1/3} \Lambda_\Theta \frac{\partial \Lambda_\Theta}{\partial D_\Theta},
\end{equation}
\begin{equation}
\label{eq:QDPhi}
Q_\Phi = Q^1_3 = \rho_0 \frac{\partial \psi}{\partial D_\Phi} = \frac{\bar{\alpha} \omega_s
\mu_0}{2}  \exp [k_s (\hat{I}_1 -3)] \frac{ \partial \hat{I}_1} {\partial D_\Phi}
+2  \bar{\alpha} \hat{\Upsilon} l_0 \biggr{(} \frac{ \diff \eta}{\diff R} \biggr{)}^2 G_\eta^{-1/3} \Lambda_\Phi \frac{\partial \Lambda_\Phi}{\partial D_\Phi},
\end{equation}
\begin{equation}
\label{eq:DI1DR}
\frac{\partial \hat{I}_1}{\partial D_R} =  \bar{J}^{-2/3} k_D \biggr{[} \frac{r ^2 }{R^2} (\Lambda_\Theta^2 D^{\hat{n}-1}_\Theta + \Lambda_\Phi^2 D^{\hat{n}-1}_\Phi)  - 2 \biggr{(} \frac{ \diff r}{\diff R} \biggr{)}^2 
\Lambda_R^2 D^{\hat{n}-1}_R \biggr{]} ,
\end{equation}
\begin{equation}
\label{eq:DI1DTheta}
\frac{\partial \hat{I}_1}{\partial D_\Theta} =  \bar{J}^{-2/3} k_D \biggr{[} \frac{r ^2 }{R^2} ( \Lambda_\Phi^2 D^{\hat{n}-1}_\Phi - 2 \Lambda_\Theta^2 D^{\hat{n}-1}_\Theta)  + \biggr{(} \frac{ \diff r}{\diff R} \biggr{)}^2 
\Lambda_R^2 D^{\hat{n}-1}_R \biggr{]} ,
\end{equation}
\begin{equation}
\label{eq:DI1DPhi}
\frac{\partial \hat{I}_1}{\partial D_\Phi} =  \bar{J}^{-2/3} k_D \biggr{[} \frac{r ^2 }{R^2} ( \Lambda_\Theta^2 D^{\hat{n}-1}_\Theta - 2 \Lambda_\Phi^2 D^{\hat{n}-1}_\Phi)  + \biggr{(} \frac{ \diff r}{\diff R} \biggr{)}^2 
\Lambda_R^2 D^{\hat{n}-1}_R \biggr{]} ,
\end{equation}
\begin{equation}
\label{eq:DL}
\frac{\partial \Lambda_R}{\partial D_R} = -k_D D_R^{\hat{n}-1} \Lambda_R,
\quad
\frac{\partial \Lambda_\Theta}{\partial D_\Theta} = -k_D D_\Theta^{\hat{n}-1} \Lambda_\Theta,
\quad
\frac{\partial \Lambda_\Phi}{\partial D_\Phi} = -k_D D_\Phi^{\hat{n}-1} \Lambda_\Phi.
\end{equation}
When symmetry conditions \eqref{eq:Phisymm} are invoked,
$Q_\Phi = Q_\Theta$ and \eqref{eq:QDPhi}, \eqref{eq:DI1DPhi}, and the last of \eqref{eq:DL} are redundant.
Preserving such symmetry via prescription of external micro-forces $E_\Phi(R) = E_\Theta(R)$, the micro-force balance \eqref{eq:micromomsph} for $\alpha = 1$ reduces to the two differential equations $Q^\alpha_C = E^\alpha_C$:
\begin{equation}
\label{eq:mfR}
Q_R =  \frac{\bar{\alpha} \omega_s
\mu_0}{2}  \exp [k_s (\hat{I}_1 -3)] \frac{ \partial \hat{I}_1} {\partial D_R}
+2  \bar{\alpha} \hat{\Upsilon} l_0 \biggr{(} \frac{ \diff \eta}{\diff R} \biggr{)}^2  \exp \biggr{(} \frac{-2k_\eta \eta^{\hat{m}}}{3\hat{m}} \biggr{)} \Lambda_R \frac{\partial \Lambda_R}{\partial D_R} = E_R,
\end{equation}
\begin{equation}
\label{eq:mfT}
Q_\Theta =  \frac{\bar{\alpha} \omega_s
\mu_0}{2}  \exp [k_s (\hat{I}_1 -3)] \frac{ \partial \hat{I}_1} {\partial D_\Theta}
+2  \bar{\alpha} \hat{\Upsilon} l_0 \biggr{(} \frac{ \diff \eta}{\diff R} \biggr{)}^2  \exp \biggr{(} \frac{-2k_\eta \eta^{\hat{m}}}{3\hat{m}} \biggr{)} \Lambda_\Theta \frac{\partial \Lambda_\Theta}{\partial D_\Theta} = E_\Theta.
\end{equation}
In the nearly incompressible limit, \eqref{eq:DI1DR}, \eqref{eq:DI1DTheta}, \eqref{eq:mfR}, and \eqref{eq:mfT} simplify to
\begin{equation}
\label{eq:DIi}
\frac{\partial \hat{I}_1}{\partial D_R} \approx  2 k_D \biggr{[} \frac{r ^2 \Lambda_\Theta^2 }{R^2}  D^{\hat{n}-1}_\Theta   -  \frac{R^4}{r^4 \Lambda_\Theta^4}  D^{\hat{n}-1}_R \biggr{]} ,
\quad
\frac{\partial \hat{I}_1}{\partial D_\Theta} \approx   k_D \biggr{[} 
 \frac{R^4}{r^4 \Lambda_\Theta^4}  D^{\hat{n}-1}_R - \frac{r ^2 \Lambda_\Theta^2 }{R^2}  D^{\hat{n}-1}_\Theta 
 \biggr{]},
\end{equation}
\begin{equation}
\label{eq:mfRi}
{k_D \omega_s \mu_0}  \exp [k_s (\hat{I}_1 -3)]
 \biggr{[} \frac{r ^2 \Lambda_\Theta^2 }{R^2}  D^{\hat{n}-1}_\Theta   -  \frac{R^4}{r^4 \Lambda_\Theta^4}  D^{\hat{n}-1}_R \biggr{]} 
- 2  k_D \hat{\Upsilon} l_0  D^{\hat{n}-1}_R \biggr{(} \frac{ \diff \eta}{\diff R} \biggr{)}^2  \frac{1}{\Lambda_
\Theta^4}  \approx E_R,
\end{equation}
\begin{equation}
\label{eq:mfTi}
 \frac{  k_D \omega_s \mu_0}{2}  \exp [k_s (\hat{I}_1 -3)] 
 \biggr{[} 
 \frac{R^4}{r^4 \Lambda_\Theta^4}  D^{\hat{n}-1}_R - \frac{r ^2 \Lambda_\Theta^2 }{R^2}  D^{\hat{n}-1}_\Theta 
 \biggr{]}
- 2 k_D \hat{\Upsilon} l_0 D_\Theta^{\hat{n}-1} \biggr{(} \frac{ \diff \eta}{\diff R} \biggr{)}^2   \Lambda_\Theta^2  \approx E_\Theta.
\end{equation}

Though not necessary, Lagrangian strain tensors can be defined to aid physical interpretation of deformation 
components $C_{AB} = F^a_A g_{ab} F^b_B$ and material metric components $G_{AB}, \bar{G}_{AB}$:
\begin{equation}
\label{eq:straintens}
\epsilon^t_{AB} = {\textstyle{\frac{1}{2}}} (C_{AB} - \bar{G}_{AB}) = \epsilon^e_{AB} + \epsilon^r_{AB},
\quad 
\epsilon^e_{AB} = {\textstyle{\frac{1}{2}}} (C_{AB} - {G}_{AB}),
\quad
\epsilon^r_{AB} = {\textstyle{\frac{1}{2}}} (G_{AB} - \bar{G}_{AB}).
\end{equation}
Total, elastic, and residual (i.e., anelastic) strains are $\bm{\epsilon}^t$, $\bm{\epsilon}^e$, and $\bm{\epsilon}^r$.
For the current geometry,
\begin{equation}
\label{eq:epsr}
\epsilon^r_{RR} = {\textstyle{\frac{1}{2}}}(G_\eta^{1/3} / \Lambda_R^2 - 1), \quad
\epsilon^r_{\Theta \Theta} = {\textstyle{\frac{1}{2}}} R^2 (G_\eta^{1/3} / \Lambda_\Theta^2 - 1), \quad
\epsilon^r_{\Phi \Phi} = {\textstyle{\frac{1}{2}}} R^2 \sin^2 \Theta (G_\eta^{1/3} / \Lambda_\Phi^2 - 1).
\end{equation}
With symmetry \eqref{eq:Phisymm} and near incompressibility \eqref{eq:incomp1}, $G_\eta \approx 1$, and mixed  components become
\begin{equation}
\label{eq:epsrc}
\begin{split}
(\epsilon^r)^R_{R} & \approx {\textstyle{\frac{1}{2}}}(1 - 1/\Lambda_\Theta^4), \qquad
(\epsilon^r)^\Theta_{\Theta} =
(\epsilon^r)^\Phi_{\Phi} \approx
 {\textstyle{\frac{1}{2}}}  (1- \Lambda_\Theta^2); \\
 (\epsilon^e)^R_{R} & \approx {\textstyle{\frac{1}{2}}}(1/\lambda_\theta^4 - 1), \qquad
 (\epsilon^e)^\Theta_{\Theta} =
(\epsilon^e)^\Phi_{\Phi} \approx
 {\textstyle{\frac{1}{2}}}  (\lambda_\theta^2 - 1).
 \end{split}
\end{equation}
From \eqref{eq:epsrc}, $\Lambda_\Theta > 1$ gives tensile remnant radial strain and compressive remnant transverse strains.

\subsection{Boundary conditions and solution procedure}
Let $\tau \geq 0$ be a loading parameter, where $\tau = \tau_0$ (which may be nonzero) corresponds to a defined reference configuration.
The unit normal at inner and outer radii, $r_i$ and $r_o$ on $\partial \mathfrak{m}$, has the lone nonzero component $|n_r| =1$. Mechanical boundary conditions are prescribed displacement of the inner radius
and prescribed external pressure $p_o$:
\begin{equation}
\label{eq:bcs1}
r_i = r_i(\tau), \qquad p_o(\tau) = -\sigma^r_r(r_o (\tau)).
\end{equation}
In the reference configuration $r_i (0) = R_i$ and $r_o(0) = R_o$, where $R_i$ and $R_o$ are
known constants (i.e., measured dimensions of the specimen). 
Different choices of reference configuration, like different referential metrics $G_{AB}$, can produce different model solutions.
Nearly incompressible assumption \eqref{eq:incomp1} is used with approximations ($\approx$) replaced henceforth by equalities (=) in notation:
\begin{equation}
\label{eq:incomp1bc}
C_i(\tau) = r^3_i(\tau) - R_i^3;
\qquad r(R,\tau) = [R^3 + C_i(\tau)]^{1/3}, \quad r_o(\tau) = [R_o^3 + C_i(\tau)]^{1/3}.
\end{equation}
The internal pressure at $r = r_i$ is $p_i(\tau) = - \sigma^r_r(r_i(\tau))$. Given $r_i$, integration of \eqref{eq:linmomrad} produces, with \eqref{eq:sigc2}, $\bar{\alpha} =1$ and $\omega_s = (1-\eta)^2$ of \eqref{eq:Qxinc}, internal pressure at $\tau$ needed to achieve motion $r_i(\tau)$:
\begin{equation}
\label{eq:pi}
\int_{r_i}^{r_o} \frac{ \diff \sigma^r_r}{\diff r} = 
 p_i - p_o =2 \mu_0 \int_{r_i}^{r_o} (1 - \eta)^2  \exp [k_s (\hat{I}_1 -3)] (\lambda^2_\theta - \lambda^2_r) \frac{\diff r}{r},
 \end{equation}
 where $\eta$, $\hat{I}_1$, $\lambda_\theta$, and $\lambda_r$ all generally depend on $r(R)$. Specifically from  \eqref{eq:incomp1}, \eqref{eq:incomplam}, and \eqref{eq:incomp1bc},
 \begin{equation}
 \label{eq:rR1}
 \lambda^2_\theta(r) = \frac{1}{ \lambda_r(r)}  = \frac{r^2}{(r^3-C_i)^{2/3}} \Lambda_\Theta^2(D_R(r),D_\Theta(r)),
 \qquad \hat{I}_1(r) = 2 \lambda_\theta^2(r) + \frac{1}{\lambda_\theta^4(r)}.
 \end{equation} 
 Integration of \eqref{eq:linmomrad} from $r_i$ to any radial location $r \in (r_i,r_o)$ gives the radial stress field $\sigma^r_r(r)$, followed by transverse stress fields and Cauchy pressure field $p(r)$ in the incompressible limit, for prescribed $r_i(\tau)$ and $p_i(\tau)$, where the latter depends on $p_o(\tau)$ via \eqref{eq:pi}:
\begin{equation}
\label{eq:sigmari}
 \sigma^r_r (r) = 
 - p_i + 2 \mu_0 \int_{r_i}^{r} (1 - \eta)^2  \exp [k_s (\hat{I}_1 -3)] (\lambda^2_\theta - \lambda^2_r) \frac{\diff \mathsf{r}}{\mathsf{r}},
 \end{equation}
 \begin{equation}
\label{eq:sigmati}
 \sigma^\theta_\theta  = \sigma^\phi_\phi = \sigma^r_r 
+ \mu_0 (1 - \eta)^2  \exp [k_s (\hat{I}_1 -3)] (\lambda^2_\theta - 1/ \lambda^4_\theta),
\qquad p(r) = -{\textstyle{\frac{1}{3}}}(\sigma^r_r(r) + 2 \sigma^\theta_\theta(r)).
 \end{equation}
 For this problem, with foregoing incompressible assumptions, $p(r)$ depends on boundary conditions
 but not bulk modulus $B_0$ of \eqref{eq:psiv}. A similar result is obtained when the Lagrange multiplier
 method is used \cite{taka1990} because integrands on right sides of \eqref{eq:pi} and \eqref{eq:sigmari} do not depend on $p(r)$. 
 
 Vanishing natural boundary conditions are prescribed on $\partial \! \mathcal{M}$ (or equivalently, $\partial \mathfrak{m}$) for thermodynamic conjugates $z^\alpha_A = Z\ua^B_A N_B$ to gradients of internal state (or equivalently, vanishing $\zeta^\alpha_A = j F^b_B  Z\ua^B_A n_b)$:
 \begin{equation}
 \label{eq:freeNBC}
 z^\alpha_A = Z\ua^B_A N_B \rightarrow Z\ua^R_A N_R = 0, \qquad (R = R_i,R_o; \, \tau \geq 0; \, \alpha = 1,2).
 \end{equation}
 Conditions \eqref{eq:freeNBC} are trivially satisfied for $\alpha = 1$ since $Z\ua^A_B = 0$ for $\alpha =1$ from \eqref{eq:psis1}.  For $\alpha = 2$, with incompressibility $k_\eta = 0 \Rightarrow \bar{\alpha} = 1$,
 \eqref{eq:Zxi} gives
 \begin{equation}
 \label{eq:Zxincbc}
 Z^R_\eta =  2 ( \hat{\Upsilon} l_0/ \Lambda_\Theta^4) ({ \diff \eta} /{\diff R}).
 \end{equation}
 Conditions \eqref{eq:freeNBC} require $\hat{\Upsilon} l_0 = 0$ or $\diff \eta / \diff R = 0$ at $R = R_i,R_o$ for $\tau \geq 0$. For simplicity of computations, the former is assumed (i.e., $\hat{\Upsilon} l_0 \rightarrow 0$), so $Z^R_\eta(R(r)) = 0$ for all $r \in [r_i,r_o]$ and all $\tau \geq 0$.
 As such, the regularizing aspect of phase-field mechanics is dropped, giving a scale-free representation
 as used elsewhere for liquid cavitation and tearing of biologic tissues \cite{levitas2011,claytonPRE2024}. Order parameter $\eta$ is akin to a damage variable in classical damage mechanics \cite{simo1987,claytonBM2020}.
  The right side of \eqref{eq:Zxinc} vanishes, from which the implicit analytical solution for field $\eta = \eta(r(\tau))$ is, for $\tau \geq 0$,
  \begin{equation}
\label{eq:xianal}
{\eta(r)}/{(1-\eta(r))} = 
   [{\mu_0 }/{(2 k_s \hat{e}_c) }] \{  \exp [k_s (\hat{I}_1(r) -3)]  - 1 \}.
 \end{equation}
  With $\hat{\Upsilon} l_0 = 0$, micro-force laws \eqref{eq:mfRi} and \eqref{eq:mfTi} are manipulated into two independent equations:
 \begin{equation}
 \label{eq:mfRs}
 \frac{  k_D \omega_s \mu_0}{2}  \exp [k_s (\hat{I}_1 -3)] 
 \biggr{[} 
 \frac{R^4}{r^4 \Lambda_\Theta^4}  D^{\hat{n}-1}_R - \frac{r ^2 \Lambda_\Theta^2 }{R^2}  D^{\hat{n}-1}_\Theta 
 \biggr{]}
 = E_\Theta, \qquad E_R = - 2 E_\Theta.
 \end{equation}
 Given external micro-force $E_\Theta$ as a loading condition representing interactions of the tissue's microstructure with its biological environment in an open-system viewpoint \cite{kuhl2003,epstein2007,epstein2012}, \eqref{eq:mfRs} is one independent equation in two unknowns $D_R, D_\Theta$ at
 each $R(r(\tau))$ and any $\tau \geq 0$. 
 
 Recall from \S5.1 and \S5.2 that metric $G_{AB}$ and energy $\psi$
 depend only on the difference $D^{\hat{n}}_R -  D^{\hat{n}}_\Theta$ and not $D_R, D_\Theta$ independently.
 So without loss of generality, assume henceforth that $D_R(r) = 0 \, \forall \, r(\tau) \in [r_i(\tau),r_o(\tau)]$. Thus,
 the first of \eqref{eq:mfRs} and the remnant stretches $\Lambda_R$ and $\Lambda_\Theta = \Lambda_\Phi$ are
 \begin{equation}
 \label{eq:DR0}
{  k_D \mu_0} \exp \biggr{[} k_s [\hat{I}_1(r,D_\Theta^{\hat{n}}) -3 ] - \frac{k_D}{\hat{n}} D_\Theta^{\hat{n}} \biggr{ ]}
 \frac{ \{(1-\eta) r \}^2}{(r^3-C_i)^{2/3}}   D^{\hat{n}-1}_\Theta 
 = E_R, 
 \quad
 \Lambda_\Theta^2 = \frac{1}{\Lambda_R} = \exp \biggr{[} - \frac{k_D}{\hat{n}} D_\Theta^{\hat{n}} \biggr{ ]}.
 \end{equation}
If $k_D = 0$, $\psi$ becomes independent of $D_\Theta$, $Q_\Theta = 0$ identically, and the
 only physically reasonable value for the external micro-force (i.e., interaction energy) is $E_R = 0$.
A nonzero force field $E_R(r)$ is generally required to support a nonzero internal structure field $D_\Theta(r)$. This implies free energy functional $\Psi$ is not at a stationary point (e.g., local minimum) when $D_\Theta$ is nonzero. 

Three sets of external force-field conditions $E_R$ are considered in calculations of \S5.4:
 \begin{equation}
 \label{eq:extforc}
 E_R = 0, \quad E_R(r(\tau)) = \bar{E}_R (D_\Theta(r(\tau)) = \bar{D}_\Theta = \text{constant}), \quad  E_R(r(\tau)) = E_R(r(\tau_0)) = \tilde{E}_R(R).
 \end{equation}
 The first condition in \eqref{eq:extforc} gives a logical and trivial solution to \eqref{eq:DR0}: $D_\Theta(r) = 0$ when $\hat{n} > 1$. The second \textit{defines} a generally transient, heterogeneous
 force-field $\bar{E}_R$ that equals the left of \eqref{eq:extforc} when the internal structure variable
 is the constant $\bar{D}_\Theta$; the latter could be zero to enable $\hat{n} = 1$.
 This loading condition replicates the uniform internal strain assumption $\Lambda_\Theta = \text{constant}$ in Ref.~\cite{taka1990}. The third of \eqref{eq:extforc} invokes a steady field $\tilde{E}_R(R)$ at a defined reference configuration for which $\tau = \tau_0$ and $r = R$. This force field is held constant over the load history $\tau \geq 0$.  For example, one choice is $\tilde{E}_R = \bar{E}_R(\bar{D}_\Theta) = \text{constant}$, meaning $D_\Theta = \bar{D}_\Theta$ in the reference configuration $\tau = \tau_0$, but $D_\Theta$ evolves with $\tau$ to maintain micro-force equilibrium in \eqref{eq:DR0}.
 For this choice, $\hat{I}_1$ is independent of $r = R$ at $\tau = \tau_0$, so $E_R$ in \eqref{eq:DR0} becomes a constant defined by $\bar{D}_\Theta$ measured at $\tau = \tau_0$.
 
 Although the variational framework of \S3 does not include explicit time dependence, dissipation, or thermal effects, the external microforce $\bm{E}^\alpha$ can be assigned a physical meaning by appealing to dissipative kinetics in the setting of a generalized Allen-Cahn or Ginzburg-Landau equation \cite{gurtin1996,claytonCMT2018,grillo2018,claytonJMPS2021}.  Assume ${E}_R$ arises from time-dependent (e.g., viscous) interactions between the solid tissue microstructure and biologic fluids, for example.
 Introduce a possibly state-dependent relaxation (time) function $\tau_R(r,D_\Theta, \ldots)$. Then approximate
 at each point $r(R)$ the micro-force as
 \begin{equation}
 \label{eq:ERTD}
 E_R \approx k_D \mu_0 \tau_R {( \diff D_\Theta }/{ \diff \tau} ).
 \end{equation}
 The first condition in \eqref{eq:extforc} corresponds to $\tau_R = 0$: instantaneous viscous relaxation. The second of \eqref{eq:extforc} corresponds to $\tau_R \rightarrow \infty \Rightarrow \diff D_\Theta / \diff \tau \rightarrow 0$
 with initial condition $D_\Theta(\tau = 0) = \bar{D}_\Theta$, meaning viscous relaxation time is infinite relative to the duration of loading and residual strains $\bm{\epsilon}^r$ in \eqref{eq:epsrc} are permanent. The third of \eqref{eq:extforc}
 implies that both $\tau_R$ and $D_\Theta$ can be finite and transient.
 
Nonzero residual strain (i.e., $\Lambda_\Theta \neq 1$ in an externally unloaded state)
 is evidenced by opening of excised annular sections of the ventricle when cut along 
 planes $\Theta = \text{constant}$ \cite{guccione1991,grobbel2018}.
 Anisotropic and out-of-plane opening distortions have also been noted \cite{guccione1991}.
 Residual strains are thought to emerge primarily from collagen fibers, with lesser
 influences from interactions between collagen and myocytes \cite{grobbel2018}.
 The opening angle that relieves internal stress changes slowly over time,
 increasing rapidly over the first 15 min and slowly over the next 75 min in
 rat tissue \cite{grobbel2018}.
 A constant residual strain field has been shown in prior
 analysis \cite{taka1990} to reduce maximum stress magnitudes when
 the ventricle is subjected to internal pressure ranges in the physiological range.
 This confirms a popular hypothesis that remodeling occurs to minimize stress concentrations
 that otherwise would harm or degrade the tissue over time, thus prolonging its 
 functional lifespan \cite{chuong1986,rachev1997}.
 
 Real ventricular tissue likely demonstrates a heterogeneous $\Lambda_\Theta(R)$ distribution
 since microstructure and mechanical properties vary with position \cite{novak1994,walker2005,holz2009}.
  Direct measurement of $\Lambda_\Theta(R)$ in the setting of a spherical model cannot be made by cutting experiments due to intrinsic curvature of the hollow sphere, in contrast to an intrinsically flat,
 hollow cylindrical section \cite{chuong1986,guccione1991,holz2000,klarbring2007}.
 The distribution of  $\Lambda_\Theta$
 that achieves the most uniform stresses depends on the geometry (e.g., wall thickness), constitutive properties, and  loading conditions. Thus, in the absence of experimental values, when
 the second case in \eqref{eq:extforc} is implemented, $\bar{E}_R$ is imposed
 so that  $\bar{D}_\theta$ yields a constant $\Lambda_\Theta$ that produces more uniform
 stresses through the ventricular wall at the mean physiologic pressure of the diastole \cite{spotnitz1966,taka1990}.
 The reference configuration $\tau_0$ is defined to coincide with this pressure.
 The resulting value of $\bar{D}_\theta$ is then used to find $\tilde{E}_R = \text{constant}$ in other calculations invoking the
 third case in \eqref{eq:extforc}. This third case mimics transients in microstructure commensurate with changes in opening angle observed in excised annular sections, meaning
 the load duration is likely at least on the order of minutes \cite{grobbel2018}, much longer than a single cardiac cycle.
 Prior analyses of the residually stressed ventricle \cite{taka1990,guccione1991}
 did not consider possibilities that residual strains can be heterogeneous and can change in magnitude over the course of applied loading. 
 The current analysis, like Refs.~\cite{janz1976,taka1990,guccione1991}, is restricted to passive
states representative of the diastole.  Finite-element models have accounted for
active muscle tension as arising during the systole \cite{guccione1995,walker2005}.
 
 Solutions to the incremental boundary value problem are obtained as follows. Load parameter $\tau$ is
 increased from $\tau = 0$ in small steps $\diff \tau$.  The minimum radius over the load
 history is $r_i(\tau) = r_i(0)$, while the inner radius in the reference configuration is $r_i(\tau_0) = R_i$. 
Given boundary condition $r_i(\tau)$, the first of \eqref{eq:incomp1bc} is solved for $C_i (\tau)$. Then
$C_i$ is used in the third of \eqref{eq:incomp1bc} to obtain external boundary $r_o(\tau)$.
Next, \eqref{eq:xianal} and \eqref{eq:DR0} are solved simultaneously at each point $r(\tau)$ for $\eta(r)$
and $D_\Theta(r)$ using numerical iteration. Given $p_o(\tau)$, \eqref{eq:pi} is integrated numerically for internal pressure $p_i(\tau)$. Lastly, the stress fields $\sigma^r_r(r)$, $\sigma^\theta_\theta(r) = \sigma^\phi_\phi(r)$, and $p(r)$
are calculated from successive use of \eqref{eq:sigmari} and \eqref{eq:sigmati}. Suitability of incompressibility is then verified from smallness of $|p|/B_0$.
In some cases, solution paths to \eqref{eq:xianal} and \eqref{eq:DR0} are found to have two branches. In such cases, the valid branch is
chosen as the (only) one of the two producing physically realistic stress fields. 
Given incompressibility constraint \eqref{eq:incomp1bc} and spherical symmetry, deformation field $r(R)$
is fully and uniquely determined by the boundary condition $r_i(\tau)$. Equations \eqref{eq:xianal} and \eqref{eq:DR0} are local and algebraic, rather than differential. Results are independent of numerical
discretization provided a small enough increment $\diff r$ is used for integration of \eqref{eq:pi} and \eqref{eq:sigmari}. No localized deformation is possible, and no gradient regularization (e.g., $\hat{\Upsilon} l_0 > 0$) is required to obtain valid solutions.

Canine data \cite{spotnitz1966} furnish the referential dimensions of the ventricular wall at $p_i - p_o = 0$ 
(i.e., at zero net external pressure) for $\mathcal{M}$: $R_i = 14.4$ mm, $R_o = 29.6$ mm. 
When a different reference configuration is chosen to best match experimental data (e.g., in situations where remnant strains are nonzero and $\Lambda_\Theta \neq 1$), then a different value of $R_i$ is defined, with associated $R_o$ calculated under the incompressibility assumption
(i.e., volume of the spherical wall is constant, regardless of $R_i$).
Five material parameters are required: $\mu_0$, $k_s$, $\hat{e}_c$, $k_D$, 
and $\hat{n}$. Shear elastic constants $\mu_0$ and $k_s$ are calibrated to 
$p_i$ versus $r_i$ data \cite{spotnitz1966} in \S5.4.
In calibrations, $p_i (\tau)$ is obtained versus $r_i(\tau)$ by solving \eqref{eq:pi} with \eqref{eq:xianal} and \eqref{eq:DR0} repeatedly over a physically credible range of $\mu_0$ and $k_s$.
Final values of $\mu_0$ and $k_s$ minimize error with data \cite{spotnitz1966} according to the least squares method.

Finslerian metric parameters $k_D$ and $\hat{n}$ are explored parametrically in calculations reported
in \S5.4, noting choices $\hat{n} = 1$ and $\hat{n} = 2$ have shown utility in prior modeling of
hard and soft solids \cite{claytonJGP2017,claytonMMS2022,claytonSYMM2023}.
Cohesive energy per unit volume (i.e., stress)
$\hat{e}_c$ is prescribed commensurate with tensile strength.
Data on canine ventricular strength appear unavailable. Interpretations of mouse \cite{gao2005} and porcine \cite{forsell2011} data suggest $\hat{e}_c \approx$ 30 kPa -- 50 kPa, so $\hat{e}_c = 40$ kPa is used.
The current model, like that in Ref.~\cite{forsell2011}, omits remnant strains from
damage processes and damage anisotropy (but not originating from remodeling via $\Lambda_\Theta$), thus not resolving certain physics at the micron scale. Such could be incorporated via setting $k_\eta > 0$ for dilatation from cavitation or crack opening (i.e., relaxing the anelastic incompressibility assumption), adding anisotropic contributions from $\eta$ to $\hat{G}^A_B$ in \eqref{eq:spheremetR} \cite{claytonSYMM2023}, and implementing anisotropic surface energy in \eqref{eq:cohsphere} \cite{claytonIJF2014,claytonJMPS2021,gultekin2019}.

\subsection{Results}
Eight sets of calculations, labeled C0, C1 $,\ldots,$ C8 are reported, with protocols listed in Table~\ref{table1}.
In all calculations, the inner radius is incremented from $r_i = 10$ mm at $\tau = 0 < \tau_0 $ to $r_i = 30$ mm at $\tau = \tau_f > \tau_0$, where $\tau_0$ defines the reference configuration when $r = R$. Local stress fields depend on the external pressure at $r = r_o$ only through additive constant $p_o$ at each increment,
so $p_o$ need not be prescribed explicitly. In other words, if mechanical tractions were applied, solutions depend only
on the difference $p_i - p_o$ (i.e., the net applied pressure) and not $p_i$ and $p_o$ individually.

Two different reference configurations are evaluated. The first, for cases C0, C1, and C6 with $R_i = 14.4$ mm, corresponds to $p_i - p_o = 0$ in experiments \cite{spotnitz1966}.
The second, for all other cases with $R_i = 21.8$ mm, corresponds to $p_i - p_o \approx 1$ kPa (i.e., 7.5 mm Hg), the mean physiologic pressure in the diastole \cite{spotnitz1966}.  Cohesive energy $\hat{e}_c$ is infinite for special cases C0 and C5 wherein degradation is artificially suppressed; otherwise
it is set to the previously justified value of 40 kPa.  Two values $\hat{n} = 1, 2$ are considered.
For cases C0 and C1, $\hat{n} = 2$ supplies the trivial solution $D_\Theta = 0$ and $\Lambda_\Theta = 1$ with  $E_R = 0$ in \eqref{eq:DR0}. In other words, cases C0 and C1 are designed such that remnant stresses and strains vanish: $G_{AB}$ is Euclidean, with null curvature, rather than Finslerian or Riemannian.  For C2 $,\ldots,$ C7,   $\hat{n} = 1$ is chosen for simplicity and convenience, as
it permits $\ln \Lambda_\Theta$ and $D_\Theta$ to have arbitrary signs for admissible (i.e., real) solutions to \eqref{eq:DR0}.
In all cases of Table~\ref{table1}, $k_D = -1$ is imposed, where the inessential negative sign gives $D_\Theta > 0 \Leftrightarrow \Lambda_\Theta > 1$.
Magnitude $|k_D| > 0$ is arbitrary when $\hat{n} = 1$: setting $k_D \rightarrow \alpha k_D$ with $\alpha = \text{constant} \neq 0$ produces linearly transformed quantities
$D_\Theta \rightarrow D_\Theta / \alpha$ and $E_R \rightarrow \alpha E_R$ in solutions to \eqref{eq:DR0}.  The other field variables ($\eta, {\bm \sigma}, \psi$, etc.) are unaffected by this linear transformation.  Case C8 invokes $\hat{n} = 2$ out of curiosity.

\begin{table}
\footnotesize
\caption{Parameters for cases C0--C8: referential inner radius $R_i$, cohesive energy $\hat{e}_c$,
Finsler exponent $\hat{n}$, remnant stretch $\Lambda_\Theta$, external interactive micro-force $E_R$, elastic shear modulus $\mu_0$, and nonlinear elastic shear parameter $k_s$}
\label{table1}       
\centering
\begin{tabular}{lccccccc}
\hline\noalign{\smallskip}
Case (model) & $R_i$ [mm] & $\hat{e}_c$ [kPa] & $\hat{n}$ & $\Lambda_\Theta$ & $E_R$ [kPa] & $\mu_0$ [kPa] & $k_s$  \\
\noalign{\smallskip}\hline\noalign{\smallskip}
C0 & 14.4 & $\infty$ & 2 & 1 & 0 & 0.56 & 1.1 \\
C1 & 14.4 & 40 & 2 & 1 & 0 & 0.46 & 1.3 \\
C2 & 21.8 & 40 & 1 & 1.2 & $\dagger$ & 0.65 & 2.1 \\
C3 & 21.8 & 40 & 1 & 1.3 & $\dagger$ & 0.30 & 1.7 \\
C4 & 21.8 & 40 & 1 & 1.4 & $\dagger$ & 0.13 & 1.5 \\
C5 & 21.8 & $\infty$ & 1 & 1.3 & $\dagger$ & 0.28 & 1.7 \\
C6 & 14.4 & 40 & 1 & $\dagger$  & -0.46 & 0.46 & 1.3 \\
C7 & 21.8 & 40 & 1 & $\dagger$ & -1.74 & 0.30 & 1.7 \\
C8 & 21.8 & 40 & 2 & $\dagger$ & -1.78 & 0.30 & 1.7 \\
 \noalign{\smallskip}\hline
 $\dagger$: transient
\end{tabular}
\end{table}

Cases C0 and C1 implement the first (trivial) external force condition in \eqref{eq:extforc}.
Cases C2, C3, C4, and C5 invoke the second condition in \eqref{eq:extforc}, whereby $E_R$ evolves with $\tau$
to supply a constant remnant (from a priori remodeling) stretch $\Lambda_\Theta$ of 1.2, 1.3, or 1.4. Since each of cases C0 $,\ldots,$ C5
seeks to match experimental loading conditions, elastic constants $\mu_0$ and $k_s$ are calibrated independently
for each of these cases, similarly to Ref.~\cite{taka1990}.  Cases C6, C7, and C8 invoke the third 
condition of \eqref{eq:extforc}, whereby $E_R$ is a constant in each case obtained by solving \eqref{eq:DR0}
with $\Lambda_\Theta$ prescribed uniformly only in the corresponding reference configuration. Case C6 uses $R_i$, $\Lambda_\Theta = 1$
at $\tau = \tau_0$,
and elastic constants of Case C1. Cases C7 and C8 use $R_i$, $\Lambda_\Theta = 1.3$ at $\tau = \tau_0$, and elastic constants
of C3.  For Cases C6, C7, and C8, $D_\Theta$ and $\Lambda_\Theta$ evolve with the load history via solutions to \eqref{eq:DR0} and are non-uniform except at $\tau = \tau_0$. Elastic constants are not recalibrated for C6, C7, and C8 since remodeling is assumed to occur over much longer time scales than  experiments of Ref.~\cite{spotnitz1966}. Results for these cases show how remodeling, over a long hold time following each increment $\diff \tau$ in the context of \eqref{eq:ERTD}, gives anelastic strains that accommodate imposed deformation from the reference state.

\begin{figure}
\begin{center}
 \subfigure[fitted pressure range]{\includegraphics[width=0.35\textwidth]{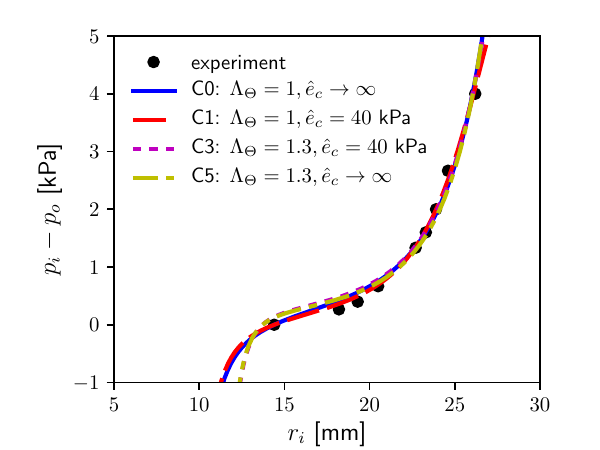} \label{fig1a}} \qquad
 \subfigure[large pressure range]{\includegraphics[width=0.35\textwidth]{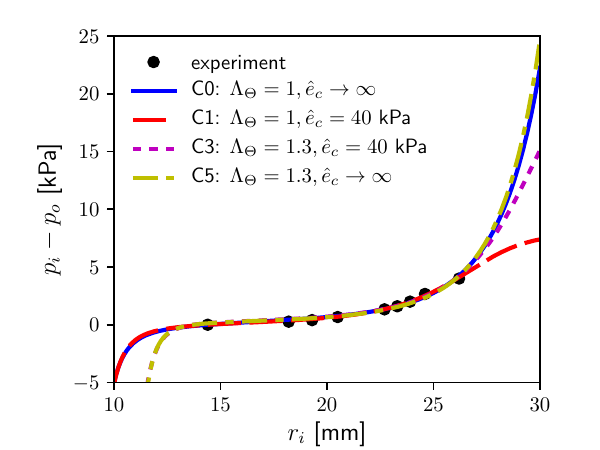}\label{fig1b}} 
  \end{center}
  \vspace{-0.5cm}
\caption{Elastic constant fits for cases C0, C1, C3, and C5 (Table~\ref{table1}, uniform remnant stretch $\Lambda_\Theta$, cohesive fracture energy $\hat{e}_c$) to ventricular pressure-volume data \cite{spotnitz1966},
inner radius $r_i$, internal pressure $p_i$, and external pressure $p_o$ (1 atm in experiments):
(a) pressure range close to experiments
(b) higher net pressures
}
\label{fig1}       
\end{figure}

Net boundary pressure $p_i - p_o$ for selected cases C0, C1, C3, and C5 is compared in Fig.~\ref{fig1} with experimental pressure-volume data \cite{spotnitz1966}, the latter likewise assuming spherical symmetry.
Two-parameter fits of $\mu_0$ and $k_s$ are deemed satisfactory in Fig.~\ref{fig1a}. 
Relative effects of damage softening are evident for cases C1 and C3 at large expansion, $r_i \gtrsim  27$ mm, in Fig.~\ref{fig1b}.
As will be shown later, fracture order parameter $\eta$ tends toward larger values in case C1 than C3, resulting in more overall softening.
Data for negative net boundary pressures are not tabulated in Ref.~\cite{spotnitz1966}, so model accuracy in the extreme compression
regime is unknown.

\begin{figure}
\begin{center}
 \subfigure[transverse, $r_i = 14.4$ mm]{\includegraphics[width=0.32\textwidth]{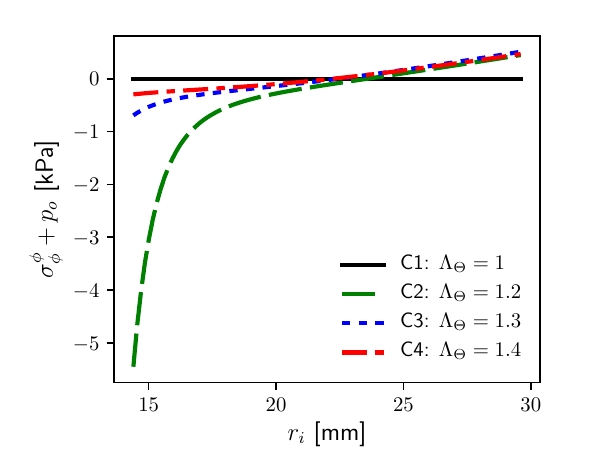} \label{fig2a}} 
 \subfigure[transverse, $r_i = 21.8$ mm]{\includegraphics[width=0.32\textwidth]{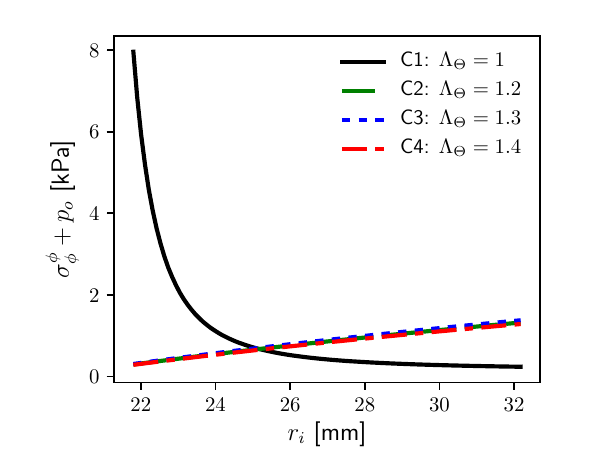}\label{fig2b}} 
 \subfigure[radial, $r_i = 21.8$ mm]{\includegraphics[width=0.32\textwidth]{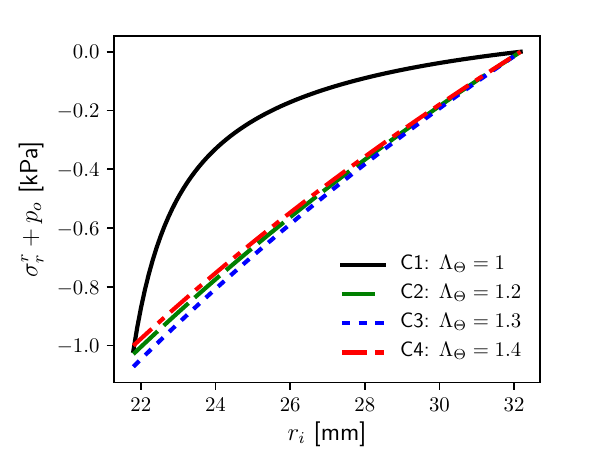}\label{fig2c}} 
  \end{center}
  \vspace{-0.5cm}
\caption{Stress versus inner radius $r_i$ for cases C1, C2, C3, and C4 (Table~\ref{table1}, uniform remnant stretch $\Lambda_\Theta$):
(a) transverse $\sigma^\phi_\phi = \sigma^\theta_\theta$ at null applied pressure
(b) transverse at $\approx 1$ kPa pressure
(c) radial $\sigma^r_r$  at $\approx 1$ kPa pressure
}
\label{fig2}       
\end{figure}

\begin{figure}
\begin{center}
 \subfigure[transverse stress, $r_i = 26.1$ mm]{\includegraphics[width=0.32\textwidth]{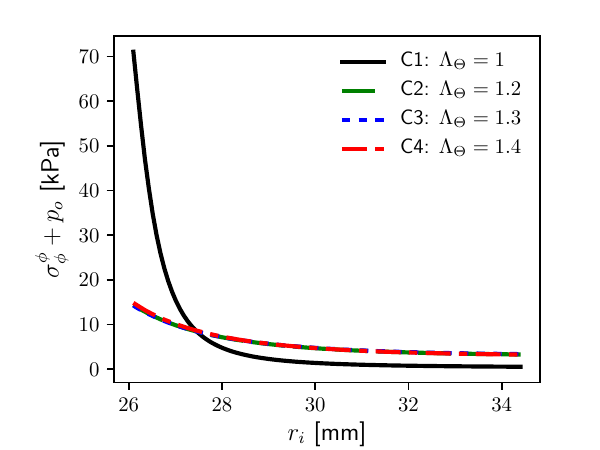} \label{fig3a}} \qquad
 \subfigure[damage state, $r_i = 26.1$ mm]{\includegraphics[width=0.32\textwidth]{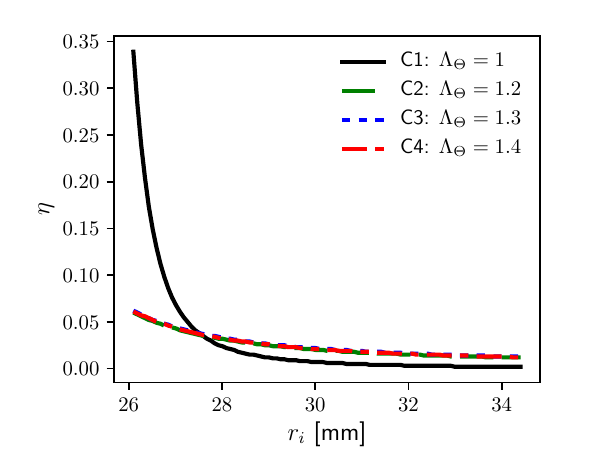}\label{fig3b}} 
  \end{center}
  \vspace{-0.5cm}
\caption{Stress and degradation versus inner radius $r_i$ for cases C1, C2, C3, and C4 (Table~\ref{table1}, uniform remnant stretch $\Lambda_\Theta$):
(a) transverse stress $\sigma^\phi_\phi = \sigma^\theta_\theta$ at $\approx 4$ kPa net applied pressure
(b) fracture order parameter $\eta$ at $\approx 4$ kPa pressure
}
\label{fig3}       
\end{figure}

\begin{figure}
\begin{center}
 \subfigure[transverse stress, $r_i = 29.2$ mm]{\includegraphics[width=0.32\textwidth]{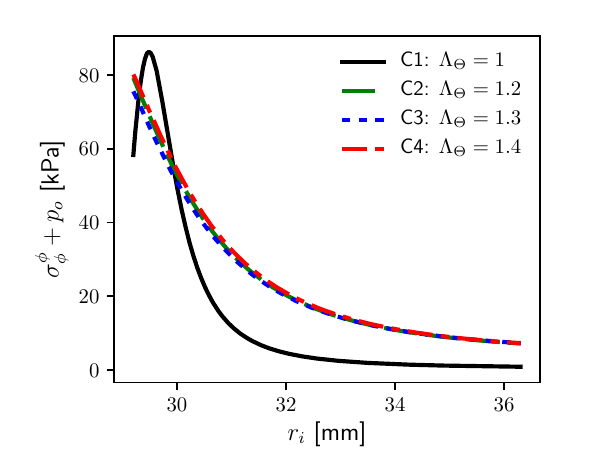} \label{fig4a}} \qquad
 \subfigure[damage state, $r_i = 29.2$ mm]{\includegraphics[width=0.32\textwidth]{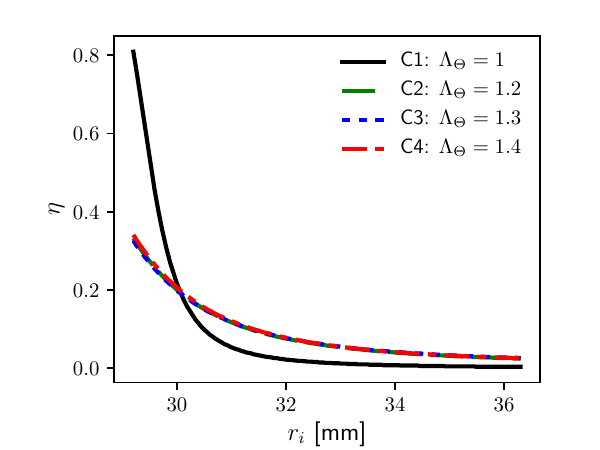}\label{fig4b}} 
  \end{center}
  \vspace{-0.5cm}
\caption{Stress and degradation versus inner radius $r_i$ for cases C1, C2, C3, and C4 (Table~\ref{table1}, uniform remnant stretch $\Lambda_\Theta$):
(a) transverse stress $\sigma^\phi_\phi = \sigma^\theta_\theta$ at $\gtrsim 7$ kPa net applied pressure
(b) fracture order parameter $\eta$ at $\gtrsim 7$ kPa pressure
}
\label{fig4}       
\end{figure}

Predictions in Figs.~\ref{fig2}, \ref{fig3}, and \ref{fig4} highlight influences of constant prescribed remnant strain from remodeling, where $\Lambda_\Theta = 1$ for C1 (i.e., null remnant strain), $\Lambda_\Theta = 1.2$ for C2, $\Lambda_\Theta = 1.3$ for C3, and $\Lambda_\Theta = 1.4$ for C4.
For cases C2, C3, and C4, with $R_i$ defined at the state where $p_i - p_o \approx 1$ kPa,
values of $\Lambda_\Theta$ much smaller than 1.2 or larger than 1.4 do not produce credible fits to data in Fig.~\ref{fig1a} for physically
reasonable values of $\mu_0$ and $k_s$. The best fit is achieved for C3 with $\Lambda_\Theta = 1.3$. 
At $r_i = 14.4$ mm in Fig.~\ref{fig2a}, $\sigma^\phi_\phi = \sigma^\theta_\theta = 0$ for C1, $\sigma^\phi_\phi $ is small ($|\sigma^\phi_\phi| < 1$ kPa) for C2 and C3, and
somewhat larger in magnitude toward the inner surface for C4.
At $r_i$ = 21.8 mm in Fig.~\ref{fig2b}, the maximum transverse stress approaches 8 kPa for C1 but is less than 2 kPa for C2, C3, and C4.
Here, remnant strains modeled with a nontrivial Finsler metric $G_{AB}$ reduce maximum transverse stress under
physiological loading. Radial stress $\sigma^r_r$ in Fig.~\ref{fig2c} is less interesting since minimum and maximum values are dictated by internal and external pressures at $r_i$ and $r_o$. Interpolation between boundary values is
more linear for cases C2, C3, and C4 than C1. 

For conditions of Fig.~\ref{fig2}, degradation from $\eta$ is
small to negligible, with maximum local values of $\eta(r)$ not exceeding 0.04 in the physiologic range.
At larger $r_i$ corresponding to much higher $p_i - p_o$, local strain energy density is sufficient
to induce more severe damage and softening behaviors, especially in case C1.
In Fig.~\ref{fig3a}, transverse stress exceeds 70 kPa at $r = r_i = 26.1$ mm for C1, with damage $\eta$ approaching 0.35 in Fig.~\ref{fig3b}.
In Fig.~\ref{fig4a}, transverse stress exceeds 85 kPa for C1, with damage $\eta$ exceeding 0.8 at $r = r_i = 29.2$ mm in Fig.~\ref{fig4b}.
Stress $\sigma^\phi_\phi$ decreases with decreasing $r$ as $r \rightarrow r_i$ from above due to tangent modulus degradation $\omega_s = (1 -\eta)^2$.
In contrast, for cases with remnant strains (C2, C3, C4), maximum $\sigma^\phi_\phi \approx 15$ kPa and maximum $\eta \approx 0.06$ at
$r_i = 26.1$ mm. Constant remnant strains from remodeling alleviate stress concentrations and damage at larger over-pressures.
Similar trends arise in Fig.~\ref{fig4b}: $\eta < 0.35$ for C2, C3, and C4.

\begin{figure}
\begin{center}
 \subfigure[C1: transverse stress]{\includegraphics[width=0.32\textwidth]{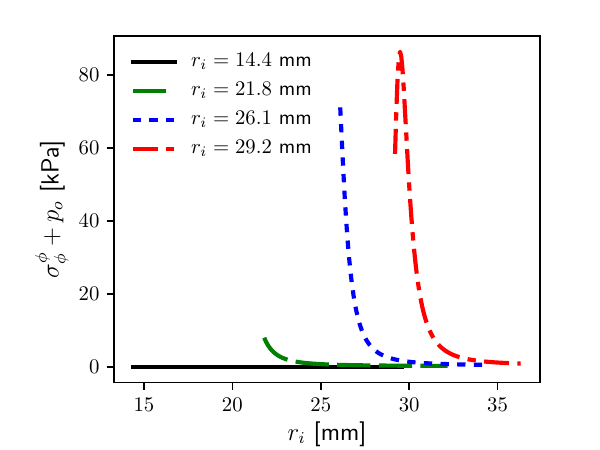} \label{fig5a}} \qquad
 \subfigure[C1: radial stress]{\includegraphics[width=0.32\textwidth]{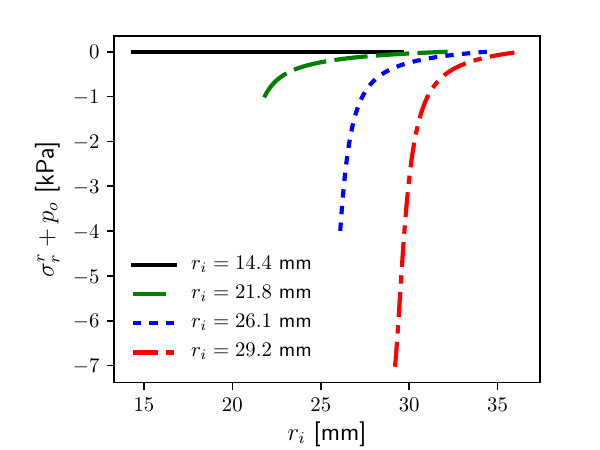}\label{fig5b}} \\
  \subfigure[C1: damage state]{\includegraphics[width=0.32\textwidth]{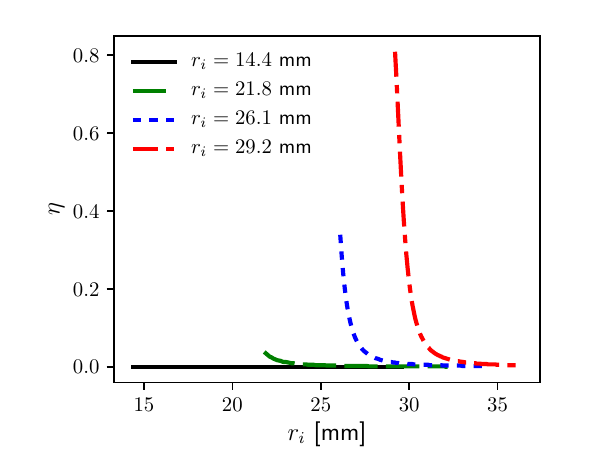} \label{fig5c}} \qquad
 \subfigure[C1: remodeling state]{\includegraphics[width=0.32\textwidth]{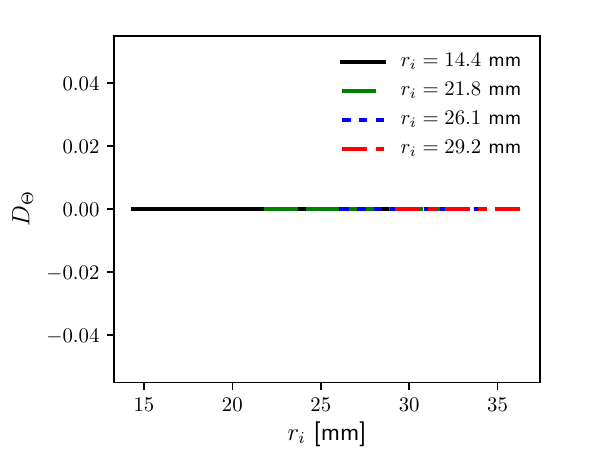}\label{fig5d}} \\
  \end{center}
  \vspace{-0.5cm}
\caption{Stress and internal state components versus inner radius $r_i$ for case C1 (Table~\ref{table1}, $E_R = 0$, $\hat{n} = 2$):
(a) transverse stress $\sigma^\phi_\phi = \sigma^\theta_\theta$ 
(b) radial stress $\sigma^r_r$
(c) fracture variable $\eta$ 
(d) remodeling variable $D_\Theta$
}
\label{fig5}       
\end{figure}

\begin{figure}
\begin{center}
 \subfigure[C6: stresses and damage]{\includegraphics[width=0.32\textwidth]{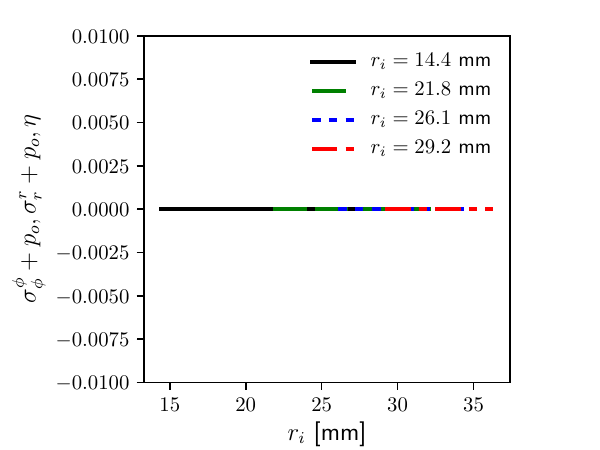} \label{fig6a}} \qquad
 \subfigure[C6: remodeling state]{\includegraphics[width=0.32\textwidth]{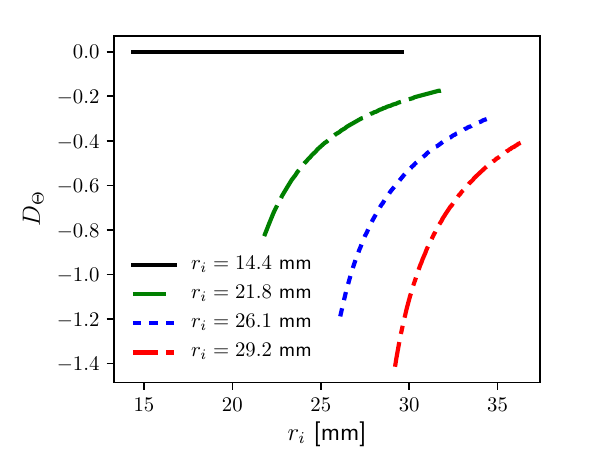}\label{fig6b}} 
  \end{center}
  \vspace{-0.5cm}
\caption{Stress and internal state versus inner radius $r_i$ for case C6 (Table~\ref{table1}, $E_R = -0.46$ kPa, $\hat{n} = 1$):
(a) stresses and fracture variable $\eta$ (all zero)
(b) remodeling variable $D_\Theta$
}
\label{fig6}       
\end{figure}

\begin{figure}
\begin{center}
 \subfigure[C7: transverse stress]{\includegraphics[width=0.3\textwidth]{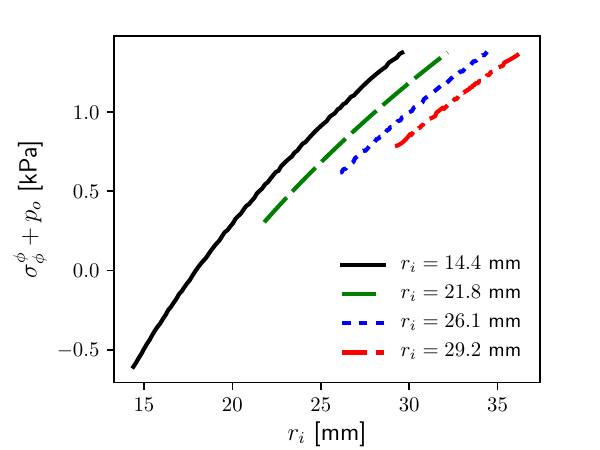} \label{fig7a}} \qquad
 \subfigure[C7: radial stress]{\includegraphics[width=0.3\textwidth]{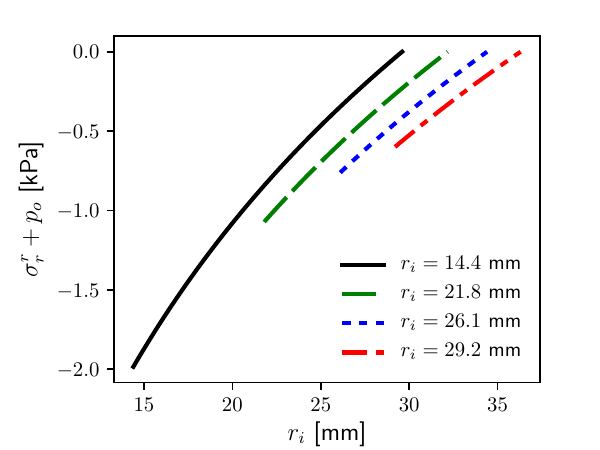}\label{fig7b}} \\
  \subfigure[C7: damage state]{\includegraphics[width=0.3\textwidth]{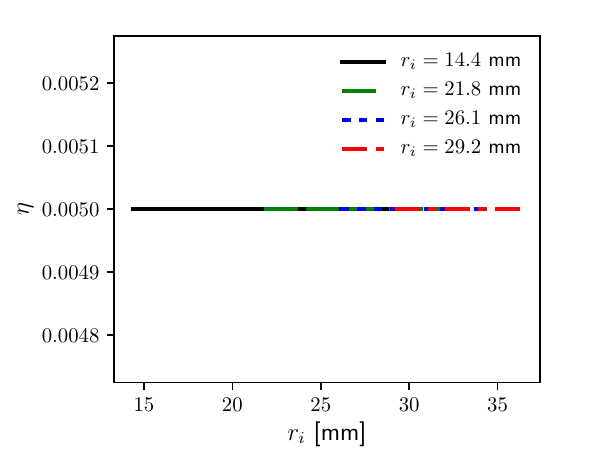} \label{fig7c}} \qquad
 \subfigure[C7: remodeling state]{\includegraphics[width=0.3\textwidth]{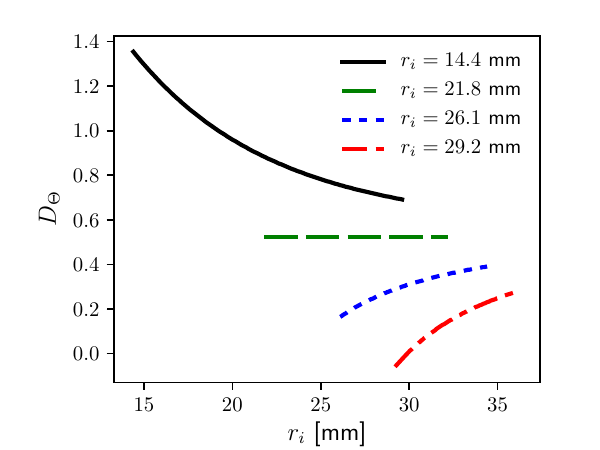}\label{fig7d}} \\
  \end{center}
  \vspace{-0.5cm}
\caption{Stress and internal state versus inner radius $r_i$ for case C7 (Table~\ref{table1}, $E_R = -1.74$ kPa, $\hat{n} = 1$):
(a) transverse stress $\sigma^\phi_\phi = \sigma^\theta_\theta$ 
(b) radial stress $\sigma^r_r$
(c) fracture variable $\eta$ 
(d) remodeling variable $D_\Theta$
}
\label{fig7}       
\end{figure}

\begin{figure}
\begin{center}
 \subfigure[C8: transverse stress]{\includegraphics[width=0.3\textwidth]{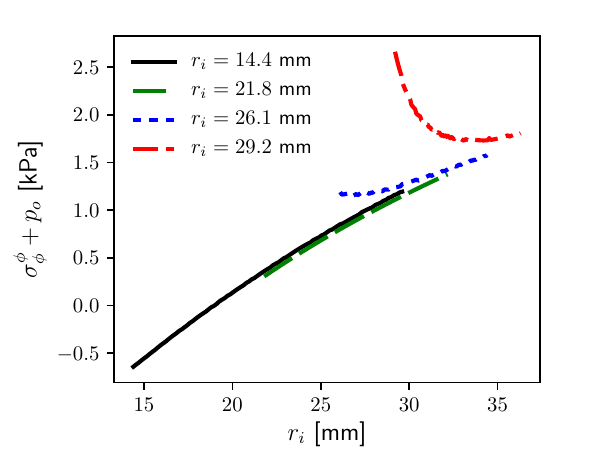} \label{fig8a}} \qquad
 \subfigure[C8: radial stress]{\includegraphics[width=0.3\textwidth]{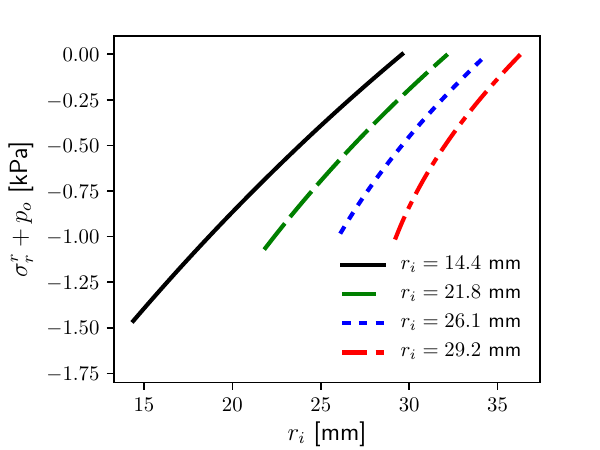}\label{fig8b}} \\
  \subfigure[C8: damage state]{\includegraphics[width=0.3\textwidth]{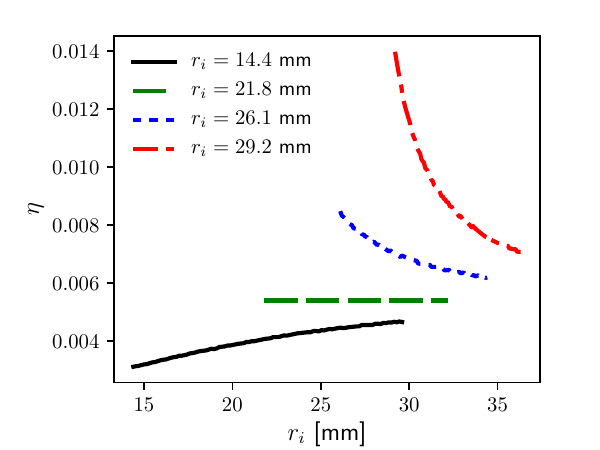} \label{fig8c}} \qquad
 \subfigure[C8: remodeling state]{\includegraphics[width=0.3\textwidth]{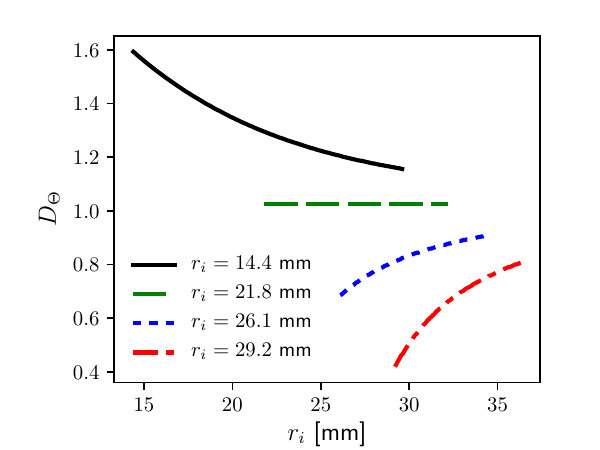}\label{fig8d}} \\
  \end{center}
  \vspace{-0.5cm}
\caption{Stress and internal state versus inner radius $r_i$ for case C8 (Table~\ref{table1}, $E_R = -1.74$ kPa, $\hat{n} = 2$):
(a) transverse stress $\sigma^\phi_\phi = \sigma^\theta_\theta$ 
(b) radial stress $\sigma^r_r$
(c) fracture variable $\eta$ 
(d) remodeling variable $D_\Theta$
}
\label{fig8}       
\end{figure}

Predictions in Figs.~\ref{fig5}, \ref{fig6}, \ref{fig7}, and \ref{fig8} compare stress and
internal state fields for degradation ($\eta$) and remodeling ($D_\Theta$) in respective
cases C1, C6, C7, and C8 at prescribed load increments $r_i = 14.4, \, 21.8, \, 26$, and $29.2$ mm.  For baseline case C1 in Fig.~\ref{fig5}, $\Lambda_\Theta = 1$ identically and no remodeling occurs.
Maximum stress and $\eta$ increase with increasing $r_i$, the latter attaining local maximum $\eta > 0.8$ at
29.2 mm.
Maximum local Cauchy pressure is $p = 55.7$ kPa, so the incompressibility assumption is credible with $B_0 \geq 2$ GPa, even at this state of extreme expansion.

For case C6 in Fig.~\ref{fig6}, $D_\Theta$ becomes negative and hence $\Lambda_\Theta$ decreases heterogeneously with increasing $r_i$
in Fig.~\ref{fig6b}. This relieves all stresses and all damage in Fig.~\ref{fig6a}.
This solution\footnote{For C6, a closed-form analytical solution to \eqref{eq:xianal} and \eqref{eq:DR0} exists:
$E_R = k_D \mu_0, \, D_\Theta = \frac{2}{k_D} \ln \frac{r}{R}, \, \eta = 0, \, \sigma_r^r = \sigma^\phi_\phi =  - p_i = -p_o$.
This solution does not apply for all possible $\hat{\psi}$ in \eqref{eq:psis1} (i.e., arbitrary free energy functions).}, 
an outcome of \eqref{eq:DR0}, predicts that the solid behaves as if it were ``perfectly plastic''.
In the context of \eqref{eq:ERTD}, at each value of $\tau$, the material relaxes through remodeling over a (long) hold time so that
strain energy $\psi_s \rightarrow 0$ after each increment $\diff \tau$.
Physically, results in Fig.~\ref{fig6} are interpreted as follows, letting $p_o= \text{constant}$. A balloon is inserted into the ventricular chamber.  Pressure $p_i$ in the balloon is increased to attain a certain radius $r_i (\tau) > r_i (\tau_0) = R_i $. 
The pressure is adjusted so $r_i (\tau)$ stays constant in ``real'' time while load parameter $\tau$ is constant. Over a sufficient hold time,
tissues remodel so the balloon pressure can be reduced while maintaining the same value of $r_i$,
eventually reaching $p_i =  p_o$.  After the hold time, the balloon can be removed and the ventricle maintains
its enlarged state.  This thought experiment for case C6, and for C7 and C8 that follow, is similar to the angioplasty procedure (e.g., for expanding blocked arteries\footnote{Besides arterial expansion, a balloon catheter enables
compression and dissection of atherosclerotic plaque \cite{byrne2017}.}) in medicine \cite{byrne2017}. If 
the environment (e.g., $p_i, p_o, E_R$) changes when the balloon
is removed due to biologic response, then the radius of the ventricle (or artery) can change (e.g., recoil) from further remodeling or elastic response.
In medicine, a stent is often, but not always, inserted to inhibit unwanted changes.

Behaviors are similar for cases C7 and C8, which also enable ``transient'' remodeling to relieve local stress, strain energy, and
thus damage.  In Figs.~\ref{fig7a} and \ref{fig7b}, local stress magnitudes do not exceed a few kPa, and $\eta = 0.005 = \text{constant}$
in Fig.~\ref{fig7c}. 
In Figs.~\ref{fig8a} and \ref{fig8b}, local stress magnitudes likewise remain low, and
degradation measure $\eta = \eta(r) < 0.015$ is heterogeneous yet small in Fig.~\ref{fig7c}.  
These conditions are contrasted with tangential stress concentrations exceeding 80 kPa and
damage $\eta$ exceeding 0.8 for C1 in Fig.~\ref{fig6}.
Modest differences in predictions between C7 and C8 in Figs.~\ref{fig7} and \ref{fig8} are due to respective prescriptions of $\hat{n} = 1$ and $\hat{n} = 2$. In both cases C7 and C8, the hollow sphere again behaves in a near perfectly plastic manner in departures from $\tau_0$, with very low overall
resistance to expansion.
Stresses do not relax completely to zero because reference configuration $\tau_0$ at which force $E_R$ is defined is not stress-free in C7 and C8 (e.g., $p_i - p_0 \approx 1$ kPa at $\tau = \tau_0$).
For all of cases C6, C7, and C8, the constant interaction force $E_R$ calculated from equilibrium conditions in the reference configuration
causes the material to remodel at each $\diff \tau$ to reduce or minimize strain energy density under imposed deformation at the inner radius $r_i(\tau)$.

The present calculations assume constant elastic properties $\mu_0$ and $k_s$ through the thickness of the ventricular wall.
Elastic properties and anisotropy vary with location in real tissues.
Analysis of a spherical model in Ref.~\cite{janz1976} shows how spatially dependent elastic stiffness through the wall thickness
might achieve uniform tangential stress, in the absence of residual strain.
In real tissue, remnant strains must exist as evidenced by cut-and-relax experiments \cite{guccione1991,grobbel2018}.
Thus, stress fields, and degradation and rupture under extreme conditions, are likely affected by both
 residual strains (from prior remodeling and growth) and heterogeneous anisotropic properties.
 The current analysis considers only the former, providing a basis for comparison with numerical simulations that might consider more physics simultaneously. Experimental quantification of heterogeneous material properties and microstructures required by realistic simulations is a persistent challenge.

\section{Conclusions}

A geometrically nonlinear theory has been advanced to describe mechanics of biologic solids and other materials with fibrous microstructures. Effects of biologic growth and remodeling, as well as resorption and degradation, are addressed in a quasi-static variational setting.
The Finslerian approach includes some distinguishing features over
theories based purely in Riemannian geometry. 
Internal state vectors parameterizing fiber families in a geometric sense are introduced that relate directly to features of the evolving fibrous microstructures in a physical sense.
Microstructure effects are explicitly delineated in Finslerian (i.e., internal state) contributions
to metric tensors and connection coefficients: Levi-Civita, Chern-Rund, Cartan, and nonlinear connections. 
Residual stresses originate from non-vanishing Cartan's tensor and spatial gradients of internal
state. 

An application to the left ventricle has shown importance of remnant strains
on stress distributions and strength degradation for loading outside the physiological range. Remodeling, whether occurring before imposed loading or during the load history, produces residual strains that can mitigate stress concentrations. Calculations have quantified how damage evolves (i.e., shear strength degrades)
when the ventricle is subjected to extreme deformation resulting from a large difference in
internal and external pressures, as might be witnessed in a traumatic injurious event.
Results have shown how internal strains can inhibit degradation from local fracture (i.e., decohesion) driven by strain energy.
An implication is that remnant strain from prior growth and remodeling should not be ignored when modeling mechanical damage to the ventricle under subsequent severe loading.
Effects of remodeling occurring over time scales commensurate with mechanical loading
have also been predicted. For complete remodeling in each load increment (e.g., infinitely slow mechanical
loading), a non-uniform remnant strain field measured by a non-trivial Finsler metric accommodates imposed deformation, nullifies strain energy, and prevents tissue damage.

The present solutions are limited by idealizations of spherical symmetry, isotropic nonlinear elasticity with uniform elastic constants,
and incompressible elastic and inelastic behaviors. 
These idealizations are needed for obtaining governing equations solvable by semi-analytical methods.
Results are thought adequate for demonstrating utility of the theory and suggesting physical trends.
Calculations can be extended to account for elastic anisotropy from multiple fiber families and for finite volume changes from large pressures and local damage processes (e.g., dilatation from cavitation).
A more sophisticated treatment would introduce distinguished state vectors $\bm{D}_\alpha$
for distinct fiber families such as muscle fibers (i.e., contractile cells) and transverse collagen fibers in the myocardium or multiple, dispersed in- and out-of-plane collagen fiber families for skin tissue.

%
\appendix

\setcounter{figure}{0} \renewcommand{\thefigure}{A.\arabic{figure}}
\setcounter{table}{0} \renewcommand{\thetable}{A.\arabic{table}}
\setcounter{section}{0} \renewcommand{\thesection}{A.\arabic{section}}
\renewcommand{\theequation}{A.\arabic{equation}}

\section*{Appendix A: Rund's theorem extended to multiple fiber families}

Denote by $\mathcal{M}$ a manifold of dimension $n$, with boundary $\partial \! \mathcal{M}$ of class $C^1$,
the latter a positively oriented hypersurface of $\dim \partial \! \mathcal{M} = n -1$.
Stokes' theorem in terms of a class $C^1$ differentiable $(n-1)$ form $\pmb{\alpha}$ on $\mathcal{M}$ is (e.g., \cite{rund1975,marsden1983})
\begin{equation}
\label{eq:stokesforms}
\int_{\mathcal{M}}  d \pmb{\alpha} = \int_{\partial \mathcal{M}} \pmb{\alpha}.
\end{equation}
\noindent {\bf Theorem A.1.}
Let $\mathcal{M}$, $\dim \mathcal{M} = n$, be the base manifold of a generalized Finsler bundle of total space $\mathcal{Z}$ having
positively oriented boundary $\partial \! \mathcal{M}$ of class $C^1$ and $\dim \partial \! \mathcal{M} = n-1$. 
Denote by $\pmb{\alpha}(X,D_\alpha) = V^A(X,D_\alpha) N_A(X,D_\alpha) \Omega(X,D_\alpha)$ a differentiable $(n-1)$-form, and let $V^A$ be contravariant
components of vector field $\bm{V} = V^A \frac{\delta}{\delta X^A} \in H \mathcal{Z}$.  
The field of covariant components of the metric tensor referred to the horizontal distribution is $G_{AB} (X,D_\alpha)$ with $G= \det (G_{AB}) > 0$.
Assign a symmetric horizontal linear connection $H^A_{BC} = H^A_{CB}$ such
that $(\sqrt{G})_{|A} = 0$.  Presume that $C^1$ functional relations $D_\alpha = D_\alpha (X)$ exist for representation of vertical fiber coordinates
at each $X \in \mathcal{M}$ and for each fiber family $\alpha = 1, \ldots, r$.
Introduce nonlinear connection coefficients $N\ua^A_B$ and Cartan's coefficients $C^{\alpha A}_{BC}$ for each $\alpha$.
Then in a coordinate chart $\{ X^A,D^A_\alpha \}$, \eqref{eq:stokesforms} is explicitly, with volume and area forms defined in \eqref{eq:volforms} and \eqref{eq:areaform} and $C^{\alpha A}_{BC}$ in \eqref{eq:Cartan1},
\begin{equation}
\label{eq:stokes}
\int_{\mathcal{M}} \bigr{[} V^A_{|A} + \sum_{\alpha = 1}^r (V^A C^{\alpha C}_{BC} + \bar{\partial}^\alpha_B V^A) D^B_{\alpha; \, A} \bigr{]}  \, d\Omega = \int_{\partial \!\mathcal{M}} V^A N_A \,\Omega ,
\end{equation}
where $N_A$ is the unit outward normal on $\partial \! \mathcal{M}$, $V^A_{|A} = \delta_A V^A + V^A H^B_{BA}$,  and $D^B_{\alpha; \, A} = \partial_A D^B_\alpha + N\ua^B_A$.

A proof of a divergence theorem in Ref.~\cite{claytonMMS2022}, implied but not explicitly derived in Ref.~\cite{claytonJGP2017}, extended that of Rund \cite{rund1975} to a generalized Finsler space with arbitrary positive-definite metric $G_{AB}(X,D)$ and arbitrary nonlinear connection $N^A_B(X,D)$. All former theorems and proofs only addressed the case of a single fiber family, that is, $r = 1$, per the usual setting of Finsler geometry. Expression \eqref{eq:stokes} advances the former theorems \cite{rund1975,claytonMMS2022} to fibers $\alpha = 1, \ldots ,r$ where $r \geq 1$.
\\

\noindent {\bf Proof.}
The integrand on the right of \eqref{eq:stokes} (i.e., the  $(n-1)$-form in \eqref{eq:stokesforms}), is
transformed as \cite{rund1975}
\begin{equation}
\label{eq:aformp}
\begin{split}
\pmb{\alpha} & = \langle \bm{V},\bm{N} \rangle \, \Omega  = V^A N_A 
\sqrt{B} \, d Y^1 \wedge \ldots \wedge d Y^{n-1} = V^A  \delta^j_A p_j \sqrt{G}  \, d Y^1 \wedge \ldots \wedge d Y^{n-1} \\ & =
\sum_{j=1}^n (-1)^{j-1} V^A \delta^j_A \sqrt{G} \, d X^1 \ldots \wedge d X^{j-1} \wedge d X^{j+1} \ldots \wedge d X^n,
\end{split}
\end{equation}
where $p_j = p \delta_j^A N_A = (-1)^{j+1} \frac{\partial (X^1, \ldots ,X^{j-1},X^{j+1},\ldots,X^n)}{\partial (Y^1,\ldots, Y^{n-1})}$ is an outward normal to $\mathcal{M}$ with magnitude $p = | G^{AB} \delta^i_A \delta^j_B p_i p_j|^{1/2}$ and direction $\bm{N} = N_A \, d X^A$ when referred to the same basis as $\bm{V}$.
The exterior derivative of \eqref{eq:aformp} is, given existence of $C^1$ functions $D_\alpha=D_\alpha (X) \, \forall \, X \in \mathcal{M}$,
\begin{equation}
\label{eq:dalpha}
\begin{split}
d \pmb{\alpha} & =  \sum_{j=1}^n (-1)^{j-1} d \{ V^A \delta^j_A \sqrt{G} \} \wedge d X^1 \ldots \wedge d X^{j-1} \wedge d X^{j+1} \ldots \wedge d X^n 
\\  = &  \sum_{j=1}^n (-1)^{j-1} \delta^j_A \{ \partial_B ( V^A  \sqrt{G})  + \sum_{\alpha=1}^r \bar{\partial}^\alpha_C ( V^A  \sqrt{G}) \partial_B D^C_\alpha \} dX^B \wedge 
d X^1 \ldots \wedge d X^{j-1} \wedge d X^{j+1} \ldots \wedge d X^n
\\  = &  \sum_{j=1}^n (-1)^{j-1} \delta^j_A \{ \partial_B ( V^A  \sqrt{G})  + \sum_{\alpha=1}^r \bar{\partial}^\alpha_C ( V^A  \sqrt{G}) \partial_B D^C_\alpha \} (-1)^{j-1}
\delta^B_j d X^1 \wedge \ldots \wedge d X^n
\\  = &    \{ \partial_A ( V^A  \sqrt{G})  + \sum_{\alpha=1}^r \bar{\partial}^\alpha_B ( V^A  \sqrt{G}) \partial_A D^B_\alpha \} 
 d X^1 \wedge \ldots \wedge d X^n.
\end{split}
\end{equation}
Using \eqref{eq:diffnot}, \eqref{eq:Gids}, and the stated condition $(\sqrt{G})_{|A} = 0$, 
\begin{equation}
\label{eq:Rundfin1}
\begin{split}
\partial_A ( V^A  \sqrt{G}) & = \sqrt{G} [ \partial_A V^A + V^A  ( H^B_{AB}  + \sum_{\alpha=1}^r N\ua^B_A C^{\alpha C}_{BC})]  
\\ & = \sqrt{G} [V^A_{|A} + \sum_{\alpha=1}^r N\ua^B_A \bar{\partial}^\alpha_B V^A + V^A  ( \{H^B_{AB} - H^B_{BA}\}  + \sum_{\alpha=1}^r N\ua^B_A C^{\alpha C}_{BC})].
\end{split}
\end{equation}
From the second of \eqref{eq:Gids2}, 
\begin{equation}
\label{eq:Rundfin2}
\begin{split}
\sum_{\alpha = 1}^r \bar{\partial}^\alpha_B ( V^A  \sqrt{G}) \partial_A D^B_\alpha  & = 
\sqrt{G} \sum_{\alpha = 1}^r
[ \bar{\partial}^\alpha_B V^A + V^A  C^{\alpha C}_{BC} ] \partial_A D^B_\alpha. 
\end{split}
\end{equation}
With stated condition $H^A_{BC} = H^A_{CB}$, substituting the sum of \eqref{eq:Rundfin1} and \eqref{eq:Rundfin2} into \eqref{eq:dalpha} produces
\begin{equation}
\label{eq:Rundfin3}
d \pmb{\alpha} = \bigr{[} V^A_{|A} + \sum_{\alpha = 1}^r (V^A C^{\alpha C}_{BC} + \bar{\partial}^\alpha_B V^A) D^B_{\alpha ; \, A} \bigr{]} \sqrt{G}  d X^1 \wedge \ldots \wedge d X^n =
V^A_{||A} d \Omega,
\end{equation}
where condensed notation extending Ref.~\cite{claytonIJF2017} is $ (\cdot)_{||A} = (\cdot)_{|A} + \sum_{\alpha = 1}^r [(\cdot) C^{\alpha C}_{BC} + \bar{\partial}^\alpha_B (\cdot)] D^B_{\alpha ; \,A}$ . 
Result \eqref{eq:Rundfin3} is verified as the integrand on the left in \eqref{eq:stokesforms} and \eqref{eq:stokes}, completing the proof.$\qquad \square$
\\

Under stipulations of Stokes' theorem \eqref{eq:stokesforms}, \eqref{eq:stokes} holds if $\mathcal{M}$ and $\partial \! \mathcal{M}$ are replaced with any compact region of
$\mathcal{M}' \subset \mathcal{M}$ and positively oriented boundary of that region.
A different basis and its dual over $\mathcal{M}$ could be used for $\bm{V}$ and $\bm{N}$
if $H^A_{BC}$ and $(\cdot)_{|B}$ are treated as the same operators in \eqref{eq:Rundfin1}--\eqref{eq:Rundfin3} with requisite symmetry and $G$-compatibility.
Chern-Rund-Cartan horizontal coefficients $H^A_{BC} \rightarrow \Gamma^A_{BC}$ via \eqref{eq:CR1} uniquely fulfill symmetry and metric-compatibility requirements.

Geometric interpretation of covariant differentiation on the left of \eqref{eq:stokes}
suggests $\{ \frac{\delta}{\delta X^A} \} $ be used for $\bm{V}$, by which dual basis $\{ d X^B \} $ should be used for 
$\bm{N}$ to ensure invariance: $\langle \bm{V}, \bm{N} \rangle \rightarrow V^A N_B \langle \frac{\delta}{\delta X^A}, dX^B  \rangle$.
If $\bm{V}$ is referred to the holonomic basis $ \{ \frac{\partial}{\partial X^A} \} $ instead,
then $N\ua^A_B = 0$ should be imposed for invariance of its inner product with $N_B d X^B$.

When functions $D_\alpha = D_\alpha (X)$ explicitly exist, metric components can be treated as absolute functions of $X$.  In this setting, the generalized Finsler metric with components $G_{AB}(X,D_\alpha(X))$ can be reclassified as a locally osculating Riemannian metric \cite{rund1959,amari1962} written as $\tilde{G}_{AB}(X)$.  Equations of generalized Finsler geometry remain valid, but more relationships emerge \cite{claytonSYMM2023}, for example:
\\

\noindent{\bf Corollary A.1.} Given $C^1$ functions $D_\alpha = D_\alpha(X)$ with $\alpha = 1,\ldots,r$, let $\tilde{G}_{AB}(X) = G_{AB}(X,D_\alpha(X))$ be components of the osculating Riemannian metric derived from $\bm{G} = G_{AB} \, dX^A \otimes dX^B$. Then \eqref{eq:stokes} is equivalent to 
\begin{equation}
\label{eq:stokeso}
\int_{\mathcal{M}} \tilde{V}^A_{: \, A} \, d\Omega = \int_{\partial \! \mathcal{M}} \tilde{V}^A \tilde{N}_A \,\Omega,
\end{equation}
where vector $\tilde{V}^A(X) = V^A(X,D_\alpha (X))$, unit normal $\tilde{N}_A(X) = N_A(X,D_\alpha (X))$, and covariant derivative 
$\tilde{V}^A_{: \, A} = \partial_A \tilde{V}^A + \tilde{\gamma}^B_{BA} \tilde{V}^A$ with connection 
 $\tilde{\gamma}^B_{BA} (X) = \partial_A (\ln \sqrt {\tilde{G}(X)}) =  \tilde{\gamma}^B_{AB} (X)$
and $\tilde{G} = \det (\tilde{G}_{AB})$.
\\

\noindent{\bf Proof.} From the prescribed change of independent arguments, the right of \eqref{eq:stokeso} matches the right of \eqref{eq:stokes}.
Differentiation in the integrand on the left of \eqref{eq:stokeso}, with vanishing \eqref{eq:Gids}, and  \eqref{eq:Gids2}, is
\begin{equation}
\label{eq:pfnew}
\partial_A \tilde{V}^A = \partial_A V^A + \sum_{\alpha = 1}^r \bar{\partial}^\alpha_B V^A \partial_A D^B_\alpha ,  
\end{equation}
\begin{equation}
\label{eq:pfnew2}
\begin{split}
\tilde{V}^A \tilde{\gamma}^B_{BA}  = \tilde{V}^A \partial_A ( \ln \sqrt{\tilde{G}} ) &  = 
V^A [ \partial_A ( \ln \sqrt{G}) + \sum_{\alpha = 1}^r \bar{\partial}^\alpha_B ( \ln \sqrt{G}) \partial_A D^B_\alpha ] \\ & 
= V^A [ \delta_A ( \ln \sqrt{G}) + \sum_{\alpha = 1}^r C^{\alpha C}_{BC} (N\ua^B_A +  \partial_A D^B_\alpha)] 
 \\ &
 =V^A [H^B_{AB} + \sum_{\alpha = 1}^r C^{\alpha C}_{BC} D^B_{\alpha ; \, A} ]
 = V^A [ H^B_{BA} + \sum_{\alpha = 1}^r C^{\alpha C}_{BC} D^B_{\alpha ; \, A} ].
\end{split}
\end{equation}
Summing \eqref{eq:pfnew} and \eqref{eq:pfnew2} and inserting $\pm \sum_{\alpha = 1}^r N\ua^B_A  \bar{\partial}^\alpha_B V^A$ terms gives
\begin{equation}
\label{eq:pfnew3}
\begin{split}
\tilde{V}^A_{: \, A} & = 
\{ \partial_A V^A +  \sum_{\alpha = 1}^r  ( \bar{\partial}^\alpha_B V^A \partial_A D^B_\alpha - N\ua^B_A  \bar{\partial}^\alpha_B V^A ) \} + 
\{  \sum_{\alpha = 1}^r  N\ua^B_A  \bar{\partial}^\alpha_B V^A + V^A [ H^B_{BA} +  \sum_{\alpha = 1}^r  C^{\alpha C}_{BC} D^B_{\alpha ; \, A} ] \} \\
& = \delta_A V^A +  V^A H^B_{BA}  + \sum_{\alpha = 1}^r [ \bar{\partial}^\alpha _B V^A (\partial_A D^B_\alpha  + N\ua^B_A) 
+ V^A C^{\alpha C}_{BC} D^B_{\alpha ; \, A} ]\\
& = V^A_{|A} + \sum_{\alpha = 1}^r (\bar{\partial}^\alpha_B V^A +  V^A C^{\alpha C}_{BC})  D^B_{ \alpha ; \, A}.
\end{split}
\end{equation}
Comparing left sides of \eqref{eq:stokes} and \eqref{eq:stokeso}, integrands are verified to match,
completing the proof.$\quad \square$
\\

From Riemannian geometry, the Levi-Civita connection of $\tilde{G}_{AB}(X)$, written as $\tilde{\gamma}^A_{BC}(X)$, uniquely fulfills the symmetry and metric-compatibility requirements used to prove \eqref{eq:stokeso}:
 \begin{equation}
\label{eq:LC1o}
\tilde{\gamma}^A_{BC}={\textstyle{\frac{1}{2}}} \tilde{G}^{AD} (\partial_C \tilde{G}_{BD} + \partial_B \tilde{G}_{CD} - \partial_D \tilde{G}_{BC})
=\tilde{G}^{AD} \tilde{\gamma}_{BCD} = \tilde{\gamma}^A_{CB}.
\end{equation}
From \eqref{eq:LC1o}, \eqref{eq:stokeso} is analogous to that
for a Riemannian manifold with boundary.
It is not identical since non-holonomic basis  $ \{ \frac{\delta}{\delta X^A} \} $ is used for $\bm{V}$. 
The holonomic basis $\{ \frac{\partial}{\partial X^A} \} $ could be used in a preferred chart $\{ X, D_\alpha(X) \}$ constrained by $N\ua^A_B = 0$, whereby the distinction would vanish.


%
\appendix

\setcounter{figure}{0} \renewcommand{\thefigure}{B.\arabic{figure}}
\setcounter{table}{0} \renewcommand{\thetable}{B.\arabic{table}}
\setcounter{section}{0} \renewcommand{\thesection}{B.\arabic{section}}
\renewcommand{\theequation}{B.\arabic{equation}}
\setcounter{equation}{0}

\section*{Appendix B: Variational derivatives}
Variational derivative $\delta(\cdot)$
implements $(\varphi^a,D^A_\alpha)$ with $a,A = 1, \ldots, n$ and $\alpha = 1, \ldots, r$ the total set of $n(1+r)$ varied independent parameters or degrees-of-freedom\footnote{Terms ``variational derivative'', ``first variation'', and ``variation'' of $f$ are used here interchangeably for the variation $\delta f$ of a function $f(\cdot)$.
This variational derivative need not be a ``functional derivative'' of a functional $F[\cdot]$ classically defined in an integral sense.}. Variations of parameters are written $\delta \varphi^a(X)$ and $\delta (D^A_\alpha)(X)$, the latter in parentheses to delineate from non-holonomic bases $\delta D^A_\alpha$ of \eqref{eq:nonhol}. Material particle $X$ is held fixed during variational differentiation. 
Specifically, let $\hat{\varphi} = \hat{\varphi}(X,\varepsilon)$ be a one-parameter family of motions when $X$ is fixed,
where $\varepsilon$ is a (small) parameter controlling the magnitude of the variation.
Let $\hat{D}_\alpha = \hat{D}_\alpha (X,\varepsilon)$ be a one-parameter family of directors $\alpha$ when $X$ is constant. Per classical definitions (e.g., Ref.~\cite{sokolnikoff1951}),
\begin{equation}
\label{eq:varphivarD}
\delta \varphi^a (X) = \varepsilon \frac{\partial \hat{\varphi}^a}{\partial \varepsilon} \biggr{\rvert}_{\varepsilon = 0} (X), \qquad
\delta (D^A_\alpha) (X) = \varepsilon \frac{\partial \hat{D}^A_\alpha}{\partial \varepsilon} \biggr{\rvert}_{\varepsilon = 0} (X).
\end{equation}

Let $f(x,X,D_\alpha)$ be generic differentiable function, for example of scalar or tensor character, of arguments 
$\{x^a,X^A,D^A_\alpha\}$
in coordinate charts on $\mathfrak{z}$ and $\mathcal{Z}$. The first 
variation of $f(x,X,D_\alpha)$ is defined as follows, extending concepts from the Riemannian setting~\cite{sokolnikoff1951} to a Finslerian one, and dropping functional arguments on the right for brevity:
\begin{equation}
\label{eq:varderiv}
\delta f(x,X,D_\alpha) = (\nabla_{\delta / \delta x^a} f)|_X \delta \varphi^a 
+ \sum_{\alpha = 1}^r ( \nabla_{\partial / \partial D^A_\alpha} f) \delta (D^A_\alpha)
= (f_{|a})_X  \delta \varphi^a +  \sum_{\alpha = 1}^r (f |^\alpha_A) \delta (D^A_\alpha).
\end{equation}
For the choices $V^A_{BC} = V^a_{bc} = 0$ and $Y^{\alpha A}_{BC} = Y^{\alpha a}_{bc} = 0$ of \eqref{eq:connrec}, 
$f |^\alpha_A = \bar{\partial}^\alpha_A f$ such that \eqref{eq:varderiv} reduces to
\begin{equation}
\label{eq:varderivsimp}
\delta f(x,X,D_\alpha) = (f_{|a})_X  \delta \varphi^a +  \sum_{\alpha = 1}^r (\bar{\partial}^\alpha_A f) \delta (D^A_\alpha).
\end{equation}

Consider deformation gradient $\bm{F} = F^a_A \frac{\delta}{\delta x^a} \otimes d X^A $ of \eqref{eq:defgradC}.
Components $F^a_A (X) =  \delta_A \varphi^a(X) = \partial_A \varphi^a (X) $ depend only on $X$, as
do components of the mapping $\varphi^a(X)$ in \eqref{eq:varphiC}.
However, bases $\{ \frac{\delta}{\delta x^a} \}$ most generally depend on $(x,d_\alpha)$, and $d X^A$ most generally depend on $(X,D_\alpha)$. Application of \eqref{eq:horiztranspf} and assumptions in \eqref{eq:varderivsimp}, with $H^a_{bc} = \Gamma^a_{bc}$ of \eqref{eq:connrec}, produces $\delta (dX^A) = 0$ and
\begin{equation}
\label{eq:varF}
\begin{split}
\delta {\bm F} & = \delta (\delta_A \varphi^a) \frac{\delta}{\delta x^a} \otimes d X^A + F^a_A \delta \bigr{(} \frac{\delta}{\delta x^a} \bigr{)} \otimes dX^A 
= \delta (\partial_A \varphi^a) \frac{\delta}{\delta x^a} \otimes d X^A 
+ F^a_A \delta \varphi^b \nabla_{\delta / \delta x^b}  \frac{\delta}{\delta x^a}   \otimes d X^A 
\\ 
& = \partial_A (\delta \varphi^a) \frac{\delta}{\delta x^a} \otimes d X^A
+ F^a_A \delta \varphi^b H^c_{ba}  \frac{\delta}{\delta x^c} \otimes d X^A
= \delta_A (\delta \varphi^a) \frac{\delta}{\delta x^a} \otimes d X^A + 
 F^c_A \delta \varphi^b \Gamma^a_{bc}  \frac{\delta}{\delta x^a} \otimes d X^A
 \\ 
 & = F^b_A  [ \delta_b (\delta \varphi^a) + \Gamma^a_{cb} \delta \varphi^c  ] \frac{\delta}{\delta x^a} \otimes d X^A
 = F^b_A (\delta \varphi^a)_{|b} \frac{\delta}{\delta x^a} \otimes d X^A
 = (\delta \varphi^a)_{|A}  \frac{\delta}{\delta x^a} \otimes d X^A.
 \end{split}
\end{equation}
In coordinates, the expression $\delta F^a_A = (\delta \varphi^a)_{|A}$ is used.
The derivation of $\delta F^a_A$ in Ref.~\cite{claytonSYMM2023} implicitly used in prior works \cite{claytonJGP2017,claytonMMS2022} defines 
$ (\delta \varphi^a)_{|A} = \delta_A (\delta \varphi^a)$, treating $\delta \varphi^a$ as a nonlinear mapping rather than components of a spatial vector field. 
The end result of \eqref{eq:varF} or eq.~(A1) of Ref.~\cite{claytonSYMM2023} used to derive Euler-Lagrange equations is the same.
The derivation in \eqref{eq:varF} is similar to Refs.~\cite{yavari2010,kumar2023}.

Next consider the covariant derivative of the director field $\bm{D}_\alpha$, where with $V^A_{BC} = 0$ from \eqref{eq:connrec},
\begin{equation}
\label{eq:gradD}
\nabla \bm{D}_\alpha  = D^A_{\alpha |B} \frac{\partial}{\partial D^A_\alpha} \otimes d X^B + 
\sum_{\beta =1}^r
D^A_\alpha|^\beta_B 
\frac{\partial}{\partial D^A_\alpha}  \otimes \delta D^B_\beta;
\end{equation}
\begin{equation}
\label{eq:gradDi}
\begin{split}
 D^A_{\alpha |B} & = \delta_B D^A_\alpha + K^{\alpha A}_{BC} D^C_\alpha = 
 \partial_B D^A_\alpha - \sum_{\beta = 1}^r N^\beta{}_B^C \bar{\partial}_C^\beta D^A_\alpha  + K^{\alpha A}_{BC} D^C_\alpha
 =  \partial_B D^A_\alpha -  N\ua_B^A  + K^{\alpha A}_{BC} D^C_\alpha,
\\
 D^A_\alpha |^\beta_B & = \bar{\partial}^\beta_B D^A_\alpha+ \delta^\beta_\alpha V ^A_{BC} D^C_\alpha  = 
  \delta^\beta_\alpha \delta^A_B.
 \end{split}
 \end{equation}
 Again invoking \eqref{eq:connrec} in \eqref{eq:varderiv}, $\delta (dX^A) = 0$, $\delta( \frac{\partial}{\partial D^A_\alpha}) =0$,
 and $\delta( \delta D^A_\alpha) =0$. Then since $\delta(\delta^\beta_\alpha \delta^A_B) = 0$, the variational derivative of \eqref{eq:gradD}, $\delta (\nabla \bm{D}_\alpha)$, is limited to the contribution from its horizontal part:
 \begin{equation}
 \label{eq:vargradD0}
 \delta (\nabla \bm{D}_\alpha) = \delta(D^A_{\alpha | B}) \frac{\partial}{\partial D^A_\alpha} \otimes d X^B
 = \delta D^A_{\alpha | B} \frac{\partial}{\partial D^A_\alpha} \otimes d X^B,
 \end{equation}
 \begin{equation}
 \label{eq:vargradD}
\begin{split}
\delta D^A_{\alpha |B} & = \delta (\partial_B D^A_\alpha) - \delta N\ua^A_B + \delta (K^{\alpha A}_{BC}) D^C_\alpha + K^{\alpha A}_{BC} \delta (D^C_\alpha) \\ & =
[\partial_B  \delta (D^A_\alpha) - \sum_{\beta=1}^r N^\beta{}^C_B \bar{\partial}^\beta_C \delta (D^A_\alpha) + K^{\alpha A}_{BC} \delta (D^C_\alpha)]
- \sum_{\beta=1}^r [ \bar{\partial}^\beta_C N\ua^A_B \delta (D^C_\beta) 
 - \bar{\partial}^\beta_D K^{\alpha A}_{BC} D^C_\alpha \delta(D^D_\beta) ] \\
& = [\delta(D^A_\alpha)]_{|B} - [ \bar{\partial}^\alpha_C N\ua^A_B - \bar{\partial}^\alpha_C K^{\alpha A}_{BD} D^D_\alpha ] \delta(D^C_\alpha).
\end{split}
\end{equation}
In \eqref{eq:vargradD}, it is assumed per \eqref{eq:funcC} that $\bar{\partial}^\beta _C \delta[D^A_\alpha(X)] = \bar{\partial}^\beta_C [\delta(D^A_\alpha) (X)] = 0$, and $N\ua^A_B (X,D_\alpha)$ and $ K^{\alpha A}_{BC} (X,D_\alpha)$ are independent of $D_\beta$ for $\beta \neq \alpha$, consistent with rules for connections in \S2.1.

Lastly, consider the volume form $d \Omega (X,D_\alpha)$. Two definitions are conceived for its variational derivative.
The first generalizes the $r = 1$ case in Refs.~\cite{claytonCMT2018,claytonMMS2022,claytonSYMM2023}, where from
\eqref{eq:varderiv} then \eqref{eq:varderivsimp},
\begin{equation}
\label{eq:omegavar1}
\begin{split}
\delta (d \Omega) &  = [{\delta{\sqrt G}}/{\sqrt{G}}] d \Omega = \sum_{\alpha = 1}^r (\ln \sqrt{G} )|^\alpha_A \delta (D^A_\alpha) d \Omega =
\sum_{\alpha = 1}^r  {\textstyle{\frac{1}{2}}} G^{BC} G_{CB}|^\alpha_A  \delta (D^A_\alpha) d \Omega \\ & = 
\sum_{\alpha = 1}^r  (C^{\alpha B}_{AB} - Y^{\alpha B}_{AB})  \delta (D^A_\alpha) d \Omega = 
\sum_{\alpha = 1}^r C^{\alpha B}_{AB} \delta (D^A_\alpha) d \Omega,
\end{split}
\end{equation}
where the first equality is a definition and \eqref{eq:Gids2} is used subsequently.
According to \eqref{eq:omegavar1}, the referential volume form is varied according to the local variation of the metric over the base manifold $\mathcal{M}$ of dimension $n$.
The second definition advances the $r =1$ case of Refs.~\cite{claytonTR2016,claytonJGP2017}, where \eqref{eq:detGS} and \eqref{eq:Gids2} are used with \eqref{eq:varderiv} then \eqref{eq:varderivsimp}:
\begin{equation}
\label{eq:omegavar2}
\begin{split}
\delta (d \Omega) &  = [{\delta{\sqrt \mathcal{G}}}/{\sqrt{\mathcal{G}}}] d \Omega = \sum_{\alpha = 1}^r (\ln \sqrt{ \mathcal{G}} )|^\alpha_A \delta (D^A_\alpha) d \Omega = \sum_{\alpha = 1}^r (\ln \sqrt{ G^{1+r} } )|^\alpha_A \delta (D^A_\alpha) d \Omega \\ & = 
(1+r) \sum_{\alpha = 1}^r (\ln \sqrt{ G } )|^\alpha_A \delta (D^A_\alpha) d \Omega
 = 
\frac{1+r}{2}\sum_{\alpha = 1}^r G^{BC} G_{CB}|^\alpha_A  \delta (D^A_\alpha) d \Omega
\\ &  =
(1+r) \sum_{\alpha = 1}^r  (C^{\alpha B}_{AB} - Y^{\alpha B}_{AB})  \delta (D^A_\alpha) d \Omega 
 = (1+r) \sum_{\alpha = 1}^r  C^{\alpha B}_{AB} \delta (D^A_\alpha) d \Omega.
\end{split}
\end{equation}
The definition of the variational derivative of the volume form in the first equality in \eqref{eq:omegavar2} is notionally consistent with Refs.~\cite{saczuk1996,saczuk1997,stumpf2000,saczuk2001,stumpf2002,saczuk2003}. In the setting of \eqref{eq:omegavar2}, the magnitude of the volume form is varied according to the variation of the metric in $[(1+r) \times n]$-dimensional total space $\mathcal{Z}$.


\bibliography{refs}

\end{document}